\documentclass[10pt]{article}
\pdfoutput=1
\usepackage{authblk}
\usepackage{amsmath,bbm}
\usepackage{multirow}
\usepackage{amssymb}
\usepackage{slashed}
\usepackage{soul}
\usepackage[normalem]{ulem}
\usepackage{graphicx} 
\usepackage{mathrsfs}
\usepackage{mathtools}
\usepackage[dvipsnames]{xcolor}
\usepackage{xcolor}
\usepackage{hyperref}
\usepackage{upgreek}
\usepackage{cite}
\usepackage{lineno}

\makeatletter
\let\@fnsymbol\@arabic
\makeatother

\graphicspath{{figures/}}

\title{Unveiling Hidden Physics at the LHC}
\date{}

\author[$\dagger$,1]{Oliver Fischer}
\author[$\dagger$,2,3]{Bruce Mellado}

\author[4]{\\Stefan Antusch}
\author[5]{Emanuele Bagnaschi}
\author[6]{Shankha Banerjee}
\author[2]{Geoff Beck}
\author[7,8]{Benedetta Belfatto}
\author[9]{Matthew Bellis}
\author[10,11]{Zurab Berezhiani}
\author[12,13]{Monika Blanke}
\author[14,15]{Bernat Capdevila}
\author[16]{Kingman Cheung}
\author[5,6,17] {Andreas Crivellin}
\author[18]{Nishita Desai}
\author[19]{Bhupal Dev}
\author[20]{Rohini Godbole}
\author[21]{Tao Han}
\author[22, 23]{Philip Harris}
\author[24]{Martin Hoferichter}
\author[25,26]{Matthew Kirk}
\author[27]{Suchita Kulkarni}
\author[28]{Clemens Lange}
\author[29]{Kati Lassila-Perini}
\author[30]{Zhen Liu}
\author[6,31]{Farvah Mahmoudi}
\author[5,17]{Claudio Andrea Manzari}
\author[32]{David Marzocca}
\author[33]{Biswarup Mukhopadhyaya}
\author[34]{Antonio Pich}
\author[2]{Xifeng Ruan}
\author[35, 36]{Luc Schnell}
\author[22, 23]{Jesse Thaler}
\author[37]{Susanne Westhoff}

\affil[1]{Department of Mathematical Sciences, University of Liverpool, Liverpool, L69 7ZL, UK}
\affil[2]{School of Physics and Institute for Collider Particle Physics, University of the Witwatersrand, Johannesburg, Wits 2050, South Africa.}
\affil[3]{iThemba LABS, National Research Foundation, PO Box 722, Somerset West 7129, South Africa.}
\affil[4]{Department of Physics, University of Basel, Klingelbergstr.~82, CH-4056 Basel, Switzerland}
\affil[5]{Paul Scherrer Institut, CH--5232 Villigen PSI, Switzerland}
\affil[6]{CERN Theory Division, CH--1211 Geneva 23, Switzerland}
\affil[7]{Dipartimendto di Fisica``E. Fermi", Universit\`a di Pisa, Largo Bruno Pontecorvo 3, I-56127 Pisa, Italy}
\affil[8]{INFN Sezione di Pisa, Largo Bruno Pontecorvo 3, I-56127 Pisa, Italy}
\affil[9]{Siena College, 515 Loudon Road, Loudonville, NY 12211-1462, United States}
\affil[10]{Dipartimendto di Fisica e Chimica, Universit\`a di L'Aquila, 67100 Coppito, L'Aquila, Italy}
\affil[11]{INFN, Laboratori Nazionali del Gran Sasso, 67100 Assergi, L'Aquila, Italy}
\affil[12]{Institute for Astroparticle Physics (IAP), Karlsruhe Institute of Technology, Hermann-von-Helmholtz-Platz 1, D-76344 Eggenstein-Leopoldshafen, Germany}
\affil[13]{Institute for Theoretical Particle Physics (TTP), Karlsruhe Institute of Technology, Engesserstrasse 7, D-76128 Karlsruhe, Germany}
\affil[14]{Dipartimento di Fisica, Universit\`a di Torino, Via P. Giuria 1, Torino I-10125, Italy}
\affil[15]{INFN Sezione di Torino, Via P. Giuria 1, Torino I-10125, Italy}
\affil[16]{Department of Physics, National Tsing Hua University, Hsinchu 300, Taiwan}
\affil[17]{Physik-Institut, Universit\"at Z\"urich, Winterthurerstrasse 190, CH--8057 Z\"urich, Switzerland}
\affil[18]{Tata Institute of Fundamental Research, 1 Homi Bhabha Road, Mumbai 400005, India}
\affil[19]{Department of Physics and McDonnell Center for the Space Sciences, Washington University, St. Louis, MO 63130, USA}
\affil[20]{Indian Institute of Science (IISc), Bangalore 560012, Karnataka, India}
\affil[21]{Pittsburgh Particle Physics, Astrophysics, and Cosmology Center, Department of Physics and Astronomy, University of Pittsburgh, Pittsburgh, PA 15206, USA}
\affil[22]{Center for Theoretical Physics, Massachusetts Institute of Technology, Cambridge, MA 02139, USA}
\affil[23]{The NSF AI Institute for Artificial Intelligence and Fundamental Interactions}
\affil[24]{Albert Einstein Center for Fundamental Physics, Institute for Theoretical Physics, University of Bern, Sidlerstrasse 5, 3012 Bern, Switzerland}
\affil[25]{Dipartimento di Fisica, Università di Roma ``La Sapienza'', Piazzale Aldo Moro 2, 00185 Roma, Italy}
\affil[26]{INFN Sezione di Roma, Piazzale Aldo Moro 2, 00185 Roma, Italy}
\affil[27]{Institute of Physics, NAWI Graz, University of Graz, Universit\"atsplatz 5, A-8010 Graz, Austria}
\affil[28]{Experimental Physics Department, CERN, Switzerland}
\affil[29]{Helsinki Institute of Physics (HIP), P.O. Box 64, 00014 University of Helsinki, Finland}
\affil[30]{School of Physics and Astronomy, University of Minnesota, Minneapolis, MN 55455, USA}
\affil[31]{Univ Lyon, Univ Lyon 1, CNRS/IN2P3, Institut de Physique des 2 Infinis de Lyon, UMR 5822, F-69622, Villeurbanne, France}
\affil[32]{INFN, Sezione di Trieste, SISSA, Via Bonomea 265, 34136, Trieste, Italy}
\affil[33]{Indian Institute of Science Education and Research, Kolkata, India}
\affil[34]{IFIC, Universitat de Val\`encia -- CSIC, E-46980 Paterna, Spain}
\affil[35]{Departement Physik, ETH Z\"urich, Otto-Stern-Weg 1, CH-8093 Z\"urich, Switzerland}
\affil[36]{D\'epartement de Physique, \'Ecole Polytechnique, Route de Saclay, FR-91128 Palaiseau Cedex, France}
\affil[37]{Institute for Theoretical Physics, Heidelberg University, 69120 Heidelberg, Germany}
\affil[$\dagger$]{{\bf Editors}}

\newcommand{\stau}{\tilde{\tau}}

\begin{document}

\maketitle

\newpage

\begin{abstract}

\noindent
The field of particle physics is at the crossroads. The discovery of a Higgs-like boson completed the Standard Model (SM), but the lacking observation of convincing resonances Beyond the SM (BSM) offers no guidance for the future of particle physics. On the other hand, the motivation for New Physics has not diminished and is, in fact, reinforced by several striking anomalous results in many experiments. Here we summarise the status of the most significant anomalies, including the most recent results for the flavour anomalies, the multi-lepton anomalies at the LHC, the Higgs-like excess at around 96 GeV, and anomalies in neutrino physics, astrophysics, cosmology, and cosmic rays.

While the LHC promises up to 4 ab$^{-1}$ of integrated luminosity and far-reaching physics programmes to unveil BSM physics, we consider the possibility that the latter could be tested with present data, but that systemic shortcomings of the experiments and their search strategies may preclude their discovery for several reasons, including: final states consisting in soft particles only, associated production processes, QCD-like final states, close-by SM resonances, and SUSY scenarios where no missing energy is produced.

New search strategies could help to unveil the hidden BSM signatures, devised by making use of the CERN open data as a new testing ground. We discuss the CERN open data with its policies, challenges, and
potential usefulness for the community. We showcase the example of the CMS collaboration, which is the only collaboration regularly releasing some of its data. We find it important to stress that individuals using public data for their own research does not imply competition with experimental efforts, but rather provides unique opportunities to give guidance for further BSM searches by the collaborations.

\begin{center}
	Wide access to open data is paramount to fully exploit the LHCs potential.
\end{center}

\end{abstract}

\newpage

\tableofcontents

\section{Introduction} 
{\it\small Section editors: Oliver Fischer and Bruce Mellado}

\medskip\noindent
The discovery of a scalar resonance that resembles the Higgs boson of the Standard Model (SM)~\cite{Higgs:1964ia,Englert:1964et,Higgs:1964pj,Guralnik:1964eu} at the Large Hadron Collider (LHC) by the
ATLAS~\cite{Aad:2012tfa} and CMS~\cite{Chatrchyan:2012ufa} collaborations has opened a new chapter in particle physics. 
The combined measurements show that that this discovered particle has properties that are compatible with those predicted by the SM~\cite{ATLAS:2016neq}, which makes its discovery a great triumph for experiment and theory.
Under the assumption that the discovered scalar particle is indeed the predicted Higgs boson, the SM has exhausted all its predictions pertaining to fundamental particles.
While the LHC collaborations continue to measure properties of the Higgs boson and other known particles and processes, the chief focus is now on the observation of new phenomena beyond the SM. 

The motivation for the existence of New Physics (NP) is no less than before the discovery of the Higgs boson.
The SM itself raises the question of naturalness, i.e.\ why the Electroweak scale is so much smaller than the Planck scale, which is addressed by Supersymmetry (SUSY) in an elegant way.
The observation of Dark Matter (DM) in the Universe, interpreted as a fundamental particle, can be addressed with minimal frameworks beyond the SM (BSM) and also with theories that often introduce an entire dark sector with new particles and forces. 
The observation of Neutrino Oscillations implies that the neutrinos are massive, which requires a mass-generating mechanism and therefore an extension of the SM. 
The above list of arguments is incomplete but suggests convincingly that the existence of NP is an established fact. 
No indication exists, however, in what form the NP manifests itself.

As the nature and energy scale of NP remains unknown, new phenomena could emerge at any experiment.
In recent years the field of particle physics has experienced a growing litany of anomalous experimental results. 
Many of them are statistically significant and continue to grow, and remain unexplained by state-of-the-art calculations based on the SM. 
In most cases the latter have become increasingly precise and reliable, where vast data sets have been used to provide extensive testing grounds. 
While some anomalies might eventually find explanations within the framework of the SM, persisting ones could contain hints of NP and may serve as a guide to model building and experimental searches.

While there is overwhelming evidence that the SM is incomplete, guidance is required to resolve the issue as to how the SM will conclusively breakdown in laboratory conditions.
Hundreds of BSM models have been proposed over the years, motivated by the big open questions as well as by the various combinations of experimental anomalies.
At the same time, extensive exploration of LHC data with respect to inclusive and model dependent signatures performed to date indicates that no striking resonances have been observed in the accessible dynamic range. 
The absence of clear BSM signatures in LHC data indicates that NP is either inaccessible at the LHC, or that it is driven by more subtle topologies and therefore hidden in the backgrounds.

The question arises what the absence of BSM resonances at the LHC implies for our search strategies.
Many models have been identified that are not captured by current searches, such that reinterpretation of experimental limits for different models became an important topic of discussion~\cite{LHCReinterpretationForum:2020xtr,Kraml:2012sg}, and analysis strategies were developed that are less model-dependent.
Since model building is the driving force behind gaining insight into new signatures, the model-centric and the model-independent approach are both necessary. The community is elaborating on data analysis methodologies that display less model dependencies. The use of Machine Learning may play a significant role.

It is high time to scrutinise existing LHC data with respect to clues on NP in the non-strongly interacting sector in order to prepare for the high-luminosity era and beyond.
When the High Luminosity LHC starts operating, the data coming out of it has to be used to glean every possible information about existing NP models.
Clear guiding principles need to come from a combination of both experimental and theoretical inquiries.
Given the present discussions on future colliders all over the world, this guidance is now more crucial than ever before.
Guidance for the future could come from NP that is currently hidden in LHC data, but might be accessible with new search strategies.

New search strategies can only be developed from the communication between theorists and experimentalists.
One remarkable example are the new means to search for hypothetical new long-lived particles, where discussions between experimentalists and theorists led to the development of new triggers, new external detectors, and influenced the planning of future experiments.
To stimulate similar discussions, a testing ground is needed, where physicists can develop new strategies with a quick turnaround.
Open data might constitute such a testing ground, as it provides a platform for knowledge and data exchange that has the potential to unleash the discovery potential of the LHC data, where experiments can be given pointers of what corners of the phase-space need particular scrutiny. 

CERN has committed to an open data policy in support of open science, promising to ``make scientific research more accessible to the community''. %
Open data is a relatively new experience in the field of particle physics.
As open data policies are not intended to compete with experimental efforts, it is essential to establish in this context a well-defined framework, within which knowledge and data are exchanged. 
In this community effort the experimentalists should remain the competent authority to set down the data access guidelines, while the role of theorists would be to provide directions with new insights and ideas.
The theorist's insight thus may keep the most crucial questions in front of the eyes of everyone involved in the effort. Of course, such questions are numerous and multi-dimensional.

In the workshop ``Unveiling Hidden Physics beyond the Standard Model at the LHC''\footnote{\url{https://indico.tlabs.ac.za/event/100/}} the existing motivation and insight into possible manifestations of NP were discussed.
This included a state-of-the-art overview of significant anomalous experimental results, lessons from theories including model classes that are challenging to detect at the LHC, computing methods, and the CERN open data.
The central aim of the workshop was to highlight the fact that the usage of open LHC data allows the community to test a much larger range of NP than ever before.

The present document includes a review of most significant anomalies in particle physics and a review of model classes that can be hidden in LHC data.
The anomalies are grouped appropriately and potential explanations are summarised, where possible interconnections are explored.
The review of hidden model classes is non-exhaustive and constitutes an example of the opportunities that model building continues to provide to the physics programme of the LHC.

\subsection*{Workshop discussion}
{\it\small Contributions: Nishita Desai, Biswarup Mukhopadhyaya}

\medskip\noindent 
The discussions at the workshop were held after individual talks, in dedicated discussion sessions, and a plenary discussion.
There was also some exchange on a dedicated mattermost channel.

The main points were collected and brought into this paper in many different places. 
Below is a list of few illustrative questions that emerged during the discussions in the opening session of the workshop, which may serve as rudders for future discussion.

\begin{itemize}
	\item Hundreds of models have been explored in our quest for new physics,
	and we have little guidelines yet as to how many additional directions
	are worthy of exploration. Till such guidelines emerge in firmer outlines, it may be advisable to carry on prediction and analysis of new physics signatures at the LHC in  model-independent ways as far as possible, so that we do not miss interesting possibilities due to any bias.
	
	At the same time, a virtue of model-based studies also looms up. The pros and cons of the theoretical characteristics of certain scenarios often become clear when their consequences are pitted against experimental data, and enrich us with wisdom that  goes beyond the ambit of those specific scenarios. Thus we derive our `lessons from theory' by occasionally resorting to models as well. 
	
	\item The 125-GeV scalar has been observed, and  whether it is `the Higgs' or `a Higgs' is still an open question. On the other hand, the electroweak symmetry-breaking sector has led to a good many questions about limitations of the standard model. 
	When the High Luminosity LHC starts operating, the data coming out of it should be used to glean every speck of additional information about this scalar, and look for effects that may serve to unveil physics beyond the standard model.
	
	\item Euclidean continuation is important in understanding global Higgs behaviour. In this context, both time-like and space-like probes at high energies should be complementary.
	
	$h \rightarrow ZZ$ would be a good start in this connection, because it interferes destructively with the SM box diagram. If the intervention of new physics makes on-shell $h \rightarrow ZZ$ small, it will enhance sensitivity 
	of off-shell Higgs signal. This is also applicable to di-Higgs production via triple Higgs coupling. 
	
	\item High-$p_T$ Higgs boson physics is complementary to the off-shell Higgs boson signal. Momentum transfer to 
	the Higgs boson production vertex in such events is space-like. So experiments should pay attention to 
	the relatively small number of events in the high-$p_T$ range, which may accentuate the
	role of off-shell Higgs boson, and any trace of BSM physics contained there.
	
	\item The running effect of $m_t(\mu)$, too, depends on features of the deep Euclidean region related to the top Yukawa coupling, although it is difficult to observe the effects at the LHC, because of the uncertainty
	in the measurement of top quark mass. This, however, can be taken up as a challenge at the high-luminosity
	runs, since (a) the top Yukawa coupling is related to the issue of naturalness, and (b) the precise relationship
	of the top pole mass with the running mass at some energy can reveal information of BSM contribution
	to the relevant renormalisation group equations.
	
	\item Since theoretical scenarios may exist just a little beyond the on-shell reach of the LHC, it is important to think not only about off-shell effects but also in terms of higher-dimensional effective operators which,
	after all, may turn out to be our major handles. The scale of such operators gets reflected in high-$p_T$ events which should therefore be probed with great emphasis during the high-luminosity runs. High $p_T$ events can also serve as probes of top physics in the deep Euclidean region.
	
	\item Observations on dark matter indicate potent new physics options. This includes, as major components, theoretical scenarios with symmetries such as $Z_2$, or those containing long-lived particles (either DM candidates themselves or others belonging to the dark sector). Since dark 
	matter is a concrete reality, such scenarios should constitute high-priority search areas.
	
	\item On a more theoretical note, the issue of naturalness of the electroweak scale is yet unresolved, especially when no evidence of supersymmetry is found yet, in regions of the parameter space with `sensible'  values of naturalness criteria. It is high time to investigate whether the high-luminosity data contain any clue on this in the non-strongly interacting sector.
\end{itemize}

\section{Anomalies}  
\label{sec:anomalies}
{\it\small Section editors: Andreas Crivellin, Oliver Fischer and Bruce Mellado\\
Contributions: Emanuele Bagnaschi, Geoffrey Beck, Benedetta Belfatto, Zurab Berezhiani, Monika Blanke, Bernat Capdevila, Bhupal Dev, Oliver Fischer, Martin Hoferichter, Matthew Kirk, Farvah Mahmoudi, Claudio Andrea Manzari,  David Marzocca, Bruce Mellado, Antonio Pich and Luc Schnell
}

\medskip\noindent
This section summarises a cohort of anomalies in the data that currently do not appear to be explained by the SM. See Ref.~\cite{anomalies} for an up-to-date summary of all existing anomalies. The section is structured as follows: Sec.~\ref{sec:flavoranomalies} gives an overview of flavour anomalies; Sec~\ref{sec:multilepton} details the multi-lepton anomalies at the LHC; Sec.~\ref{sec:96} describes the Higgs-like excess at 96\,GeV; Sec.~\ref{sec:neutrino} discusses anomalies in the neutrino sector; finally, Secs.~\ref{sec:astrophysics}, ~\ref{sec:cosmology} and~\ref{sec:cosmics} touch upon anomalies in astrophysics, cosmology and in Ultra-High energy cosmic rays, respectively.

\subsection{Flavour Anomalies} 
\label{sec:flavoranomalies}
\begin{figure}
\centering
\includegraphics[width=0.7\textwidth]{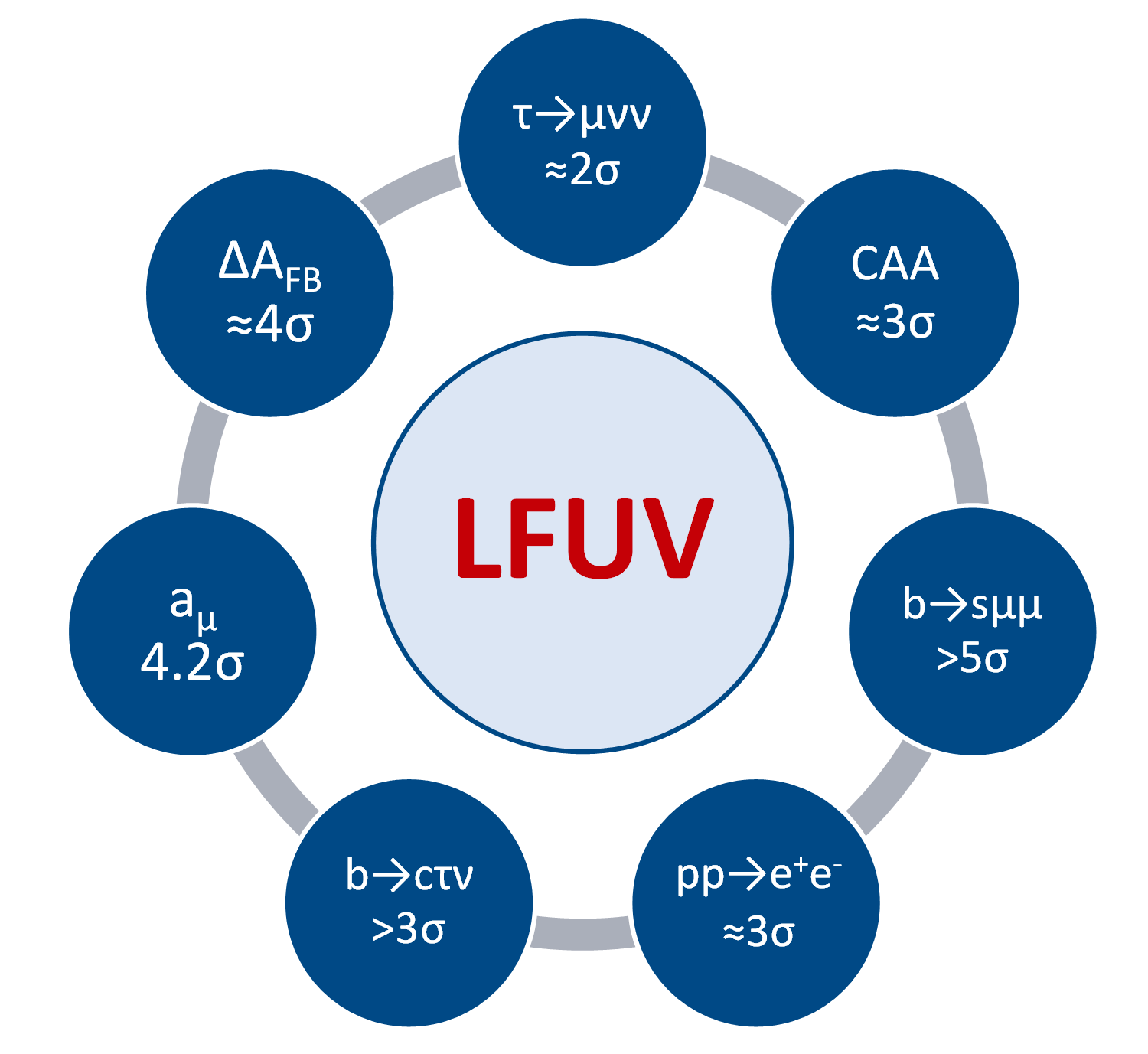}
  \caption{Summary of the experimental hints for LFUV beyond the SM.}
  \label{fig:LFUV}
\end{figure}

Intriguing indirect hints for BSM physics have been accumulated in flavour observables within recent years:
Semi-leptonic bottom quark decays ($b\to s\ell^+\ell^-$);
Tauonic $B$ meson decays ($b\to c\tau\nu$);
The anomalous magnetic moment of the muon ($a_\mu$);
The Cabibbo angle anomaly (CAA);
Non-resonant di-electrons ($q\bar q \to e^+e^-$);
The difference of the forward-backward asymmetry in $B\to D^*\mu\nu$ vs $B\to D^*e\nu$ ($\Delta A_{\rm FB}$);
Low-energy lepton flavour universality violation (LFUV) in the charged current, including leptonic tau decays ($\tau\to\mu\nu\nu$).
Interestingly, all these observables admit an interpretation in terms of LFUV, i.e., NP that distinguishes between muon, electrons and tau leptons. While some of the anomalies are by construction measures of LFUV, also the other observables can be interpreted in this context (see Fig.~\ref{fig:LFUV}). This unified view suggests a common origin of the anomalies in terms of BSM physics, which reinforces the case for LFUV with important theoretical and experimental implications. In the following, we will review these flavour anomalies and related processes.

\subsubsection[$b\to s\ell^+\ell^-$]{\boldmath $b\to s\ell^+\ell^-$}
% Bernat Capdevila and Farvah Mahmoudi
%
Consistent hints of New Physics have been observed in semileptonic $B$-meson decays involving $b\to s\ell^+\ell^-$ transitions. Different experimental collaborations at the LHC, with LHCb playing the leading role, and the Belle experiment have reported deviations from SM expectations at the $2$--$3\sigma$ level in several channels mediated by these transitions. The most relevant discrepancies include observables characterising the $B^0\to K^{*0}\mu^+\mu^-$~\cite{LHCb:2020lmf} and $B^+\to K^{*+}\mu^+\mu^-$~\cite{LHCb:2020gog} decay distributions, in particular the so-called $P_5^\prime$ observable in two adjacent anomalous bins in the low-$q^2$ region,
\begin{align}
P_5^\prime(B^0\to K^{*0}\mu^+\mu^-)^{[4.0, 6.0]}_\text{LHCb} &= -0.439 \pm 0.111 \pm 0.036 \quad (2.5\sigma),\\
P_5^\prime(B^0\to K^{*0}\mu^+\mu^-)^{[6.0, 8.0]}_\text{LHCb} &= -0.583 \pm 0.090 \pm 0.030 \quad (2.9\sigma),
\end{align}
the $R_K$~\cite{LHCb:2021trn} and $R_{K^*}$~\cite{Aaij:2017vbb} ratios, defined as~\cite{Hiller:2003js}
\begin{equation}
R_{K^{(*)}} = \frac{{\rm Br}(B^{+(0*)} \to K^{+(0*)} \mu^+ \mu^-)}{{\rm Br}(B^{+(0*)} \to K^{+(0*)} e^+ e^-)},
\end{equation}
which measure LFUV in the $B\to K\ell^+\ell^-$ and $B\to K^*\ell^+\ell^-$ modes,
\begin{align}
{R_K}^{[1.1,6]}_\text{LHCb} &= 0.846^{+0.042+0.013}_{-0.039-0.012} \quad (3.1\sigma),\\
{R_{K^*}}^{[0.045,1.1]}_\text{LHCb} &= 0.66^{+0.11}_{-0.07} \pm 0.03 \quad (2.2\sigma),\\
{R_{K^*}}^{[1.1,6]}_\text{LHCb} &= 0.69^{+0.11}_{-0.07} \pm 0.05 \quad (2.6\sigma),
\end{align}
and the $B_s \to \phi \mu^+ \mu^-$ branching ratio~\cite{Aaij:2015esa,LHCb:2021zwz},
\begin{equation}
{\rm \left\langle \frac{dBr(B_s \to \phi \mu^+ \mu^-)}{dq^2} \right\rangle}^{[1.1,6]}_\text{LHCb}= (2.88 \pm 0.15 \pm 0.05 \pm 0.14) \times 10^{-8} \quad (3.6\sigma).
\end{equation}
In addition, the branching ratio of the leptonic decay $B_s\to\mu^+\mu^-$ shows some tension with respect to its SM prediction:
\begin{equation}
{\rm Br}(B_s\to\mu^+\mu^-)=2.85^{+0.34}_{-0.31}\times 10^{-9} \quad (2.15\sigma),
\end{equation}
where the quoted value corresponds to the average of the latest LHCb measurement~\cite{LHCb:2021awg,LHCb:2021vsc} with the results from CMS~\cite{CMS:2019bbr} and ATLAS~\cite{Aaboud:2018mst} (see Refs.~\cite{Alguero:2021anc,Altmannshofer:2021qrr,Hurth:2021nsi,Geng:2021nhg} for further details).

These tensions, together with the discrepancies observed in $b\to c\ell\nu$ modes, see Sec.~\ref{sec:RD}, are commonly referred to in the literature as ``$B$ anomalies''. The current situation is exceptional since all deviations in $b\to s\ell^+\ell^-$ channels are consistent with a deficit in muonic modes and form coherent patterns in global fits, some of which are preferred over the SM with a very high significance. State-of-the-art global analyses of $b\to s\ell^+\ell^-$ data can be found in Refs.~\cite{Altmannshofer:2021qrr,Alguero:2021anc,Hurth:2021nsi,Cornella:2021sby,Geng:2021nhg,Ciuchini:2020gvn,Alda:2020okk,Hurth:2020ehu}. These global fits differ in the treatment of theoretical uncertainties, with the most important differences being the choice of form factors~\cite{Khodjamirian:2010vf,Straub:2015ica,Gubernari:2018wyi}, the parametrisation used to include factorisable and non-factorisable hadronic uncertainties~\cite{Jager:2012uw,Capdevila:2017ert,Ciuchini:2015qxb,Chobanova:2017ghn} and the approach used in the statistical analysis itself~\cite{DescotesGenon:2012zf,Altmannshofer:2014rta,Hurth:2016fbr,Albrecht:2018vsa,Ciuchini:2019usw,Bhom:2020lmk}.

However, all the abovementioned global analyses share the same model-independent framework based on the effective Hamiltonian of the Weak Effective Theory (WET), in which heavy degrees of freedom with characteristic scales above the $W$ boson mass -- including any potential heavy new particles -- are integrated out in short-distance Wilson coefficients $\mathcal{C}_i$,
\begin{eqnarray}
{\cal H}_{\rm eff} = -\frac{4G_F}{\sqrt{2}}V_{tb}^{} V^\star_{ts}\sum_i\mathcal{C}_i{\cal O}_i\,.
\label{effH}
\end{eqnarray}
Even though NP could generate further effective operators with structures not present in the SM, because of the processes included in the global fits, most analyses focus their attention to the electromagnetic and semileptonic operators (including their chirally-flipped counterparts):\footnote{Global fits with a more general set of 20 operators can be found in Refs.~\cite{Arbey:2018ics,Arbey:2019duh,Hurth:2021nsi}.}
\begin{align}
{\cal O}_7 &= \frac{e}{16\pi^2}m_b(\bar{s}\sigma_{\mu\nu}P_Rb)F^{\mu\nu}, &{\cal O}_{7^\prime} &= \frac{e}{16\pi^2}m_b(\bar{s}\sigma_{\mu\nu}P_Lb)F^{\mu\nu}, \nonumber \\
{\cal O}_{9\ell} &= \frac{e^2}{16\pi^2}(\bar{s}\gamma_{\mu}P_Lb)(\bar\ell \gamma^\mu\ell), &{\cal O}_{9^\prime\ell} &= \frac{e^2}{16\pi^2}(\bar{s}\gamma_{\mu}P_Rb)(\bar\ell \gamma^\mu\ell), \nonumber \\
{\cal O}_{10\ell} &= \frac{e^2}{16\pi^2}(\bar{s}\gamma_{\mu}P_Lb)(\bar\ell \gamma^\mu\gamma_5\ell), &{\cal O}_{10^\prime\ell} &= \frac{e^2}{16\pi^2}(\bar{s}\gamma_{\mu}P_Rb)(\bar\ell \gamma^\mu\gamma_5\ell),
\end{align}
where $\ell=\mu$, $e$, $P_{L,R}=(1\mp \gamma_5)/2$ and $m_b=m_b(\mu_b)$ is the running $b$-quark mass in the $\overline{\text{MS}}$ scheme at the characteristic scale of the process $\mu_b\sim 4.8\,\text{GeV}$. The SM values of the relevant Wilson coefficients are $\mathcal{C}^\text{SM}_{7,9\ell,10\ell}(\mu_b)=-0.29,4.07,-4.31$ and $\mathcal{C}^\text{SM}_{7^\prime,9^\prime\ell,10^\prime\ell}(\mu_b)\sim 0$, for both $\ell=\mu$ and $\ell=e$. In this language, NP effects are parametrised as shifts from their SM values $\mathcal{C}_{i\ell} = \mathcal{C}_{i\ell}^\text{SM}+\mathcal{C}_{i\ell}^\text{NP}$.

Since no deviations have been observed in channels with electrons in the final state, NP contributions to the electronic Wilson coefficients are assumed to be negligible. Then, the most updated global fits to the muonic coefficients reveal the vectorial $\mathcal{C}_{9\mu}^\text{NP}$ and left-handed $\mathcal{C}_{9\mu}^\text{NP}=-\mathcal{C}_{10\mu}^\text{NP}$ structures as the favourite NP scenarios according to current $b\to s\ell^+\ell^-$ data~\cite{Altmannshofer:2021qrr,Alguero:2021anc,Hurth:2021nsi,Cornella:2021sby,Geng:2021nhg,Ciuchini:2020gvn}. Additionally, restricted fits to LFUV observables and $B_s\to\mu^+\mu^-$ show a NP signal in  $\mathcal{C}_{10\mu}^\text{NP}$ with high significance~\cite{Altmannshofer:2021qrr,Hurth:2021nsi,Geng:2021nhg}. Also, scenarios including right-handed couplings (RHC) have been recently found to provide very competitive descriptions of the data~\cite{Alguero:2019ptt,Alguero:2021anc}. The statistical significance of these scenarios, as measured by the so-called $\text{Pull}_\text{SM}$, ranges from roughly to well-above $5\sigma$ depending on the particular details of each analysis.

Notice that the NP scenarios discussed so far are all based on the underlying assumption of LFUV NP, where the NP is entirely attached to the muons. However, some analyses have also started exploring scenarios with lepton flavour universal (LFU) NP effects in addition to LFUV contributions to muons only~\cite{Alguero:2018nvb,Aebischer:2019mlg,Alguero:2019ptt,Cornella:2021sby}. In order to account for these contributions, one possible parametrisation reads,
\begin{equation}\label{eq:LFUbsll}
\mathcal{C}_{ie}^\text{NP}=\mathcal{C}_i^\text{U}, \qquad \mathcal{C}_{i\mu}^\text{NP}=\mathcal{C}_i^\text{U}+\mathcal{C}_{i\mu}^\text{V},
\end{equation}
with $i=9^{(\prime)}$, $10^{(\prime)}$. The basis redefinition in Eq.~\eqref{eq:LFUbsll} provides a new description of the data with a concrete NP structure, namely, that $b\to s\ell^+\ell^-$ transitions get a common LFU NP contribution for all charged leptons (electrons, muons and tau leptons), opening new directions and extending the possible interpretations of the global fits. Interestingly, when allowing for LFU NP, the scenario $(\mathcal{C}_{9\mu}^{\rm V}=-\mathcal{C}_{10\mu}^{\rm V},\mathcal{C}_{9}^{\rm U})$ with an $SU(2)_L$ LFUV structure emerges as an acceptable NP solution~\cite{Alguero:2021anc,Cornella:2021sby}. Also, scenarios with $\mathcal{C}_{10(')}^{\rm U}$, like $(\mathcal{C}_{9\mu}^{\rm V},\mathcal{C}_{10}^{\rm U})$ and $(\mathcal{C}_{9\mu}^{\rm V},\mathcal{C}_{10'}^{\rm U})$, get selected with very high significance~\cite{Alguero:2019ptt,Alguero:2021anc}.

It is also important to discuss the implications of the global $b\to s\ell^+\ell^-$ fits on popular NP models. Now we briefly review those that are able to generate the preferred structures suggested by the global fits.

{\begin{boldmath} $\mathcal{C}_{9}^{\rm NP}$: \end{boldmath}} $Z^\prime$ models with vectorial couplings to leptons preferably yield $\mathcal{C}_{9\mu}^{\rm NP}$-like solutions in order to avoid gauge anomalies. In this context, $L_\mu-L_\tau$ models~\cite{Altmannshofer:2014cfa,Crivellin:2015mga,Altmannshofer:2016oaq,Crivellin:2015lwa,Crivellin:2016ejn} are popular since they do not generate effects in electron channels. Fits including $R_{K^*}$ are also very favourable to models predicting $\mathcal{C}_{9\mu}^{\rm NP}=-3\mathcal{C}_{9e}^{\rm NP}$~\cite{Bhatia:2017tgo}. Concerning leptoquarks (LQs), a $\mathcal{C}_{9\mu}^{\rm NP}$ solution can only be generated by adding two scalar (an $SU(2)_L$ triplet and an $SU(2)_L$ doublet with $Y=7/6$) or two vector representations (an $SU(2)_L$ singlet with $Y=2/3$ and an $SU(2)_L$ doublet with $Y=5/6$).

{\begin{boldmath} $\mathcal{C}_{9\mu}^{\rm NP}=-\mathcal{C}_{10\mu}^{\rm NP}$: \end{boldmath}} This pattern can be achieved in $Z^\prime$ models with loop-induced couplings~\cite{Belanger:2015nma} or with heavy vector-like fermions~\cite{Boucenna:2016wpr,Boucenna:2016qad}. Regarding LQ models, here a single representation (the scalar $SU(2)_L$ triplet or the vector $SU(2)_L$ singlet with $Y=2/3$) can generate a $\mathcal{C}_{9\mu}^{\rm NP}=-\mathcal{C}_{10\mu}^{\rm NP}$ solution~\cite{Gripaios:2014tna,Fajfer:2015ycq,Varzielas:2015iva,Alonso:2015sja,Calibbi:2015kma,Barbieri:2015yvd,Sahoo:2016pet}. This pattern can also be obtained in models with loop contributions from three heavy new scalars and fermions~\cite{Gripaios:2015gra,Arnan:2016cpy,Mahmoudi:2014mja,Grinstein:2018fgb,Arnan:2019uhr} and in composite Higgs models~\cite{Niehoff:2015bfa}.

{\bf RHC:} with a value of $R_K$ closer to one, scenarios with right-handed currents, namely $\mathcal{C}_{9\mu}^{\rm NP}=-\mathcal{C}_{9^\prime\mu}$, $(\mathcal{C}_{9\mu}^{\rm NP}, \mathcal{C}_{9'\mu})$ and $(\mathcal{C}_{9\mu}^{\rm NP}, \mathcal{C}_{10'\mu})$, seem to emerge. The first two scenarios are naturally generated in $Z^\prime$ models with certain assumptions on its couplings to right-handed and left-handed quarks, as it was shown in Ref.~\cite{Altmannshofer:2014cfa} within the context of a gauged $L_\mu-L_\tau$ symmetry with vector-like quarks. One could also obtain $\mathcal{C}_{9\mu}^{\rm NP}=-\mathcal{C}_{9^\prime\mu}$ by adding a third Higgs doublet to the model of Ref.~\cite{Crivellin:2015lwa} with opposite $U(1)$ charge. On the other hand, generating the aforementioned contribution in LQ models requires one to add four scalar representations or three vector ones.

{\begin{boldmath} $(\mathcal{C}_{9\mu}^{\rm V}=-\mathcal{C}_{10\mu}^{\rm V},\mathcal{C}_{9}^{\rm U})$: \end{boldmath}} this scenario can be realised via off-shell photon penguins in a LQ model explaining also $b\to c\tau\nu$ data~\cite{Crivellin:2018yvo} (see Sec.~\ref{sec:combined_expl}). Remarkably, as we will discuss below, a NP contribution with this structure allows for a model-independent combined explanation of $b\to s\ell^+\ell^-$ and $b\to c\tau\nu$ data with very high statistical significance~\cite{Crivellin:2018yvo,Alguero:2019ptt,Alguero:2021anc}.

{\begin{boldmath} $\mathcal{C}_{10(')}^{\rm U}$: \end{boldmath}} NP solutions with $\mathcal{C}_{10(')}^{\rm U}$ (see scenarios 9--13 from Refs.~\cite{Alguero:2019ptt,Alguero:2021anc}) arise naturally in models with modified $Z$ couplings. In this case, $\mathcal{C}_{9(')}^{\rm U}$ contributions are also generated but to a good approximation can be neglected. The $(\mathcal{C}_{9\mu}^{\rm V}=-\mathcal{C}_{10\mu}^{\rm V},\mathcal{C}_{10}^{\rm U})$ pattern also occurs in Two-Higgs-Doublet models~\cite{Crivellin:2019dun}. For scenarios $(\mathcal{C}_{9\mu}^{\rm V},\mathcal{C}_{10}^{\rm U})$ and $(\mathcal{C}_{9\mu}^{\rm V},\mathcal{C}_{10'}^{\rm U})$, one can also invoke models with vector-like quarks, where modified $Z$ couplings are even induced at tree-level. The LFU effect in $\mathcal{C}_{10(')}^{\rm U}$ can be accompanied by a $\mathcal{C}_{9,10(')}^{\rm V}$ effect from $Z^\prime$ exchanges~\cite{Bobeth:2016llm}. Vector-like quarks with the quantum numbers of right-handed down quarks (left-handed quarks doublets) generate effects in $\mathcal{C}_{10}^{\rm U}$ and $\mathcal{C}_{9'}^{\rm V}$ ($\mathcal{C}_{10(')}^{\rm U}$ and $\mathcal{C}_{9}^{\rm V}$) for a $Z^\prime$ boson with vector couplings to muons~\cite{Bobeth:2016llm}.

Given that LQs should possess very small couplings to electrons in order to avoid dangerous effects in $\mu\to e\gamma$, they naturally violate LFU~\cite{Crivellin:2017dsk}. While $Z^\prime$ models can easily accommodate LFUV data~\cite{Falkowski:2015zwa}, variants based on the assumption of only LFU NP~\cite{Gauld:2013qba,Buras:2013dea} are now disfavoured. The same is true if one aims at explaining $P_5'$ via NP in four-quark operators leading to a NP ($q^2$-dependent) contribution from charm loops~\cite{Jager:2017gal}.

\begin{figure}
\centering
\includegraphics[width=0.55\textwidth]{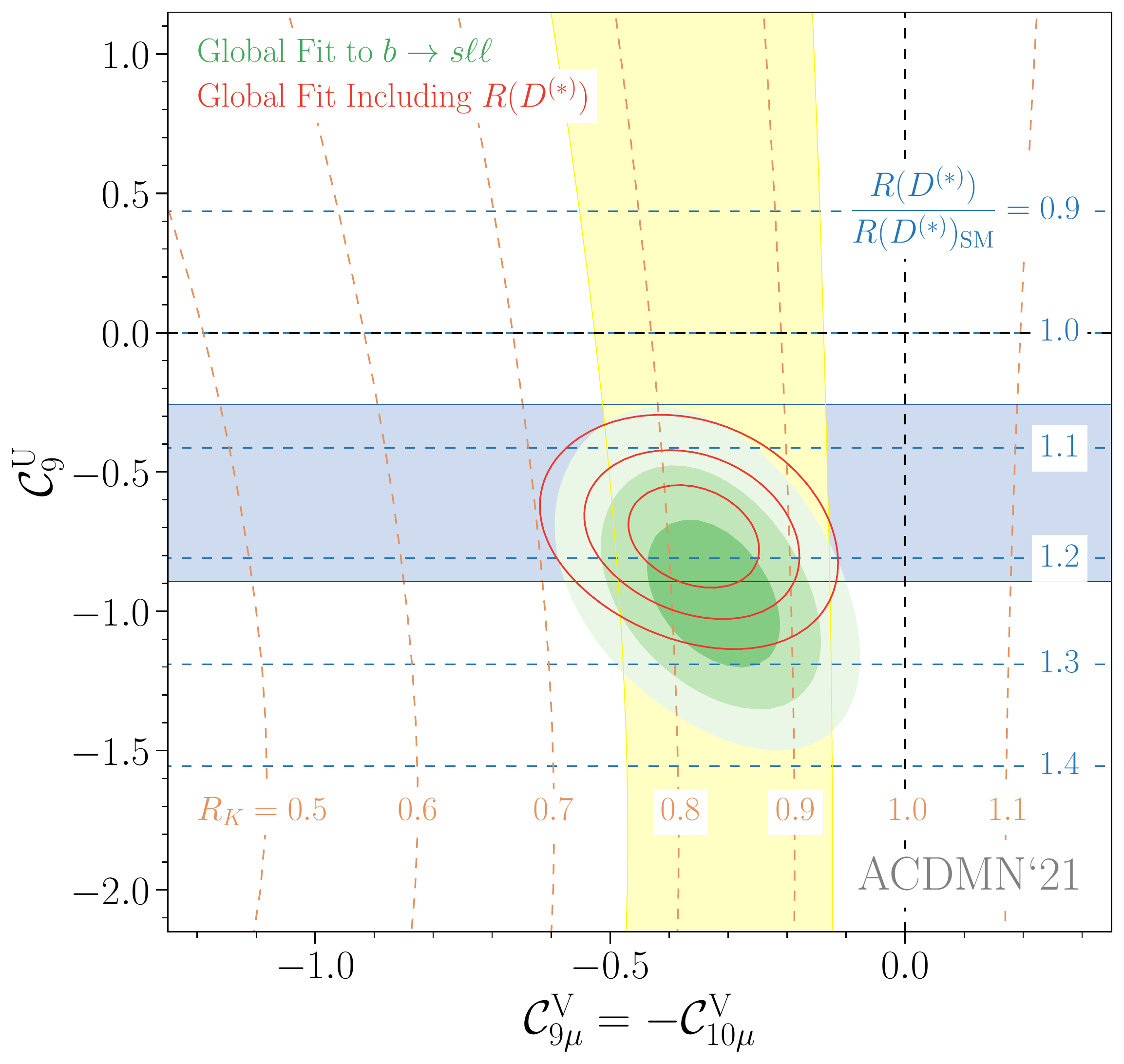}
\caption{Preferred regions at the 1, 2 and 3$\,\sigma$ level (green) in the $(\mathcal{C}_{9\mu}^{\rm V}=-\mathcal{C}_{10\mu}^{\rm V},\,\mathcal{C}_{9}^{\rm U})$ plane from $b\to s\ell^+\ell^-$ data. The red contour lines show the corresponding regions once $R_{D^{(*)}}$ is included in the fit (for $\Lambda=2$~TeV). The horizontal blue (vertical yellow) band is consistent with $R_{D^{(*)}}$ ($R_{K}$) at the $2\,\sigma$ level and the contour lines show the predicted values for these ratios.}
\label{fig:bsllRDRD*}
\end{figure}

Finally, we further discuss the scenario $(\mathcal{C}_{9\mu}^{\rm V}=-\mathcal{C}_{10\mu}^{\rm V},\mathcal{C}_{9}^{\rm U})$ and how its structure allows for a model-independent connection between the $b\to s\ell^+\ell^-$ anomalies and the deviations in $b\to c\tau\nu$ transitions~\cite{Amhis:2019ckw}. This connection arises in the SMEFT scenario where $\mathcal{C}^{(1)}=\mathcal{C}^{(3)}$ expressed in terms of gauge-invariant dimension-6 operators~\cite{Grzadkowski:2010es,Capdevila:2017iqn}. The operator involving third-generation leptons explains $R_{D^{(*)}}$ and the one involving the second generation gives a LFUV effect in $b\to s\mu^+\mu^-$ processes. The constraint from $b\to c\tau\nu$ and $SU(2)_L$ invariance leads to large contributions enhancing $b\to s\tau^+\tau^-$ processes~\cite{Capdevila:2017iqn}, whereas the mixing into ${\cal O}_{9\ell}$ generates $\mathcal{C}_{9}^{\rm U}$ at $\mu=m_b$~\cite{Crivellin:2018yvo}.
Therefore, this NP structure correlates $\mathcal{C}_9^{\rm U}$ and $R_{D^{(*)}}$ in the following way~\cite{Capdevila:2017iqn,Crivellin:2018yvo}:
\begin{equation}
\mathcal{C}_{9}^{\rm U}\! \approx \! 7.5\left(1-\sqrt{\frac{R_{D^{(*)}}}{R_{D^{(*)}{\rm SM}}}}\right)\!\! \left(1+\frac{\log(\Lambda^2/(1{\rm TeV}^2))}{10.5}\right), \\  \\
\end{equation}
where $\Lambda$ is the typical scale of NP involved. In Fig.~\ref{fig:bsllRDRD*}, we show the global fit of the pattern $(\mathcal{C}_{9\mu}^{\rm V}=-\mathcal{C}_{10\mu}^{\rm V},\mathcal{C}_{9}^{\rm U})$ without and with the additional input on $R_{D(^*)}$ from Ref.~\cite{Amhis:2019ckw}, taking the scale $\Lambda=2$ TeV. This connection between neutral and charged anomalies is remarkable as it offers a NP solution that is able to accommodate both sets of data simultaneously, and hence one finds a very high $\text{Pull}_\text{SM}$ of 8.1$\sigma$ for the combined fit~\cite{Alguero:2021anc}.

\subsubsection[Tauonic $B$-meson decays]{\boldmath Tauonic $B$-meson decays}
\begin{figure}
\centering{\includegraphics[width=.6\textwidth]{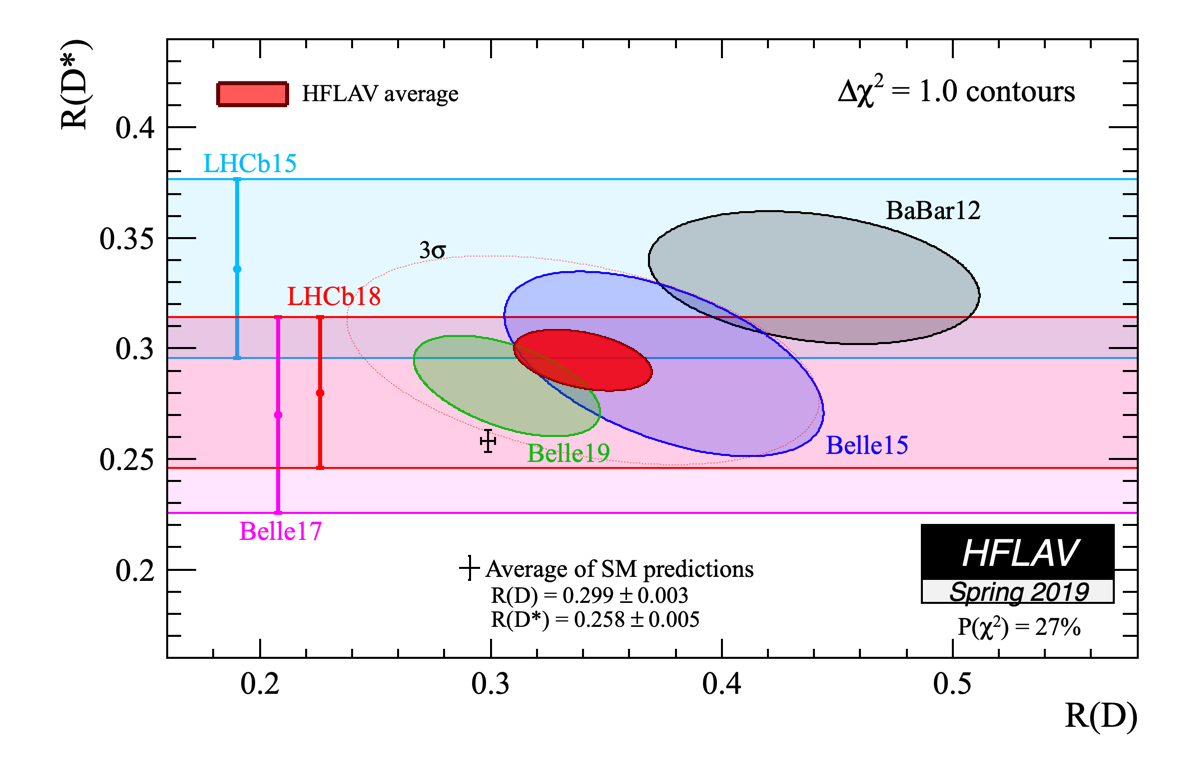}}
\caption{Experimental results for the LFU ratios $R(D^{(*)})$ and their average, provided by the HFLAV collaboration. The SM prediction is indicated by the black cross. From Ref.~\cite{Amhis:2016xyh}.}
\label{fig:rdrds}
\end{figure}

% Monika Blanke
\label{sec:RD}
In addition to the neutral-current $b\to s\ell^+\ell^-$ transitions discussed above, also charged-current $b\to c\tau\nu$ data exhibit tensions with the SM predictions. Of particular interest are the lepton flavour universality (LFU) ratios
\begin{equation}
R(D^{(*)})=\frac{\text{BR}(B\to D^{(*)} \tau
  \nu)}{\text{BR}(B\to D^{(*)} \ell \nu)} \qquad (\ell=e,\mu)\,,
\end{equation}
for which measurements from BaBar~\cite{Lees:2012xj,Lees:2013uzd}, Belle~\cite{Huschle:2015rga,Hirose:2016wfn,Hirose:2017dxl,Abdesselam:2019dgh} and LHCb~\cite{Aaij:2015yra,Aaij:2017uff,Aaij:2017deq} exist, see Fig.~\ref{fig:rdrds}.
The latest HFLAV average combining these data~\cite{Amhis:2016xyh}
\begin{eqnarray}
\begin{aligned}
R(D)\,=\,{0.340\pm0.027 \pm  0.013}\,, \label{new_average}\\
R(D^*)\,=\,{0.295\pm0.011  \pm 0.008 }\,,
\end{aligned}
\end{eqnarray}
deviates by $3.1\sigma$ from the SM prediction. Furthermore, the data for the analogous ratio $R(J/\psi)$ also seem to hint at an enhancement relative to the SM~\cite{Aaij:2017tyk}.

NP contributions to $b\to c\tau\nu$ transitions can be parametrised by the Wilson coefficients $C_i$ in the effective Hamiltonian
\begin{equation}
 {\cal H}_{\rm eff}=  2\sqrt{2} G_{F} V^{}_{cb} \big[(1+C_{V}^{L}) O_{V}^L + C_{S}^{L} O_{S}^L+  C_{S}^{R} O_{S}^{R} 
   +   C_{T} O_{T}\big] \,,
\label{eq:Heff}
\end{equation}
assuming the absence of light right-handed neutrinos. Here $O_{V}^L$ is the left-handed current-current operator present already in the SM, $O_{S}^L$ and $O_{S}^R$ are the left- and right-handed scalar operators, and $O_T$ is the tensor operator, as defined, e.g., in Ref.~\cite{Blanke:2018yud}. 

Global fits to the data, including polarisation observables in $B\to D^*\tau\nu$, have been performed in Refs.~\cite{Murgui:2019czp,Shi:2019gxi,Aebischer:2019mlg,Blanke:2019qrx}. From the results of these fits, several simplified NP models can be identified as potential candidates for an explanation of the $b\to c\tau\nu$ anomalies. Due to the rather large size of the required NP contribution with respect to the SM, in all cases new particles contribute to $b\to c\tau\nu$ at the tree level, and for the sake of simplicity we restrict our attention to models with a single new state:

{\bf\boldmath Charged $W'$ bosons:} A good fit to the available $b\to c\tau\nu$ data is obtained by a shift $C_V^L\ne 0$ of the SM $(V-A)\otimes (V-A)$ contribution, which could originate from a heavy charged $W'$ gauge boson coupling to left-handed quarks and leptons~\cite{He:2012zp,Greljo:2015mma}. This model, however, is challenged by LHC high-$p_T$ di-$\tau$ data~\cite{Faroughy:2016osc} as well as by precision measurements of $Z$-pole observables~\cite{Feruglio:2016gvd}. 

{\bf\boldmath Charged Higgs boson $H^\pm$:} 
This scenario~\cite{Kalinowski:1990ba,Hou:1992sy,Crivellin:2012ye,Crivellin:2015hha}, leading to non-zero $C_S^{L,R}$, currently provides the best fit to the low-energy $b\to c\tau\nu$ data, as -- in contrast to the other simplified models -- it allows one to accommodate the measured $D^*$ polarisation, $F_L(D^*)$~\cite{Abdesselam:2019wbt}, at the $1\sigma$ level. However, this solution is in tension with the LHC mono-$\tau$ data~\cite{Greljo:2018tzh}, and it induces a large branching ratio $\text{BR}(B_c\to\tau\nu)>50\%$. While no direct experimental bound on the latter exists, upper limits of 30\%~\cite{Alonso:2016oyd} and even 10\%~\cite{Akeroyd:2017mhr} have been estimated in the literature. On the other hand, a critical reassessment reached the conclusion that values even as large as 60\% cannot be excluded at present~\cite{Blanke:2018yud,Blanke:2019qrx}. A recent update on the SM prediction of the $B_c$ lifetime supports the latter reasoning~\cite{Aebischer:2021ilm}.

{\bf\boldmath Scalar leptoquarks:} The scalar $SU(2)_L$-singlet leptoquark $S_1$~\cite{Deshpande:2012rr,Tanaka:2012nw,Sakaki:2013bfa}, giving rise to the scenario $C_V^L, C_S^L=-4C_T\ne0$, offers a good fit to the $b\to c\tau\nu$ data, predicts only modest contributions to the decay $B_c\to\tau\nu$, and passes the mono-$\tau$ test. The scalar $SU(2)_L$-doublet leptoquark $S_2$, inducing $C_S^L=4C_T$, on the other hand, can be brought in agreement with the $b\to c\tau\nu$ data only in the presence of complex, i.e., CP-violating couplings~\cite{Becirevic:2018afm}. The latter scenario predicts a significant contribution to $\text{BR}(B_c\to\tau\nu)\sim 20\%$, and its best-fit point is on the verge of being tested by the mono-$\tau$ searches. There are also stringent LHC constraints on these LQs from their pair-production and $t$-channel mediated dilepton processes~\cite{Angelescu:2018tyl, Babu:2020hun}.   

{\bf\boldmath Vector leptoquark:} Last but not least also an $SU(2)_L$-singlet vector leptoquark $U_1$~\cite{Alonso:2015sja,Calibbi:2015kma,Fajfer:2015ycq,Bordone:2017bld,DiLuzio:2017vat,Calibbi:2017qbu,Blanke:2018sro, BhupalDev:2020zcy} provides a good fit to the $b\to c\tau\nu$ data, both with only left-handed couplings ($C_V^L \ne 0$) and in the presence of an additional small right-handed $b\tau$ coupling ($C_V^L, C_S^R \ne 0$). As in the case of the scalar $SU(2)_L$-singlet leptoquark, also here the contributions to $B_c\to\tau\nu$ are small and the model evades the current LHC mono-$\tau$ searches. One of the most stringent constraints on models with an $SU(2)_L$-singlet vector leptoquark stems instead from LHC searches for colour-octet resonances, which are often introduced together with the leptoquark in UV-complete models~\cite{Buttazzo:2017ixm,Diaz:2017lit,Baker:2019sli}.

\begin{sloppypar}
In addition to these simplified models parametrised by the effective interactions in Eq.~\eqref{eq:Heff}, models with light right-handed neutrinos have been examined in the literature~\cite{Iguro:2018qzf,Greljo:2018ogz,Robinson:2018gza,Azatov:2018kzb,Mandal:2020htr}. While it is possible to accommodate the low-energy $b\to c\tau\nu$ data in this case, a very large NP contribution is required due to the absence of interference with the SM contribution. Consequently the constraints from direct LHC searches, particularly mono-$\tau$, tend to be even more severe.
\end{sloppypar}

To further disentangle the NP structure at work, a major role will be played by the measurement  of differential and angular observables~\cite{Nierste:2008qe,Becirevic:2016hea,Celis:2016azn,Iguro:2018vqb,Blanke:2018yud,Becirevic:2019tpx,Asadi:2020fdo}, such as the $D^*$ and $\tau$ polarisations $F_L(D^*)$ and $P_\tau (D^{(*)})$, whose correlations turn out to discriminate well between the different scenarios. To fully exploit their model-discriminating potential, both precise measurements and a better theoretical understanding of the underlying form factors are necessary. A measurement of the baryonic LFU ratio
\begin{equation}
R(\Lambda_c)=\frac{\text{BR}(\Lambda_b \to \Lambda_c \tau
  \nu)}{\text{BR}(\Lambda_b \to \Lambda_c \ell \nu)} \qquad (\ell=e,\mu)\,,
\end{equation}
will instead provide an experimental consistency check for the $R(D^{(*)})$ anomaly, thanks to a model-independent sum-rule~\cite{Blanke:2018yud} relating $R(\Lambda_c)$ to $R(D)$ and $R(D^*)$, with the current prediction~\cite{Blanke:2019qrx}
\begin{equation}
R(\Lambda_c) = R_\text{SM}(\Lambda_c)(1.15\pm 0.04)
=0.38\pm0.01\pm0.01\,.
\end{equation}

In addition to the constraints mentioned above, further tensions may arise in concrete UV completions. For example,  in certain models  electroweak $SU(2)_L$ symmetry implies large contributions to the decays $B\to K^{(*)}\nu\bar\nu$, $B_s\to\tau^+\tau^-$ and $B\to K^{(*)}\tau^+\tau^-$~\cite{Calibbi:2015kma,Crivellin:2017zlb}, and significant rates for  $\Upsilon\to\tau^+\tau^-$ or $\psi\to\tau^+\tau^-$ are expected~\cite{Aloni:2017eny}. In summary it is fair to say that stringent constraints on all NP scenarios for the $R(D^{(*)})$ anomaly exist, challenging a full resolution of the latter in the context of NP.

\subsubsection{Anomalous magnetic moments of charged leptons }
% Emanuele Bagnaschi and Martin Hoferichter
%
Ever since Schwinger's famous prediction $a_\ell=(g-2)_\ell/2=\alpha/(2\pi)$~\cite{Schwinger:1948iu} (and its experimental verification~\cite{Kusch:1948mvb}),
the anomalous magnetic moments of the electron and muon have been critical precision tests of the SM. For the electron, the current best direct measurement~\cite{Hanneke:2008tm}
\begin{equation}
\label{aedirect}
 a_e^\text{exp}=1\,159\,652\,180.73(28)\times 10^{-12}
\end{equation}
can be contrasted with its SM prediction once independent input for the fine-structure constant $\alpha$ is specified. With the mass-independent $4$-loop QED coefficient known semi-analytically~\cite{Laporta:2017okg}, the dominant uncertainties now arise from the numerical evaluation of the $5$-loop coefficient~\cite{Aoyama:2019ryr} and hadronic corrections~\cite{Keshavarzi:2019abf}, both of which enter at the level of $10^{-14}$ (for the $5$-loop QED coefficient there is a $4.8\sigma$ tension between Refs.~\cite{Aoyama:2019ryr,Volkov:2019phy} regarding the contribution of diagrams without closed lepton loops). However, the current most precise measurements of $\alpha$ in atom interferometry, using Cs~\cite{Parker:2018vye} and Rb~\cite{Morel:2020dww} atoms, respectively, differ by $5.4\sigma$,
\begin{align}
a_e^\text{SM}[\text{Cs}]&=1\,159\,652\,181.61(23)\times 10^{-12},\notag\\
 a_e^\text{SM}[\text{Rb}]&=1\,159\,652\,180.25(10)\times 10^{-12},
 \label{gm2:electrons}
\end{align}
resulting in a difference to Eq.~\eqref{aedirect} of $-2.5\sigma$ and $+1.6\sigma$.

The world average of the muon $g-2$ is determined by the Run 1 results from the Fermilab experiment~\cite{Abi:2021gix,Albahri:2021ixb,Albahri:2021kmg,Albahri:2021mtf} and the Brookhaven measurement~\cite{Bennett:2006fi}
\begin{equation}
 a_\mu^\text{exp}=116\,592\,061(41)\times 10^{-11},
\end{equation}
with a combined precision of $0.35\,\text{ppm}$. Comparison with the current SM prediction~\cite{Aoyama:2020ynm}
\begin{align}
\label{amuSM}
a_\mu^\text{SM}=116\,591\,810(43)\times 10^{-11}
\end{align}
then reveals a $4.2\sigma$ tension. Experimental efforts to corroborate or refute this tension are underway at subsequent runs at Fermilab~\cite{Grange:2015fou} and at J-PARC~\cite{Abe:2019thb}, with a precision goal of $0.14\,\text{ppm}$ and $0.45\,\text{ppm}$, respectively, and the J-PARC experiment pioneering a new experimental technique that does not rely on the magic momentum in a storage ring, see Ref.~\cite{Gorringe:2015cma} for a more detailed comparison of the two methods. 
The SM prediction in Eq.~\eqref{amuSM}, currently at $0.37\,\text{ppm}$, represents a coherent theory effort organised in the Muon $g-2$ Theory Initiative~\cite{Aoyama:2020ynm}, and is mainly based on the underlying work from 
Refs.~\cite{Aoyama:2012wk,Aoyama:2019ryr,Czarnecki:2002nt,Gnendiger:2013pva,Davier:2017zfy,Keshavarzi:2018mgv,Colangelo:2018mtw,Hoferichter:2019gzf,Davier:2019can,Keshavarzi:2019abf,Kurz:2014wya,Melnikov:2003xd,Masjuan:2017tvw,Colangelo:2017fiz,Hoferichter:2018kwz,Gerardin:2019vio,Bijnens:2019ghy,Colangelo:2019uex,Blum:2019ugy,Colangelo:2014qya}. The uncertainty is completely dominated by hadronic contributions, with hadronic vacuum polarisation (HVP) and hadronic light-by-light scattering (HLbL) at $0.34\,\text{ppm}$ and $0.15\,\text{ppm}$, respectively. Improvements on both HVP and HLbL will continue over the next years, including new $e^+e^-\to \text{hadrons}$ data, lattice-QCD calculations at a similar level of precision, and direct input on space-like HVP from the proposed MUonE experiment~\cite{MUonE:LoI,Banerjee:2020tdt}. Some recent developments include the first lattice calculation of HVP reporting subpercent precision~\cite{Borsanyi:2020mff}, with subsequent work exploring the consequences of the emerging $2.1\sigma$ tension with the data-driven determination~\cite{Lehner:2020crt,Crivellin:2020zul,Keshavarzi:2020bfy,Malaescu:2020zuc,Colangelo:2020lcg}, new $e^+e^-\to \pi^+\pi^-$ data from SND~\cite{Achasov:2020iys}, improved radiative corrections~\cite{Campanario:2019mjh}, a lattice-QCD calculation of HLbL at a similar level of precision as the phenomenological evaluation~\cite{Chao:2021tvp}, and work aimed at refining the subleading contributions to HLbL~\cite{Hoferichter:2020lap,Bijnens:2020xnl,Bijnens:2021jqo,Zanke:2021wiq,Danilkin:2021icn,Colangelo:2021nkr}.

\paragraph{New Physics explanations}
The absolute value of the difference between measurement and theory prediction exceeds the size of the EW contribution of the SM.
Therefore, some form of enhancement mechanism is required to explain the current $4.2\sigma$ tension with BSM physics, and well-motivated scenario do exist. One possibility is that NP involves heavy particles at or above the EW scale, with an enhanced chirality flip originating from an interaction between new particles with the SM Higgs boson, with a coupling strength that is larger than the muon Yukawa. Depending on the model, this type of chiral enhancement allows for viable solutions for new particles with masses up to tens of TeV. Such an enhancement can be achieved in models with new scalars and fermions~\cite{Crivellin:2018qmi} with the MSSM being a specific example. Alternatively, the anomaly can be explained by new, light (or very light) weakly coupled states, such an axion-like particles (ALPs) or a dark photon $Z_d$. For a more detailed overview of various models in light of the most recent measurement, we refer the reader to Ref.~\cite{Athron:2021iuf}.

\paragraph{The Minimal Supersymmetric Standard Model}
One possible theoretical framework of NP above the electroweak scale is the MSSM. Here, the chiral enhancement is provided by the factor $\tan\beta \equiv v_u/v_d$, where
$v_u$ and $v_d$ are the vacuum expectation values of the two Higgs doublets of the model, $H_u$ and $H_d$
(which give mass to up-type and down-type fermions respectively). A large value of $\tan \beta\approx 50$ can be motivated by top--bottom Yukawa coupling unification~\cite{Ananthanarayan:1991xp,Carena:1994bv}, and thus
this would provide a natural explanation for a large enhancement factor~\cite{Lopez:1993vi,Chattopadhyay:1995ae,Dedes:2001fv}.

\begin{figure}
  \includegraphics[width=0.495\textwidth]{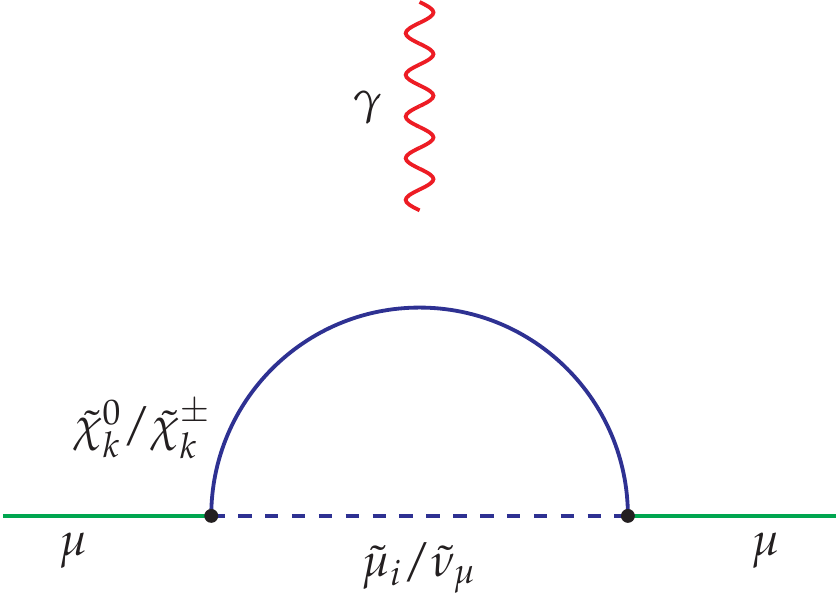}~\includegraphics[width=0.495\textwidth]{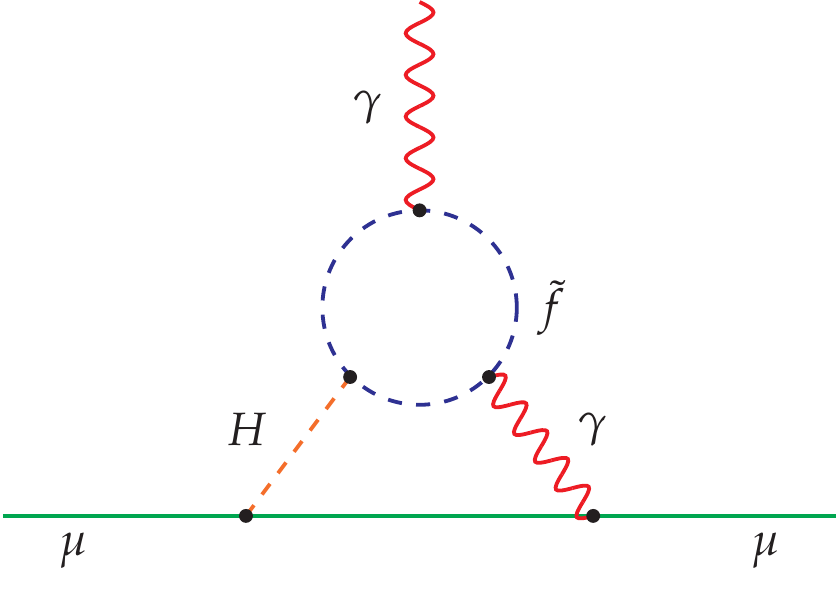}
  \caption{Left: one-loop contributions to $(g-2)_{\mu}$ from MSSM particles; the external photon line can be attached to any charged particle. Right: example of a two loop Barr-Zee contribution involving a charged sfermion.}
  \label{gm2:mssm1l}
\end{figure}

The leading MSSM contributions arise from one-loop diagrams involving a loop of either a neutralino and smuon ($a^{\tilde{\chi}^0}_{\mu}$) or a chargino and a sneutrino ($a^{\tilde{\chi}^{\pm}}_{\mu}$) (see the left plot of Fig.~\ref{gm2:mssm1l})~\cite{Fayet1980,Grifols:1982vx,Ellis:1982by,Barbieri:1982aj,Romao:1984pn,Kosower:1983yw,Yuan:1984ww,Vendramin:1988rd}.\footnote{The two-loop
contributions have been calculated in Refs.~\cite{Chen:2001kn,Arhrib:2001xx,Heinemeyer:2003dq,Heinemeyer:2004yq,Feng:2006ei,Marchetti:2008hw,vonWeitershausen:2010zr,Fargnoli:2013zda,Fargnoli:2013zia} and the resummation
of $\tan\beta$ has been achieved~\cite{Marchetti:2008hw,Crivellin:2011jt,Bach:2015doa}. These results have been implemented in codes
such as {\tt SuperIso}~\cite{Mahmoudi:2007vz,Mahmoudi:2008tp,Mahmoudi:2009zz}, {\tt micrOMEGAs}~\cite{Belanger:2018ccd}, {\tt FeynHiggs}~\cite{Heinemeyer:2004yq,Heinemeyer:2003dq}, {\tt
(SARAH-)Spheno}~\cite{Porod:2011nf} and {\tt FlexibleSUSY}~\cite{Athron:2017fvs} to the dedicated software such as {\tt SUSY\_FLAVOR}~\cite{Crivellin:2012jv} and {\tt GM2Calc}~\cite{Athron:2015rva}.}

The relevant phenomenological question is how to account for relatively light SUSY particles (and explain $(g-2)_{\mu}$) while meeting constraints from the LHC searches, DM phenomenology, and other observables. Incorporating LHC Run 2 limits ~\cite{Zhu:2016ncq,Yamaguchi:2016oqz,Choudhury:2017acn,Hagiwara:2017lse,Chakraborti:2017dpu,Yanagida:2017dao,Endo:2017zrj,Yanagida:2018eho,Pozzo:2018anw,Cox:2018qyi,Bhattacharyya:2018inr,Cox:2018vsv,Endo:2020mqz,Chakraborti:2020vjp,Horigome:2021qof,Athron:2021iuf,Tran:2018kxv,Wang:2018vrr,Abdughani:2019wai,Ibe:2019jbx,Yanagida:2020jzy,Baum:2021qzx,Ibe:2021cvf,Yin:2021mls,Han:2021ify,VanBeekveld:2021tgn,Cox:2021gqq,Endo:2021zal,Ahmed:2021htr,Wang:2021bcx,Gu:2021mjd,Chakraborti:2021bmv,Chakraborti:2021kkr,Chakraborti:2021dli,Baer:2021aax,Li:2021pnt},
with a varying degree of sophistication and diverse focuses (model building, collider, DM etc.). Furthermore, several global studies, trying to incorporate and correlate LHC Run 2 bounds with limits coming from different sectors, have been performed as well, both in the context of scenarios with universal and minimal SUSY breaking mechanisms~\cite{Bagnaschi:2016afc,Bagnaschi:2016xfg,GAMBIT:2017snp,Costa:2017gup}, and for more phenomenological oriented models~\cite{GAMBIT:2017zdo,Bagnaschi:2017tru}. In the former case, LHC constraints push the mass of the SUSY states to the TeV scale, such that the $\tan\beta$ enhancement is insufficient to provide a sufficiently large contribution to $a_{\mu}$. For the latter scenarios, if sufficiently freedom is allowed, SUSY contributions can still account for the observed discrepancy~\cite{Bagnaschi:2017tru}.

\begin{figure}[t]
  \includegraphics[width=0.495\textwidth]{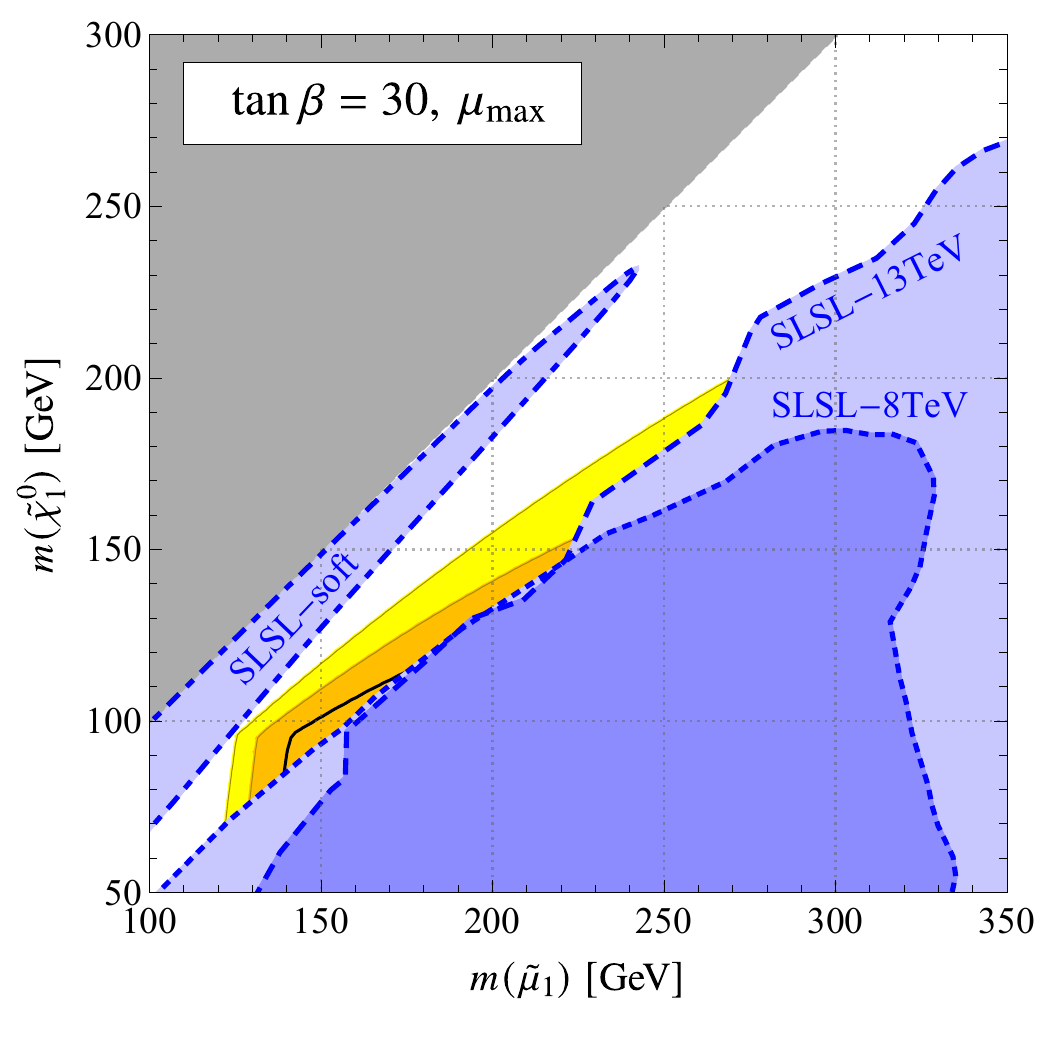}~\includegraphics[width=0.495\textwidth]{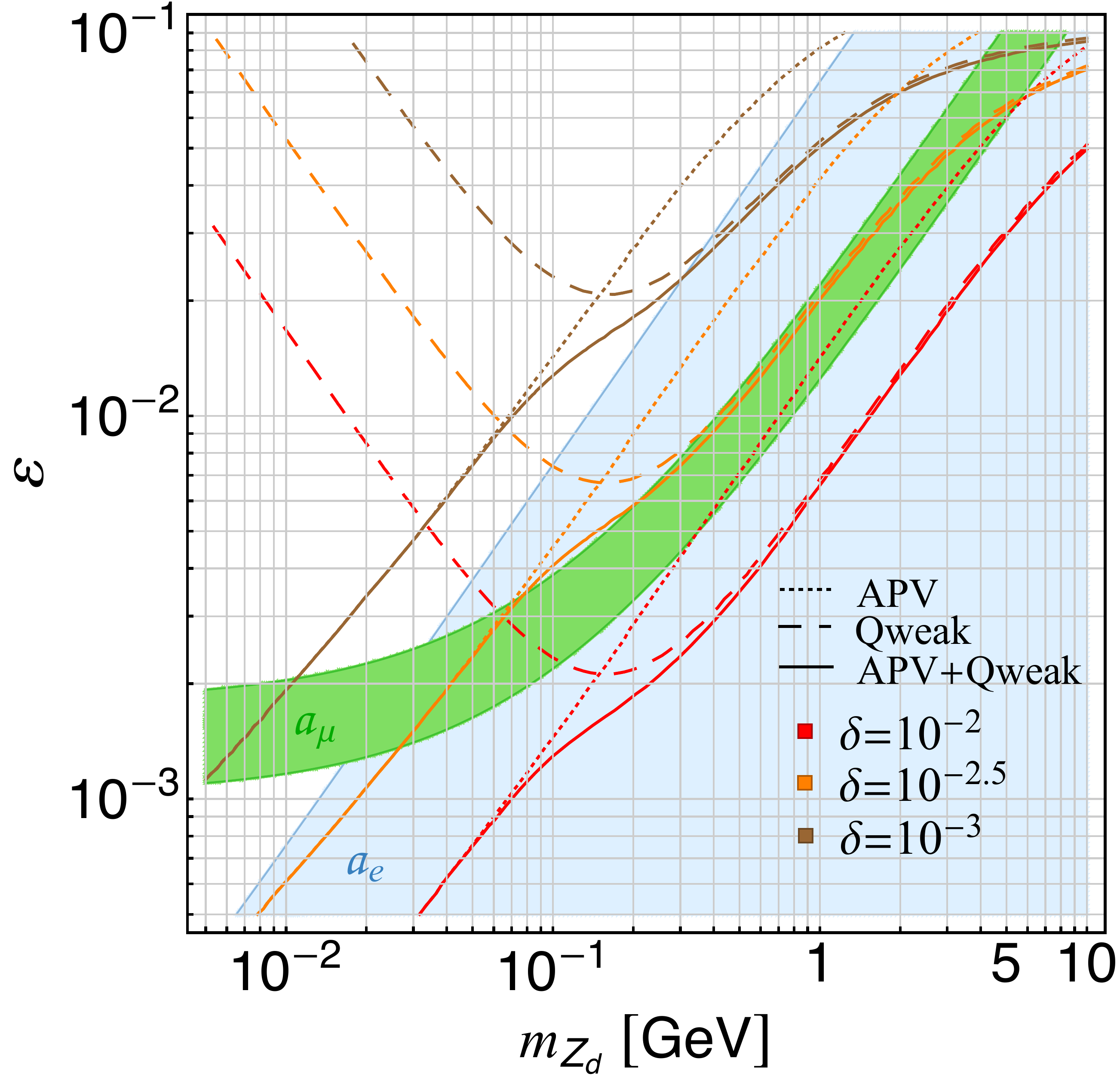}
  \caption{Left: plot from Ref.~\cite{Endo:2021zal} showing the neutralino-smuon mass range still allowed after considering LHC constraints (shaded blue areas) for a MSSM $(g-2)_{\mu}$ solution at $1\sigma$ ($2\sigma$) in orange (yellow), with a bino-like LSP, degenerate smuons and $\tan\beta=30$. Right: plot from Ref.~\cite{Cadeddu:2021dqx} showing in green (light blue) the preferred region by the current $(g-2)_{\mu}$ ($(g-2)_{e}$) measurement in the plane of the kinetic mixing parameter $\epsilon$ and of the mass the dark $Z_d$ mediator; the dotted and dashed lines corresponds to the limits from the $Q_{\mathrm{WEAK}}$ and APV experiments respectively, while the solid lines represent their combinations; the $\delta$ parameter is related to the $Z$--$Z_d$ mass mixing.}
  \label{fig:gm2plot}
\end{figure}

The possibility of explaining the anomaly in non-minimal supersymmetric extensions of the SM, in light of recent LHC constraints, has also been extensively studied
in the literature, cf., e.g., Refs.~\cite{Kim:2001se, Shimizu:2015ara,Frank:2017ohg,Ning:2017dng,Li:2017fbg,Choudhury:2017fuu,Wang:2018vxp,Yang:2018guw,Kotlarski:2019muo,Dong:2019iaf,Yang:2020bmh,Cao:2019evo,Han:2020exx,Zhang:2021gun,Abdughani:2021pdc,Yang:2021duj,Aboubrahim:2021rwz,Heinemeyer:2021zpc,Altmannshofer:2021hfu}.

\paragraph{Leptoquarks}
Another possible explanation, which also provides a viable solution to the hints for lepton flavour universality violation in semi-leptonic $B$ decays, is given by leptoquarks.
Indeed, two scalar LQ representations can provide a chiral enhancement factor of {$m_t/m_\mu \approx 1600$}~\cite{Djouadi:1989md,
Chakraverty:2001yg,Cheung:2001ip,Bauer:2015knc,ColuccioLeskow:2016dox,Crivellin:2020tsz}. This allows for a TeV scale explanation with perturbative couplings that is not in conflict with direct LHC searches. It is furthermore very predictive as it involves, besides the LQ mass only two couplings, whose product is fixed by requiring that $g-2$ be explained. Therefore, correlated effects in $h\to\mu^+\mu^-$~\cite{Crivellin:2020tsz}, and to a lesser extent in $Z\to\mu^+\mu^-$~\cite{ColuccioLeskow:2016dox,Crivellin:2020mjs}, arise which can be used to (indirectly) distinguish the two LQ representations at future colliders. In fact, correlations with $h\to \mu^+\mu^-$ and $Z\to\mu^+\mu^-$ become of interest for a wide range of chirally enhanced scenarios, see Ref.~\cite{Crivellin:2021rbq}.

\paragraph{Interplay with the lepton EDMs}
A consequence of explanations via chiral enhancement concerns the phase of the Wilson coefficient of the dipole operator, which emerges as a free parameter. In particular, such scenarios in general violate
the scaling expected from MFV~\cite{Giudice:2012ms,Crivellin:2018qmi},
which may result in a large muon EDM well above MFV projections
derived from the limit on the electron EDM~\cite{Andreev:2018ayy}. A large part of the parameter space in which the phase is ${\mathcal O}(1)$, as well possible from an EFT
perspective~\cite{Pruna:2017tif,Crivellin:2018qmi,Crivellin:2019mvj}, could be covered by a proposed dedicated muon EDM experiment at
PSI~\cite{Adelmann:2021udj}.  In fact, the corresponding non-MFV flavour structure is not at odds with naturalness arguments, since, in the limit of vanishing neutrino masses, lepton flavour is conserved, and thus it is possible to completely disentangle the muon from the electron EDM via a symmetry, meaning that no fine tuning is
necessary. This could for example be achieved via an $L_\mu-L_\tau$ symmetry~\cite{He:1990pn,Foot:1990mn,He:1991qd}, which can naturally give rise to the observed neutrino mixing
matrix~\cite{Binetruy:1996cs,Bell:2000vh,Choubey:2004hn}, and, even after its breaking, protects the electron EDM and $g-2$ from BSM contributions~\cite{Altmannshofer:2016oaq}.

\paragraph{Other solutions via heavy NP}
In addition, there exists a plethora of alternative BSM explanations
of the muon $g-2$, including composite or extra-dimensional
models~\cite{Das:2001it,Xiong:2001rt,Park:2001uc} or models with vector-like leptons~\cite{Stockinger:1900zz,Giudice:2012ms,Dermisek:2013gta,Falkowski:2013jya,Altmannshofer:2016oaq,Kowalska:2017iqv,Crivellin:2018qmi,Arnan:2019uhr,Crivellin:2021rbq}, including in addition a second Higgs doublet~\cite{Ferreira:2021gke,Chun:2020uzw,Frank:2020smf}.

Also a pure 2HDM can provide a solution. This is either possible via Barr-Zee diagrams in the 2HDM-X~\cite{Cao:2009as,Broggio:2014mna,Wang:2014sda,Ilisie:2015tra,Abe:2015oca,Crivellin:2015hha,Chun:2015hsa,Chun:2016hzs,Cherchiglia:2017uwv,Wang:2018hnw,Chun:2019oix,Chun:2019sjo,Athron:2021iuf,Ferreira:2021gke}, where the external photon
couples to an internal charged fermion loop, which then couples to the muon line via one of the new Higgs bosons and a photon (as in Fig.~\ref{gm2:mssm1l} right, but with the sfermion replaced by a fermion) or including vector-like leptons. Alternatively, a lepton flavour violating $\tau\mu$ couplings can provide a $m_\tau/m_\mu$ enhancement~\cite{Crivellin:2019dun,Iguro:2020rby,Wang:2021fkn,Hou:2021qmf}, which is however strongly constrained from $h\to\tau\mu$ searches.

\paragraph{Weakly coupled models with new light states}
Another possibility to explain the anomaly is to have weakly coupled new states (sometimes
called generally Feebly Interacting Particles -- FIPs) that however can provide a significant contribution to their small mass, a case for which a rich literature is available~\cite{Holdom:1985ag,Pospelov:2008zw,Chen:2015vqy,Davoudiasl:2018fbb,Darme:2020sjf,Cadeddu:2021dqx,Buen-Abad:2021fwq,Darme:2021qzw}. Below we mention only a small selection of these studies.
In the case of a spin-1 explanation, an example is given by dark $Z$ models. In Fig.~\ref{fig:gm2plot} we show a plot taken from Ref.~\cite{Cadeddu:2021dqx} where a dark $Z_d$ mediator model was studied. In the plot, the interplay with other current and future low-energy experiments is also shown.
Concerning the possibility of an axion-like explanation, a recent paper~\cite{Buen-Abad:2021fwq} points out that this prospect poses some problems due to the fact that it seems to require an axion decay constant of $\mathcal{O}(10)$\,GeV, which in turn implies the existence of new states at low scales, creating phenomenological issues which are not easily addressable.
Another interesting possibility is given by ALP-portal explanations, where the ALP assumes the role of mediator with a dark sector~\cite{Darme:2020sjf}. The possibility of explaining simultaneously $(g-2)_{\mu}$ and the flavour anomalies using FIPs has been recently presented in Ref.~\cite{Darme:2021qzw}.

\subsubsection{Cabibbo Angle Anomaly}
\begin{figure}
\centering
\includegraphics[width=0.8\textwidth]{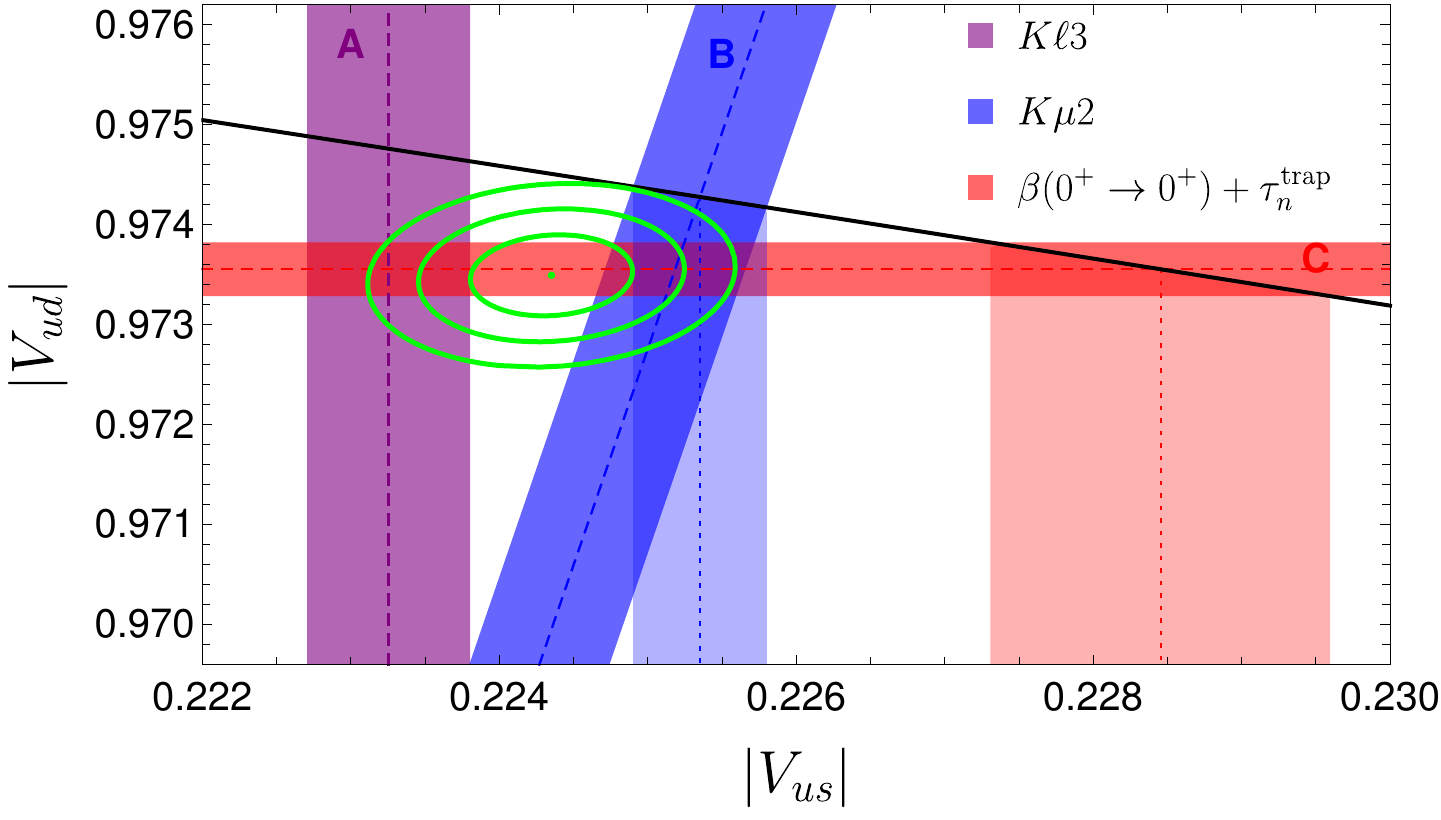}
\caption{Updated plot of Ref.~\cite{Belfatto:2019swo} for the data in Eq.~\eqref{caa:ABC} in $V_{us}$--$V_{ud}$ plane. The $1\sigma$, $2\sigma$ and $3\sigma$ contours (green circles) of the fit are in tension with CKM unitarity (black solid curve). The projections on $\vert V_{us}\vert$ axis show the values $\vert V_{us}\vert_B$ and $\vert V_{us}\vert_C$ obtained from the unitarity condition.}
\label{caa:vusvud}
\end{figure}

% Benedetta Belfatto, Zurab Berezhiani, Claudio Andrea Manzari and Matthew Kirk
\label{subsec:CAA} 
One of the fundamental predictions of the SM is the unitarity 
of the CKM matrix. In particular, for the first row of CKM elements it implies the condition
\begin{equation}\label{caa:unitarity} 
\vert V_{ud} \vert^2 + \vert V_{us} \vert^2 + \vert V_{ub} \vert^2 = 1  \,,
\end{equation}
which in practice reduces to the Cabibbo universality
($\vert V_{ud} \vert \approx \cos\theta_{12}$, $\vert V_{us} \vert \approx \sin\theta_{12}$),   
since the last entry is negligibly small: $\vert V_{ub} \vert^2 < 2\times 10^{-5}$ ~\cite{Zyla:2020zbs}. At present, with the improved control of theoretical uncertainties in the determinations of $\vert V_{us} \vert$ and $\vert V_{ud} \vert$, anomalies are emerging that could be a signal of NP at the TeV scale~\cite{Belfatto:2019swo,Coutinho:2019aiy,Crivellin:2020lzu,Manzari:2020eum,Grossman:2019bzp,Belfatto:2021jhf}. 
The present situation is shown in Fig.~\ref{caa:vusvud} and can be summarised as  
\begin{equation}\label{caa:ABC} 
\mathrm{A:} ~~ \vert V_{us} \vert = 0.22326(55), \quad 
\mathrm{B:} ~~ \vert V_{us}/V_{ud} \vert = 0.23130(49), \quad    
\mathrm{C:} ~~ \vert V_{ud} \vert = 0.97355(27)  \,.
\end{equation}
The first two results A and B are extracted from data on kaon 
semileptonic $K_{\ell 3}$ and leptonic $K_{\mu2}$  decays~\cite{Zyla:2020zbs}, respectively,
using the most accurate lattice QCD calculations for the vector form factor 
$f_+(0)$ and for the decay constants ratio $f_K/f_\pi$~\cite{Aoki:2019cca}. 
The precision of the third result C crucially depends on the knowledge of radiative corrections to be applied in $\beta$ decays~\cite{Marciano:2005ec,Seng:2018yzq,Seng:2018qru,Czarnecki:2019mwq,Seng:2020wjq,Hayen:2020cxh,Shiells:2020fqp}.  Using the value of the Fermi constant $G_F = 1.1663787(6)\times 10^{-5}$\,GeV$^{-2}$
from muon decay~\cite{Tishchenko:2012ie}, the value of $\vert V_{ud} \vert $ is then obtained from the latest update of ${\cal F}t$ values in superallowed $0^+$--$0^+$ nuclear transitions, ${\cal F}t=3072.24(1.85)$s~\cite{Hardy:2020qwl}, but affected by additional nuclear corrections~\cite{Hardy:2020qwl,Gorchtein:2018fxl}. The extracted value of $\vert V_{ud} \vert $ is consistent with a determination via neutron decay (included in C above), based on 
the average of the neutron lifetimes measured by the eight latest experiments using the neutron trap method,  $\tau_n = 879.4(6)$s, and employing the latest experimental average $g_A=1.27625(50)$ for the axial coupling.\footnote{Let us note, however, that the neutron lifetime  measured in beam experiments, $\tau_n=888.0(2.0)$s, is $4\sigma$ away from the trap value, and it is also incompatible with superallowed $0^+$--$0^+$ transitions~\cite{Czarnecki:2018okw}. This discrepancy, barring the possibility of some unfixed systematics, could be another indication for NP~\cite{Fornal:2018eol,Berezhiani:2018eds,Grinstein:2018ptl,Berezhiani:2018udo}.} Even when using the currently most precise measurements, $\tau_n =  877.75(0.28)^{+0.22}_{-0.16}$s~\cite{UCNt:2021pcg} and $g_A=1.27641(45)(33)$~\cite{Markisch:2018ndu}, the determination from neutron decay is not yet competitive with superallowed $\beta$ decays, but will provide a powerful independent determination in the future. In addition, note that there is also a deficit in the first-column CKM unitarity relation
\begin{equation}
\vert V_{ud} \vert^2 + \vert V_{cd} \vert^2 + \vert V_{td} \vert^2 = 0.9970(18)  \,,
\end{equation}
less significant than the tension in the first row, but strengthening the possibility of NP related to the determination of $V_{ud}$.

A fit of the data in Eq.~\eqref{caa:ABC} shows the deviation from unitarity at the $3\sigma$ level (see Fig.~\ref{caa:vusvud}). Alternatively, by employing unitarity, these data can be translated into three different results for the Cabibbo angle: 
$\vert V_{us} \vert_A = 0.22326(55)$, $\vert V_{us}\vert_B= 0.22535(45)$ and $\vert V_{us}\vert_C= 0.2284(11)$, which are in obvious tension with each other. It has also been shown that the discrepancy between the $K_{\ell3}$ and $K_{\mu2}$ results 
$\vert V_{us} \vert_A$ and $\vert V_{us} \vert_B$ is unlikely to be due to radiative corrections~\cite{Seng:2021wcf,DiCarlo:2019thl}. 

In fact, the three determinations of the Cabibbo angle do not necessarily correspond to quite the same values in the presence of NP. The amplitudes of the $K_{\ell3}$ and $K_{\mu2}$ decays are proportional to vector $\overline{u} \gamma^\mu s$ and axial $\overline{u} \gamma^\mu\gamma^5 s$ currents, respectively. On the other hand, superallowed nuclear transitions are only sensitive to vector current $\overline{u} \gamma^\mu d$, and the Fermi constant, which is fixed by the muon decay width, could also be affected by NP~\cite{Crivellin:2021njn}. Therefore, the Cabibbo angle anomaly (CAA) can be a signal of BSM physics, which can also have other phenomenological implications and be related to other existing anomalies. The different possible explanations can be broadly grouped into three categories: modifications of the \(W\) quark vertex, modifications to the \(W\) lepton vertex, or effects in four-fermion contact interaction operators.\footnote{The different tree-level extension of the SM, which give rise to these effects have been categorised in Ref.~\cite{deBlas:2017xtg}.} 

\paragraph{\boldmath Modifying the \(W\) quark vertex}
The $W$ couplings to quarks are modified after EW symmetry breaking by the two operators: \(Q_{\phi q}^{(3)ij}\) and \(Q_{\phi ud}^{ij}\,\) (see Ref.~\cite{Grzadkowski:2010es} for the definitions of the operators). The latter generates right-handed $W$-quarks couplings and it has been showed that the interplay between \(C_{\phi ud}^{11}\,\) and \(C_{\phi ud}^{12}\,\) can solve the tension in the CAA~\cite{Grossman:2019bzp} and bring the determinations of $|V_{us}|$ from $K_{\ell3}$ and $K_{\mu2}$ into agreement. \(Q_{\phi q}^{(3)ij}\) generates left-handed $W$-quark couplings and the CAA requires \(C_{\phi q}^{(3)11}\approx -(9\,\text{TeV})^{-2}\). These operators can be induced via the mixing of SM quarks with vector-like quarks~\cite{Belfatto:2019swo,Cheung:2020vqm,Belfatto:2021jhf,Branco:2021vhs,Crivellin:2021bkd}. 
Note however, that because of \(SU(2)\) invariance this operator generates also effects in \(\Delta F=2\) processes which would rule out this solution, unless these effects are suppressed by assuming that \(Q_{\phi q}^{(3)ij}\) respect a global \(U(2)^2\) flavour symmetry.

\paragraph{\boldmath Modifying the $W$ lepton vertex}
The SMEFT coefficient $C_{\phi \ell}^{(3)}$ corresponds to modifications of the \(W\ell\nu\) and \(Z\ell\ell\) leptonic currents after EW symmetry breaking. This is also interesting for another reason -- namely that since this coefficient carries flavour indices such that NP could be related to LFUV~\cite{Coutinho:2019aiy,Crivellin:2020lzu}, which ties in to many of the other anomalies discussed in this white paper. In order to explain the CAA, this coefficient must be approximately \(C_{\phi \ell}^{(3)22} \approx (9\,\text{TeV})^{-2}\)~\cite{Kirk:2020wdk,Crivellin:2020ebi}.

There are four different vector-like leptons (VLLs), as well as an \(SU(2)_L\) triplet vector boson, that can be responsible for generating this operator, and hence explain the CAA.
The phenomenology of VLLs for the CAA has been studied in detail~\cite{Endo:2020tkb,Crivellin:2020ebi,Kirk:2020wdk,Manzari:2021prf}
and the vector triplet idea has been examined in Refs.~\cite{Kirk:2020wdk,Capdevila:2020rrl}.

For the VLLs, extra phenomenological consequences arise since the \(SU(2)\) singlet operator \(Q_{H\ell}^{(1)}\) is generated. This operator alters leptonic \(Z\) decays, which are strongly constrained by EW precision observables from LEP and the LHC. Such effects allow us to distinguish between the different models. E.g., it has been shown that extending the SM with a single VLL~\cite{Kirk:2020wdk,Alok:2020jod} leads to tensions between the region of parameter space favoured by the CAA and EW observables. Adding multiple new representations, each coupling to a single different SM lepton can provide a better fit to data~\cite{Crivellin:2020ebi}. For the vector triplet boson, a minimal model leads to tensions with EW precision observables~\cite{Kirk:2020wdk}, which can be eased in a less minimal setup~\cite{Capdevila:2020rrl}.

\paragraph{Four-fermion operators}
There are several four-fermion operators in the SMEFT that can affect the determination of the Fermi constant or directly alter semi-leptonic decays~\cite{Crivellin:2021njn}. Starting with four-lepton operators, the severe constraints from the Michel parameter, muonium--anti-muonium oscillations and the upper bounds on LFV processes lead to the conclusion that the only viable solution to the CAA proceeds via a modification of the SM operator \(Q_{\ell\ell}^{2112}\) with a Wilson coefficient \(C_{\ell\ell}^{2112} \approx -(8\,\text{TeV})^{-2}\).
Simple models generating this contribution via a singly charged scalar have been recently proposed in Refs.~\cite{Crivellin:2020oup,Crivellin:2020klg,Marzocca:2021azj}. This option was also proposed in Ref.~\cite{Belfatto:2019swo} via a generic flavour-changing boson (see also Ref.~\cite{Buras:2021btx}), which can be induced by gauge bosons of chiral inter-family symmetry~\cite{Berezhiani:1983rk,Berezhiani:1990wn}.
All these possibilities lead to constructive interfere with the SM in muon decay such that the Fermi constant of the Lagrangian \(G_F\) is smaller than the one measured from by the muon lifetime. We note that while this type of solution resolves the tension between A/B determinations with C, it can only slightly alleviate the tension between A and B themselves and leads to additional tensions in the EW fit~\cite{Crivellin:2021njn}.

Concerning 2-quark--2-lepton operators, only \(Q_{\ell q}^{(3)1111}\) is able to give a sizable BSM effect in \(\beta\) decays via interference with the SM and the CAA requires \(C_{\ell q}^{(3)1111}\approx (11\,\text{TeV})^{-2}\). Possible extensions of the SM that induce this operator are LQs~\cite{Crivellin:2021tmz,Crivellin:2021bkd} or a colour-neutral vector triplet~\cite{Capdevila:2020rrl}. It is also worth noting that the size required to explain the CAA is compatible with the one preferred by CMS searches for $pp\to e^+e^-$, see Sec.~\ref{CMSppee}. Finally, scalar interactions are typically negligible for first-generation fermions, while most severely constrained from processes that display chiral enhancement~\cite{Bryman:2011zz,PiENu:2015seu,Hoferichter:2021lct,KTeV:2006pwx}.

\subsubsection{Lepton flavour universality in the charged current}
% Antonio Pich
%
The recently observed anomalies in $b\to c\tau\nu$ and $b\to s\mu^+\mu^-$ transitions suggest a possible violation of lepton universality in other processes where strong constraints on the universality of the leptonic $W^\pm$ couplings $g_\ell$ ($\ell = e, \mu, \tau$), emerging from the measured weak decays of the $\mu$, $\tau$, $\pi$ and $K$, exist. The most accurate phenomenological tests of the universality of the leptonic charged-current couplings are summarised in Table~\ref{tab:ccuniv}, which updates Ref.~\cite{Pich:2013lsa}.

The leptonic decays $\ell\to\ell'\bar\nu_{\ell'}\nu_\ell$ provide very clean measurements of the $W^\pm$ couplings. The $\tau\to\mu/\tau\to e$ ratio directly constrains $|g_\mu/g_e|$, while the comparison of $\tau\to e,\mu$ with $\mu\to e$ provides information on $|g_\tau/g_\mu|$ and $|g_\tau/g_e|$. 
Taking into account the different lepton masses involved and the small higher-order electroweak corrections, the current data confirm the universality of the leptonic $W^\pm$ couplings with a $0.15$\% precision.

A slightly better sensitivity on $|g_\mu/g_e|$ has been obtained from the precisely measured ratio of the $\pi^-\to e^-\bar\nu_e$ and $\pi^-\to \mu^-\bar\nu_\mu$ decay widths~\cite{Zyla:2020zbs}. At this level of precision, a good control of radiative QED corrections is compulsory~\cite{Cirigliano:2007ga,Cirigliano:2007xi}. Comparable accuracies have been also reached from the corresponding $e/\mu$ ratios in $K_{\ell 2}$ and $K_{\ell 3}$ decays~\cite{Cirigliano:2011ny}.

The comparison of the $\tau^-\to P^-\nu_\tau$ and $P^-\to\mu^-\bar\nu_\mu$ ($P=\pi,K$) decay widths allows for an independent  determination of $|g_\tau/g_\mu|$. The radiative corrections to these ratios involve low-energy hadronic effects that have been recently re-evaluated~\cite{Arroyo-Urena:2021nil}, using Chiral Perturbation Theory techniques and the large-$N_C$ expansion. While this updated calculation agrees with previous evaluations, the estimated hadronic uncertainties are found to be slightly larger. Nevertheless, one obtains a quite accurate test of universality at the $0.4\%$ ($0.8\%$) level from the $\pi$ ($K$) ratios.

\begin{table}[tb]\centering
\caption{Experimental determinations of the ratios \ $g_\ell/g_{\ell'}$~\cite{Pich:2013lsa,Zyla:2020zbs,Arroyo-Urena:2021nil,Aad:2020ayz}.}
\label{tab:ccuniv}
\begin{tabular}{llllll}
\hline\noalign{\smallskip} 
& $\!\Gamma_{\tau\to\mu}/\Gamma_{\tau\to e}$ &
 $\Gamma_{\pi\to\mu} /\Gamma_{\pi\to e}$ &
 $\Gamma_{K\to\mu} /\Gamma_{K\to e}$ &
 $\Gamma_{K\to\pi\mu} /\Gamma_{K\to\pi e}$ &
 $\Gamma_{W\to\mu} /\Gamma_{W\to e}$
\\ \noalign{\smallskip} \hline\noalign{\smallskip}
 $|g_\mu/g_e|$
 & $\!1.0017\; (16)$ & $1.0010\; (9)$ & $0.9978\; (18)$ & $1.0010\; (25)$ & $0.998\; (4)$
\\ \noalign{\smallskip} \hline\hline\noalign{\smallskip}
%%% Present constraints on $|g_\tau/g_\mu|$
& $\!\Gamma_{\tau\to e}/\Gamma_{\mu\to e}$ &
 $\Gamma_{\tau\to\pi}/\Gamma_{\pi\to\mu}$ &
 $\Gamma_{\tau\to K}/\Gamma_{K\to\mu}$ &
 $\Gamma_{W\to\tau}/\Gamma_{W\to\mu}$
\\ \noalign{\smallskip} \hline\noalign{\smallskip}
 $|g_\tau/g_\mu|$
 & $\!1.0011\; (14)$ & $0.9964\; (38)$ & $0.9857\; (78)$ & $1.004\; (16)$
\\ \noalign{\smallskip} \hline\hline\noalign{\smallskip}
%%% Present constraints on $|g_\tau/g_e|$
& $\!\Gamma_{\tau\to\mu}/\Gamma_{\mu\to e}$
 & $\Gamma_{W\to\tau}/\Gamma_{W\to e}$
\\ \noalign{\smallskip} \hline\noalign{\smallskip}
 $|g_\tau/g_e|$
 & $\!1.0028\; (15)$ & $1.021\; (12)$
\\ \noalign{\smallskip} \hline
\end{tabular}
\end{table}

The decays $W\to\ell\bar\nu_\ell$ provide a more direct access to the leptonic $W$ couplings. However, with the limited statistics collected at LEP it was only possible to reach precisions of $O(1\%)$~\cite{ALEPH:2013dgf}. The LEP data exhibited a slight excess of $W\to\tau\bar\nu_\tau$ events, implying $2.7\%$ and $2.4\%$ deviations from lepton universality  in $|g_\tau/g_\mu|$ and $|g_\tau/g_e|$, respectively. This was very difficult to reconcile with the much more precise indirect constraints from $\tau,\mu, \pi$ and $K$ decays~\cite{Filipuzzi:2012mg}. 

The large amount of data provided by the LHC has made it possible to perform more precise tests of the leptonic $W$ decays. The recent ATLAS determination of $\Gamma_{W\to\tau}/\Gamma_{W\to \mu}$~\cite{Aad:2020ayz} agrees well with the SM expectation. The ATLAS measurement alone would imply $|g_\tau/g_\mu| =0.996\pm 0.007$. The larger error quoted in Table~\ref{tab:ccuniv} reflects the sizable discrepancy with the old LEP value. A preliminary CMS measurement of the $W$ leptonic branching fractions~\cite{CMS:2021qxj}, not yet included in Table~\ref{tab:ccuniv}, fully confirms the ATLAS result, eliminating the long-standing $W\to\tau$ anomaly. The separate results from LEP and the LHC experiments are collected in Table~\ref{tab:Wtest}, which also displays the preliminary world averages including the Tevatron data. Following the PDG prescription, the errors of the $|g_\tau/g_\mu|$ and $|g_\tau/g_e|$ averages have been increased to account for the discrepancy with the LEP values.
% showing the importance of performing new precise measurements of statistically-limited observables. 

Clearly, the current data verify the universality of the leptonic $W$ couplings to the $0.15\%$ level. In Table~\ref{tab:ccuniv} one can only identify two small deviations that do not reach the $2\sigma$ level: there is a slight ($1.9\sigma$) excess of $\tau\to\mu$ versus $\mu\to e$ events, and a small deficit ($1.8\sigma$) of $\tau\to K$ versus $K\to\mu$ transitions. The relatively large hadronic uncertainty involved~\cite{Arroyo-Urena:2021nil} could easily explain the second deviation, although a systematic deficit of kaon final states seems to be present in $\tau$ decays~\cite{Pich:2013lsa}, leading to a determination of $|V_{us}|$ slightly lower than the one obtained from kaon decays~\cite{Gamiz:2004ar,Gamiz:2006xx,Pich:2020gzz}.
The slight excess of $\tau\to\mu$ events could be correlated with a possible explanation of the Cabibbo anomaly through a slight violation of lepton universality~\cite{Crivellin:2020lzu,Crivellin:2020oup,Crivellin:2020klg}.
In any case, more precise experimental studies are needed.

The different universality tests provide complementary information, since they are sensitive to different types of NP contributions. While the decays of the $W$ boson probe directly its leptonic couplings, the indirect constraints from low-energy leptonic and semileptonic decays test the potential presence of additional intermediate particles, which could modify each analysed process in a different way. From this point of view, it is worth to mention the universality test extracted from the ratio of $B\to D^{(*)}\mu\nu$ and 
$B\to D^{(*)}e\nu$ transitions: $|g_\mu/g_e| = 0.989\; (12)$~\cite{Jung:2018lfu}. Although much less precise that the other indirect determinations in Table~\ref{tab:ccuniv}, it severely restricts the type of possible explanations to the $b\to c\tau\nu$ anomaly.

\begin{table}[tb]\centering
\caption{Experimental determinations of the ratios \ $g_\ell/g_{\ell'}$
from $W\to\ell\bar\nu_\ell$ decays.}
\label{tab:Wtest}
\begin{tabular}{llllll}
\hline\noalign{\smallskip} 
& LEP~\cite{ALEPH:2013dgf} & LHCb~\cite{LHCb:2016zpq} & ATLAS~\cite{Aad:2020ayz,ATLAS:2016nqi} &
 CMS~\cite{CMS:2021qxj} & Average
\\  &&&& (prelim.) & (prelim.)
\\ \noalign{\smallskip} \hline\noalign{\smallskip}
 $|g_\mu/g_e|$
  & $0.996\; (10)$ & $0.990\; (9)$ & $1.002\; (5)$ & $1.005\; (6)$ & $1.000\; (3)$
\\ \noalign{\smallskip} \hline\noalign{\smallskip}
 $|g_\tau/g_\mu|$
 & $1.034\; (13)$ && $0.996\; (7)$ & $0.992\; (10)$ & $1.001\; (10)$
\\ \noalign{\smallskip} \hline\noalign{\smallskip}
 $|g_\tau/g_e|$
 & $1.031\; (13)$ && & $0.997\; (11)$ & $1.008\; (12)$
\\ \noalign{\smallskip} \hline
\end{tabular}
\end{table}

\subsubsection{Non-resonant di-electrons}
\label{CMSppee}
% Claudio Andrea Manzari, Luc Schnell and Andreas Crivellin
%
A search for LFUV in the non-resonant production of di-leptons was recently performed by CMS, observing a $\approx$ $4\,\sigma$ excess in electron pairs with an invariant mass greater than $1.8\,$TeV in the $pp\to e^+e^-$ channel. As the muon channel agrees with the SM expectation, this measurement points towards LFUV. Also ATLAS~\cite{Aad:2020otl} and HERA~\cite{Abramowicz:2019uti} found more di-electrons than expected in their studies of quark-lepton contact interactions. 

\begin{figure}
	\centering
  	\includegraphics[height=0.5\textwidth]{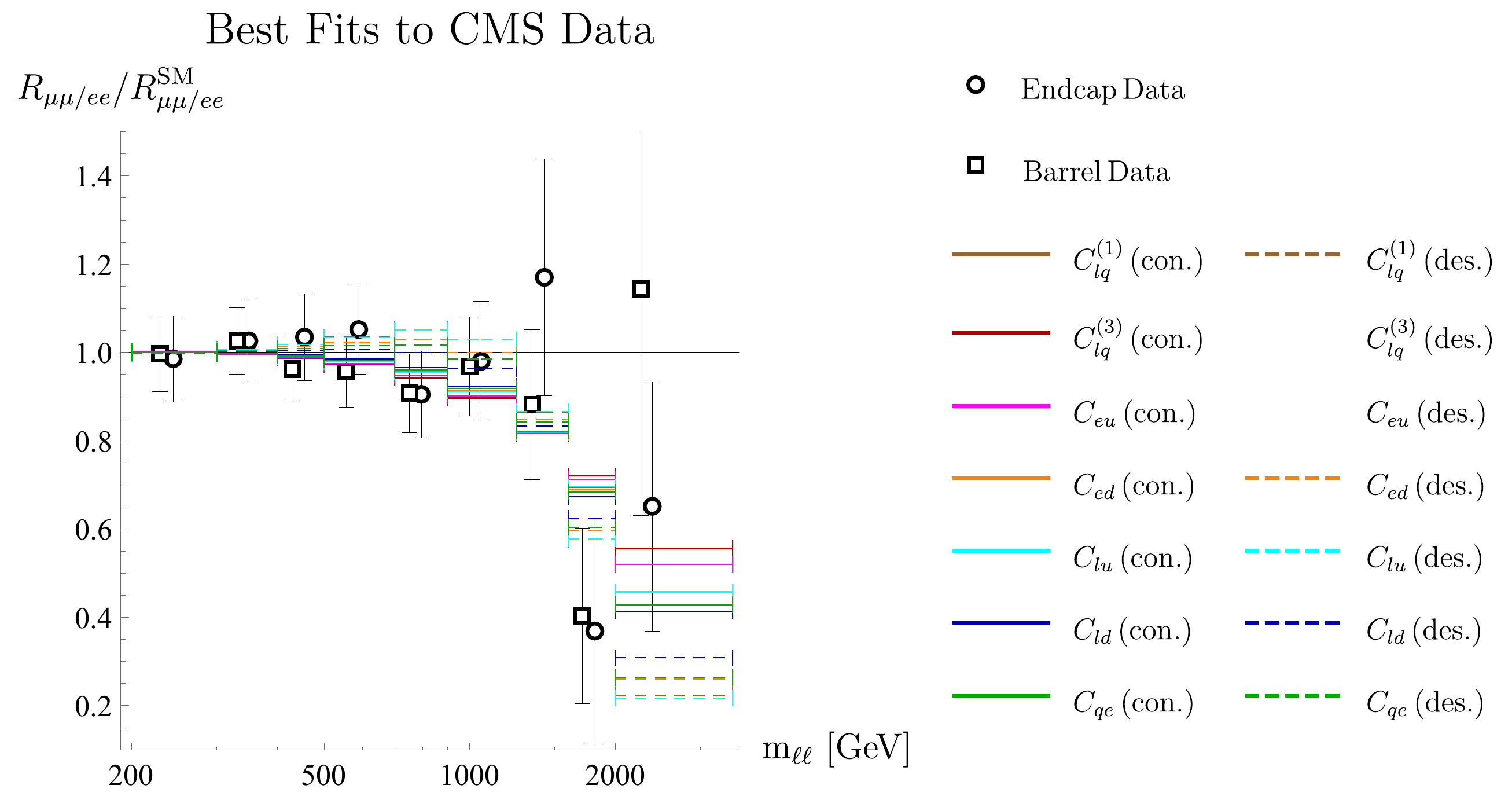}
  	\caption{The double ratio values $R_{\mu \mu/e e, ij} \big/ R_{\mu \mu/ e e, ij}^{\text{SM}}$ measured by the CMS collaboration in nine $m_{\ell \ell}$ bins between 200 and 3000\,GeV are shown (black circles and squares) together with the best fit curves for different 2-quark--2-lepton operator scenarios (coloured lines). }
  	\label{fig:CMSDY}
\end{figure}

In addition to the total cross-section, the CMS collaboration provided the differential cross-section ratio 
\begin{equation}
R_{\mu \mu/e e} \equiv \frac{d \sigma (pp \to \mu^+ \mu^-) / dm_{\mu \mu}}{d \sigma (p p \to e^+ e^-) / dm_{e e}}\,,
\end{equation}
for different $m_{\ell \ell}$ ($\ell = e, \mu$) bins. For each bin, they quoted two values, distinguishing the cases where zero (at least one) of the di-leptons were detected in the endcaps, corresponding to barrel only (endcap) measurements. These were then compared to the SM predictions obtained from Monte Carlo simulations
\begin{equation}
R_{\mu \mu/ee}^{\text{Data}} \big/ R_{\mu \mu / e e}^{\text{MC}}\,.
\end{equation}
In this double ratio, many of the experimental and theoretical uncertainties cancel~\cite{Greljo:2017vvb}. It was further normalised to one in the bin from 200 to 400\,GeV to correct for the relative sensitivity to electrons and muons. The measurements are indicated by the black squares (circles) for the barrel only (endcap) measurements in Fig.~\ref{fig:CMSDY}. A trend towards values smaller than 1 is visible for large $m_{\ell \ell}$.

ATLAS followed a different strategy, measuring the differential cross-section
\begin{equation}
d \sigma (p p \to \ell^+ \ell^-) / dm_{\ell \ell} \text{~~~~for~~~~} \ell = e, \mu\,,
\end{equation}
integrated over the signal regions [2.2, 6.0] TeV and [2.7, 6.0] TeV for searches of NP interfering constructively and destructively with the SM, respectively~\cite{Aad:2020otl}. Comparing these results to the SM predictions, they derived exclusion limits for NP coupling equally to up and down quarks of both chiralities. Interestingly, also ATLAS measured more di-electrons than expected in the constructive channel, but their measurements are still consistent with the SM prediction.

Potential explanations to the di-electron excess in the CMS measurements, including the constraints from the ATLAS analysis, have recently been discussed in Refs.~\cite{Crivellin:2021egp,Crivellin:2021rbf,Crivellin:2021bkd}. The presence of NP was studied within an EFT approach, considering only operators coupling to first-generation quarks and leptons. The Wilson coefficients $C_i$ of the relevant operators are reported in Table~\ref{tab:2q2lExplanationstoDY}. For each of them, the ratio $R_{\mu \mu / ee, ij}^{\text{SM+NP}}(C_i) \big / R_{\mu \mu / ee, ij}^{\text{SM}}$ was computed and fitted to CMS data using a $\chi^2$ statistical analysis with 
\begin{equation}
\begin{aligned}
\chi^2(C_{i}) \equiv \sum_{\substack{i=1, \dots, 9 \\ j = e,b}}\dfrac{\left(\dfrac{R_{\mu \mu / e e, ij}^{\text{Data}}}{R_{\mu \mu / e e, ij}^{\text{MC}}} - \dfrac{R_{\mu \mu / e e, ij}^{\text{SM+NP}}(C_{i})}{R_{\mu \mu / ee, ij}^{\text{SM}}} \right)^2}{\sigma_{ij}^2} \,,
\end{aligned}
\end{equation}
where $\sigma_{ij}$ are the corresponding uncertainties reported in Ref.~\cite{Sirunyan:2021khd}. Additionally, the $\text{ATLAS}$ exclusion limits were recast for the cases where the NP couples differently to up and down quarks and to the different chiralities. The preferred (excluded) values for the Wilson coefficient, given by the CMS (ATLAS) analysis are shown in Table~\ref{tab:2q2lExplanationstoDY} and the best fits to CMS data are displayed in Fig.~\ref{fig:CMSDY}. NP interfering constructively with the SM contribution is preferred and the operators considered can improve the fit with a pull of up to $3.3 \sigma$. While the corresponding Wilson coefficient values are not excluded by the ATLAS analysis, constraints coming from kaon decays,  $K^0$--$\bar{K}^0$ mixing or $D^0$--$\bar{D}^0$ mixing are more stringent and need to be considered when addressing the CMS excess. Finally, it is important to note that the operator $[Q_{\ell q}^{(3)}]_{1111}$ can also provide an explanation to the Cabibbo Angle Anomaly, discussed in Sec.~\ref{subsec:CAA}. In Ref.~\cite{Crivellin:2021rbf} it was shown that $[Q_{\ell q}^{(3)}]_{1111}$ can provide a simultaneous explanation for these two anomalies and a best-fit value of $[C_{\ell q}^{(3)}]_{1111}=1.1/(10\, \rm TeV)^2$ was extracted.
\begin{center}
	\begin{table}
		\centering
		\begin{tabular}{c|cccccccccc}
			Coefficient & $C_{lq}^{(1)}$ & $C_{lq}^{(3)}$ & $C_{eu}$ & $C_{ed}$ & $C_{lu}$ & $C_{ld}$ & $C_{qe}$\\
			\hline
			CMS (con.) & $-20^{+6}_{-7}$ & $+10^{+4}_{-3}$ & $-17^{+6}_{-6}$ & $+37^{+11}_{-11}$ & $-25^{+8}_{-8}$ & $+44^{+12}_{-12}$ & $-24^{+7}_{-7}$\\
			Pull & $3.3 \sigma$ & $3.3 \sigma$ & $3.3 \sigma$ & $3.3 \sigma$ & $3.3 \sigma$ & $3.3 \sigma$ & $3.3 \sigma$ \\
			CMS (des.) & $+50^{+10}_{-9}$ & * & * & $-74^{+15}_{-16}$ & $+61^{+12}_{-12}$ & $-63^{+14}_{-15}$ & $+43^{+9}_{-9}$ \\
			Pull & $2.3 \sigma$ & * & * & $2.8 \sigma$ & $2.3 \sigma$ & $3.1 \sigma$ & $2.9 \sigma$\\
			ATLAS (con.)& $-19$ & $+13$ & $-18$ & $+36$ & $-24$ & $+39$ & $-23$\\
			ATLAS (des.)& $+24$ & $-28$ & $+30$ & $-40$ & $+26$ & $-37$ & $+21$\\
		\end{tabular}
		\caption{Wilson coefficient values ($\times 10^{-3} \text{ TeV}^{-2}$) that are preferred (excluded) by the CMS (ATLAS) $pp \to e^+e^-$ measurements. For CMS we state the $1\sigma$ ranges as well as the pull to the SM. The asterisk labels scenarios that do not provide an improved fit compared to the SM point. For ATLAS we show the 95\% C.L. exclusion limits. }
		\label{tab:2q2lExplanationstoDY}
	\end{table}
\end{center}

\subsubsection[Forward-backward asymmetry in semi leptonic $B$ decays]{\boldmath Forward-backward asymmetry in semi leptonic $B$ decays} 
%Andreas Crivellin
\label{sec:deltaAFB}
The observable $\Delta A_{\mathrm{FB}}$ encodes the difference of the forward-backward asymmetry in $B\to D^*\mu\nu$ vs $B\to D^*e\nu$. Like for $R(K^{(*)})$ the muon and electron mass can both be neglected such that the form-factor dependence cancels and the SM prediction is, to the currently relevant precision, zero. Even though the corresponding measurements of the total branching ratios are consistent with the SM expectations~\cite{Belle:2015pkj,Belle:2017rcc}, recently Ref.~\cite{Bobeth:2021lya} unveiled a $\approx\!4\sigma$ tension in $\Delta A_{\mathrm{FB}}$, extracted from $B\to D^*\ell\bar \nu$ data of BELLE~\cite{Belle:2018ezy}.

A good fit to data requires a non-zero Wilson coefficient of the tensor operators. Importantly, among the set of renormalisable models, only two scalar LQ can generate this operator at tree level and only the $SU(2)_L$ singlet gives a good fit to data~\cite{Carvunis:2021dss}. However, even in this case, due to the constraints from other asymmetries, $\Delta A_{FB}$ cannot be fully explained but the global fit to $b\to c\mu\nu$ and $b\to c e\nu$ data can be improved by more than $3\sigma$~\cite{Carvunis:2021dss}.

\subsubsection{Combined explanations}
\label{sec:combined_expl} %David
In the presence of the deviations from the SM discussed above, it is reasonable to explore NP models that are able to address, in a coherent manner, more than one anomaly. While it is often possible to simply combine independent solutions for each anomaly, we consider here \emph{combined explanations} in the sense that the mediators are either directly connected from a UV perspective or their \emph{joint} contribution to some anomalous observable is crucial for a successful explanation.

\paragraph{\boldmath Neutral and Charged-current $B$-anomalies}
When combining $b\to s\ell^+\ell^-$ and $R(D^{(*)})$ anomalies, the viable scenarios are dictated by those that offer a good explanation of $R(D^{(*)})$, since this observable is the one that requires the lowest NP scale. As discussed in Sec.~\ref{sec:RD}, the possible solutions necessarily involve LQs, specifically the vector $U_1 \sim ({\bf 3}, {\bf 1}, 2/3)$ or the scalars singlet $S_1 \sim ({\bf \bar 3}, {\bf 1}, 1/3)$ or doublet $R_2 \sim ({\bf 3}, {\bf 2}, 7/6)$. Of these, only the vector LQ  $U_1$ also generates a viable contribution to $b\to s \ell \ell$, and is thus the only single-mediator scenario for combined explanations of $B$-anomalies~\cite{Barbieri:2015yvd,Buttazzo:2017ixm,DiLuzio:2017vat,Bordone:2017bld,Calibbi:2017qbu,Blanke:2018sro,Crivellin:2018yvo,Angelescu:2018tyl,Cornella:2019hct,Fuentes-Martin:2019ign,Fuentes-Martin:2020luw,Fuentes-Martin:2020hvc,Cornella:2021sby,Angelescu:2021lln}. Alternatively, two scalar leptoquarks can also provide a good explanation, by adding the triplet $S_3 \sim ({\bf \bar 3}, {\bf 3}, 1/3) $, that mediates successfully $b\to s \ell \ell$, to $S_1$ or $R_2$.

The $S_1 + S_3$ scenario~\cite{Crivellin:2017zlb,Buttazzo:2017ixm,Marzocca:2018wcf,Arnan:2019olv,Crivellin:2019dwb,Saad:2020ihm,Crivellin:2020ukd,Gherardi:2020qhc,DaRold:2020bib,Bordone:2020lnb} has two viable possibilities: if the $S_1$ couplings to right-handed fermions vanish, then both $S_1$ and $S_3$ contribute to $R(D^{(*)})$ via $C_V^L$; if instead also right-handed couplings are involved, then the largest contribution to $R(D^{(*)})$ arises from $S_1$ via $C_S^L \approx - 4 C_T$ coefficients. From a UV perspective, these two LQ could arise as pseudo-Nambu-Goldstone bosons from a strongly-coupled UV sector, together with the SM Higgs~\cite{Marzocca:2018wcf,DaRold:2020bib}.

In the $R_2 + S_3$ scenario the two leptoquarks contribute mostly independently to  $R(D^{(*)})$ and $b\to s \ell \ell$, respectively~\cite{Babu:2020hun, Dorsner:2017ufx, Saad:2020ucl}. However, these two scalars have been proposed as arising from the same GUT scenario~\cite{Becirevic:2018afm}.

\paragraph{\boldmath $B$-anomalies and anomalous muon magnetic moment}
This scenario can be seen as a further phenomenological requirement on the models discussed in the previous paragraphs. Of the three setups presented as combined explanations of $B$-anomalies, only those involving the scalar leptoquarks $S_1$ and $R_2$ can also provide a satisfactorily contribution to $a_\mu$, thanks to the $m_t / m_\mu$ enhancement.
In this case, however, given the presence of couplings to both left and right-handed muons and tau, a too-large contribution to $\tau \to \mu \gamma$ is induced. This can be cancelled by other contributions from the same mediator, with a tuning at the level of one part in three at the amplitude level~\cite{Crivellin:2019dwb,Saad:2020ihm,Gherardi:2020qhc}.

\paragraph{\boldmath $B$-anomalies, anomalous muon magnetic moment, and CAA}
Possible combined explanations for all these anomalies has been recently proposed involving the scalar LQ $S_1$ and a scalar charged singlet $\phi^+$~\cite{Marzocca:2021azj}. In this scenario $S_1$ contributes to $R(D^{(*)})$ and $a_\mu$ in the same way as described above, while $\phi^+$ generates a tree-level contribution to the muon decay $\mu^- \to e^- \nu_\mu \bar{\nu}_e$ that, shifting the Fermi constant, improves the fit of the Cabibbo angle~\cite{Crivellin:2020oup,Crivellin:2020klg,Felkl:2021qdn,Crivellin:2021njn}. The contribution to $b\to s \mu\mu$ is instead induced via a one-loop box diagram involving both $S_1$ and $\phi$. While this model is able to pass all present bounds, the masses for these scalars are required to be at about 5~TeV, with some couplings reaching somewhat large values of $\approx 3$, which are at the threshold of limits from perturbative unitarity~\cite{Allwicher:2021rtd}.

An alternative BSM scenario that can account for both neutral and charged-current B-anomalies, as well as the muon $g-2$ is based on a minimal $R$-parity violating (RPV) SUSY framework with relatively light third-generation sfermions~\cite{Altmannshofer:2017poe, Altmannshofer:2020axr, BhupalDev:2021ipu}. Although the $LQD$-type RPV couplings of squarks somewhat resemble the scalar LQ couplings, there are some key differences in case of RPV, such as the possibility of explaining muon $g-2$ purely via $LLE$-type couplings, same chiral gauge structure as in the SM, natural flavour violation, as well as many other attractive inbuilt features of SUSY, such as radiative stability of the Higgs boson, radiative neutrino masses, radiative electroweak symmetry breaking, stability of the electroweak vacuum, gauge coupling unification,  (gravitino) dark matter and electroweak baryogenesis.

\subsection{Multi-lepton anomalies at the LHC}
\label{sec:multilepton}
\begin{sloppypar}
One of the implications of a two-Higgs doublet model with an additional singlet scalar $S$ (2HDM+$S$), is the production of multiple-leptons through the decay chain $H\rightarrow Sh,SS$~\cite{vonBuddenbrock:2016rmr}, where $H$ is the heavy CP-even scalar and $h$ is considered as the SM Higgs boson with mass $m_h=125$\,GeV. Excesses in multi-lepton final states were reported in Ref.~\cite{vonBuddenbrock:2017gvy}. In order to further explore results with more data and new final states, while avoiding biases and look-else-where effects, the parameters of the model were fixed in 2017 according to Refs.~\cite{vonBuddenbrock:2016rmr,vonBuddenbrock:2017gvy}. This includes setting the scalar masses to $m_H=270$\,GeV, $m_S=150$\,GeV,\footnote{A possible candidate of the lighter scalar has been reported in Ref.~\cite{Crivellin:2021ubm}.} treating $S$ as a SM Higgs-like scalar and assuming the dominance of the decays $H\rightarrow Sh,SS$. Statistically compelling excesses in opposite sign di-leptons, same-sign di-leptons, and three leptons, with and without the presence of $b$-tagged hadronic jets were reported in Refs.~\cite{vonBuddenbrock:2019ajh,vonBuddenbrock:2020ter,Hernandez:2019geu}. The possible connection with the anomalous magnetic moment of the muon $g-2$ was reported in Ref.~\cite{Sabatta:2019nfg}. Interestingly, the model can explain anomalies in astro-physics (the positron excess of AMS-02~\cite{AMS:2019rhg} and the excess in gamma-ray fluxes from the galactic center measured by Fermi-LAT~\cite{Fermi-LAT:2017opo}) (see Sec.~\ref{sec:astrophysics}) if it is supplemented by a Dark Matter candidate~\cite{Beck:2021xsv}.
\end{sloppypar}

\begin{table}[t]
\begin{center}
      \begin{tabular}{c|c|c|c}
      \hline\hline
Final state & Characteristics & SM backgrounds & Significance \\
\hline
$\ell^+\ell^-$+$b$-jets & $m_{\ell\ell}<100$\,GeV, low $b$-jet mult. & $t\overline{t}, Wt$ & $>5\sigma$ \\ 
$\ell^+\ell^-$+jet veto & $m_{\ell\ell}<100$\,GeV & $W^+W^-$ & $\approx 3\sigma$ \\ 
$\ell^\pm\ell^\pm, 3\ell$ + $b$-jets & Moderate $H_T$ & $t\overline{t}W^{\pm}, t\overline{t}t\overline{t}$ & $>3\sigma$ \\ 
$\ell^\pm\ell^\pm, 3\ell, n_b=0$ & In association with $h$ & $W^{\pm}h, WWW$ & $\approx 4.5\sigma$ \\ 
$Z(\rightarrow\ell\ell)\ell, n_b=0$ & $p_{TZ}<100$\,GeV & $ZW^{\pm}$ & $> 3\sigma$ \\ 
    \hline
      \end{tabular}
      \caption{Summary of the status of the multi-lepton anomalies at the LHC, where $\ell=e,\mu$.}
      \label{tab:anatomymultilepton}
      \end{center}
\end{table}

We give a succinct description of the different final states and corners of the phase-space that are affected by the anomalies. %As discussed in the introduction 
The anomalies are reasonably well captured by a 2HDM+$S$ model. Here, $H$ is predominantly produced through gluon-gluon fusion and decays mostly into $H\rightarrow SS,Sh$ with a total cross-section in the rage 10-25\,pb~\cite{vonBuddenbrock:2019ajh}. Due to the relative large Yukawa coupling to top quarks needed to achieve the above mentioned direct production cross-section, the  production of $H$ in association with a single top-quark.\footnote{As the $HWW$ coupling is suppressed with respect to the top Yukawa coupling, the associated production of $H$ with a single top not suppressed.} These production mechanisms together with the dominance of $H\rightarrow SS,Sh$ over other decays, where $S$ behaves like a SM Higgs-like boson, lead to a number of final states that can be classified into several groups of final states. There are three groups of final states where the excesses are statistically compelling: opposite sign (OS) leptons ($\ell=e,\mu$); same sign (SS) and three leptons ($3\ell$) in association with $b$-quarks; SS and $3\ell$ without $b$-quarks. 
In the sections below a brief description of the final states is given with emphasis on the emergence of new excesses in addition to those reported in Refs.~\cite{vonBuddenbrock:2017gvy,vonBuddenbrock:2019ajh,Hernandez:2019geu}, when appropriate. It is important to reiterate that the new excesses reported here are not the result of scanning the phase-space, but the result of looking at pre-defined final states and corners of the phase-space, as predicted by the model described above.

\subsubsection{Opposite sign di-leptons}
\label{sec:OS}
The production chain $pp\rightarrow H\rightarrow SS,Sh\rightarrow\ell^+\ell^-+X$,  is the most copious multi-lepton final state. Using the benchmark parameter space in Ref.~\cite{vonBuddenbrock:2018xar}, the dominant of the singlet are $S\rightarrow W^+W^-,b\overline{b}$. This will lead to OS leptons with without $b$-quarks. The most salient characteristics of the final states are such that the di-lepton invariant mass $m_{\ell\ell}<100$\,GeV where the bulk of the signal is produced with low $b$-jet multiplicity, $n_b<2$~\cite{vonBuddenbrock:2019ajh}. The dominant SM background in events with $b$-jets is $t\overline{t}+Wt$. It is important to note that the $b$-jet and light-quark of the signal is significantly different from that of top-quark related production mechanisms. As a matter of fact, excesses are seen when applying a full jet veto, top-quark backgrounds become suppressed and where the dominant backgrounds is non-resonant $W^+W^-$ production~\cite{vonBuddenbrock:2017gvy,ATLAS:2019rob,vonBuddenbrock:2019daj}.\footnote{While the ATLAS experiment provided unfolded cross-sections, the CMS experiment presented a comparison of the yields in the data to a MC~\cite{CMS:2020mxy}. The MC does not describe simultaneously the di-lepton invariant mass and transverse momentum. As such, for the purposes of this study this data set is inconclusive, where we encourage the CMS experiment to provide differential measurements similar to those performed by ATLAS in Ref.~\cite{ATLAS:2019rob}.} A review of the NLO and EW corrections to the relevant processes can be found in Refs.~\cite{vonBuddenbrock:2019ajh,vonBuddenbrock:2019daj}, where to date the $m_{\ell\ell}$ spectra at low masses remains unexplained by MC tools. A measurement of the differential distributions in OS events with $b$-jets with Run 2 data further corroborates the inability of current MC tools to describe the $m_{\ell\ell}$ distribution~\cite{ATLAS:2019hau}. A summary of  deviations for this class of excesses is given in Table~\ref{tab:anatomymultilepton}.

\subsubsection[SS and $3\ell$ with $b$-quarks]{\boldmath SS and $3\ell$ with $b$-quarks}
\label{sec:SS3lb}
The associated production of $H$ with top quarks lead to the anomalous production of SS and $3\ell$ in association with $b$-quarks with moderate scalar sum of leptons and jets, $H_T$. The elevated $t\overline{t}W^{\pm}$ cross-section measured by the ATLAS and CMS experiments can be accommodated by the above mentioned model~\cite{vonBuddenbrock:2019ajh,vonBuddenbrock:2020ter}. Based on a number of excesses involving $Z$ bosons, in Ref.~\cite{vonBuddenbrock:2018xar} it was suggested that the $CP$-odd scalar of the 2HDM+$S$ model could be as heavy as $m_A\approx500$\,GeV, where the two leading decays would be $A\rightarrow t\overline{t},ZH$. The cross-section for the associated production $pp\rightarrow t\overline{t}A$ with $A\rightarrow t\overline{t}$ would correspond to $\approx 10$\,fb. This is consistent with the elevated $t\overline{t}t\overline{t}$ cross-section reported by ATLAS and CMS~\cite{CMS:2019rvj,ATLAS:2020hpj,ATLAS:2021kqb}. The combined significance of the excesses related to the cross-section measurements of $t\overline{t}W^{\pm}$ and $t\overline{t}t\overline{t}$ surpass 3$\sigma$, as detailed in  Table~\ref{tab:anatomymultilepton}.
It is important to note that the ATLAS collaboration has reported a small excess in the production of four leptons with a same flavour OS pair consistent with a $Z$ boson, where the four-lepton invariant mass, $m_{4\ell}<400$\,GeV~\cite{ATLAS:2021wob}. This excess can also be accommodated by the direct production of $A\rightarrow ZH$.

\subsubsection[SS and $3\ell$ without $b$-quarks]{\boldmath SS and $3\ell$ without $b$-quarks}
\label{sec:SS3lnob}
The production chain $pp\rightarrow H\rightarrow SS,Sh$ can give rise to SS and $3\ell$ events, where $b$-jet activity would be depleted compared to production mechanism considered in Sec.~\ref{sec:SS3lb}. The potential impact on the measurement of the production of the SM Higgs boson in association with a $W$ boson and other measurements in the context discussed here was reported in Ref.~\cite{Fang:2017tmh}. A survey of available measurements of the signal yield of the $Wh$ production was performed in Ref.~\cite{Hernandez:2019geu}. A deviation of 3.8$\sigma$ with respect to the $Wh$ yield in the SM in corners of the phase-space predicted by the simplified model. The CMS experiment has recently reported the signal strength of the $Vh, V=Z,W^{\pm}$ production with the $h\rightarrow W^+W^-$ decay for low and high $V$ transverse momentum~\cite{CMS:2021ixs}. The signal strength for $Vh$ with the $V$ transverse momentum, $p_{TV}<150$\,GeV, where the BSM signal is concentrated, is $2.65^{+0.69}_{-0.64}$. This deviates from the SM value by an additional 2.6$\sigma$. It is worth noting that in order to reconcile observed excesses in Secs.~\ref{sec:OS} and~\ref{sec:SS3lb} with the ones described here, it is necessary to assume the dominance of the $H\rightarrow SS$ decay over $H\rightarrow Sh$~\cite{Hernandez:2019geu}. Another important prediction of the simplified model is the elevated $WWW$ cross-section. The ATLAS experiment reports a signal strength of $1.66\pm0.28$~\cite{ATLAS-CONF-2021-039}.\footnote{CMS~\cite{CMS:2020hjs} pursues a different approach compared to the more inclusive selection performed by ATLAS. For instance, the requirement that the azimuthal separation between the vector of the three leptons and the missing transverse energy be greater than 2.5\,rad and other requirements suppress the contribution from the BSM signal considered here.} The latter includes the $Wh\rightarrow WWW^*$ production, hence it is not added to the combination due to partial double counting. Lastly, another final state of interest is the production of $ZW^{\pm}$ events where $Z$ transverse momentum, $p_{TZ}<100$\,GeV with depleted $b$-jet activity. Excesses were reported in Ref.~\cite{vonBuddenbrock:2019ajh}. The CMS experiment has recently reported an important excess in events with $3\ell$ in association with one and two jets used for the measurement of $Zh, h\rightarrow W^+W^-$ production, where $ZW^{\pm}$ is the dominant background~\cite{CMS:2021ixs}. As the analysis of the excess in the context of the simplified model described here is in progress, the significance of this excess is not added to the combination reported in Table~\ref{tab:anatomymultilepton}. 

\subsection[Higgs-like excess at $\approx96$\,GeV]{\boldmath Higgs-like excess at $\approx96$\,GeV}
\label{sec:96}

The LEP experiments reported a mild excess with a local significance of 2.3$\sigma$ in the search for a SM Higgs boson~\cite{LEPWorkingGroupforHiggsbosonsearches:2003ing} using the process $e^+e^-\rightarrow Zh(\rightarrow b\overline{b})$. 
The largest excess was observed for the $b\overline{b}$ invariant mass of 98\,GeV. Renewed interest in this excess emerged with the CMS experiment reporting similar excesses with Run 1 and 35.9\,fb$^{-1}$ of Run 2 data~\cite{CMS:2018cyk}, with a local significance of 2.8$\sigma$ at 95.3\,GeV. The ATLAS experiment has reported the results of a search with over 80\,fb$^{-1}$ of integrated luminosity~\cite{ATLAS:2018xad}. No excess was found there, but the measured limit does not exclude the results from the CMS experiment. As such, sufficient data is available in the complete Run 2 data set to understand if the above mentioned excesses are due to statistical fluctuations.    
While the excess described above is not yet statistically compelling, it has been studied by a number of authors, as we will review below.

\subsubsection{Interpretation as SM Higgs-like scalar boson}
The LEP and CMS excesses can be interpreted as signal from a SM Higgs-like boson with a mass around 96\,GeV, as was done for instance in Ref.~\cite{Fox:2017uwr}.
In this interpretation the signal strength in terms of a would-be SM Higgs boson with this mass is~\cite{Cao:2016uwt,Biekotter:2019kde}:
\begin{equation}
\mu_{LEP}= {\sigma\left(e^+e^-\rightarrow ZS \rightarrow Zb\overline{b}  \right)   \over \sigma\left(e^+e^-\rightarrow Zh_{98} \rightarrow Zb\overline{b}  \right)} = 0.117\pm 0.057
\label{eq:lep}
\end{equation}
and
\begin{equation}
\mu_{CMS}= {\sigma\left(pp\rightarrow S \rightarrow \gamma\gamma  \right)   \over \sigma\left(pp\rightarrow h_{96} \rightarrow \gamma\gamma  \right)} = 0.6\pm 0.2
\label{eq:cms}
\end{equation}
where $h_{X}$ stands for a SM Higgs-like boson with $m_h=X$\,GeV. 
A few comments are in order:
First, the measurement in Eq.~\eqref{eq:lep} is mostly sensitive to the coupling $h_{98}ZZ$, the size of which is compatible with the coupling measurements of the SM Higgs boson at the LHC~\cite{CMS:2012vby,ATLAS:2013xga,CMS:2020gsy,ATLAS:2020qdt}. 
Second, the masses are compatible when one takes the coarse resolution of the LEP measurement into account~\cite{Heinemeyer:2018wzl}.
Third, the branching ratio of a SM Higgs-like scalar, expected to be $Br(h_{96}\rightarrow ZZ^*)<1\cdot 10^{-3}$, makes the observation of an excess with $4\ell$ events at the LHC very difficult with the available data set~\cite{ATLAS:2020wny,CMS:2021ugl}.

It is important to note, that a SM Higgs-like boson with such a mass decays predominantly into $b\overline{b}$, where the branching ratio $Br(h_{96}\rightarrow b\overline{b})>0.8$, and that $\mu_\text{CMS} > \mu_\text{LEP}$. 
Therefore the two signal strengths reported by CMS and LEP suggest an underlying model that is more complicated and has more than just one additional degree of freedom, which has been addressed in Ref.~\cite{Fox:2017uwr} by including additional light fields, and considering a more extended scalar sector.

\subsubsection{Two Higgs doublet model}
A two Higgs doublet model was used in Refs.~\cite{Fox:2017uwr,Haisch:2017gql} to explain the excess, the key being a moderately-to-strongly fermiophobic CP-even scalar boson with a mass of 95\,GeV.
The scalar's fermiophobic nature suppresses the top contribution and thus enhances its di-photon rate, such that the signal strength of around 0.1 pb (cf.\ Eq.~\eqref{eq:cms}) can be explained by the combination of VBF, $Wh$ and $Zh$ production alone. 
The authors show that sizable scalar production rates can also arise from either top-quark pair and single-top production followed by decay chains involving (comparatively light) charged scalars, or associated production from the decays of a heavier CP-odd scalar bosons.

An extension of the two Higgs doublet model with an additional Higgs singlet field, the so-called ``next to minimal two Higgs doublet model'' can fit the CMS excess as demonstrated in Ref.~\cite{Biekotter:2019kde}. 
In this model the additional (potentially light) CP-even scalar resonance is largely unconstrained by other phenomenological observations and therefore a more general fit is possible.
The N2HDM of type II and type IV (flipped) can fit the two excesses, while type I and
type III (lepton specific) cannot. 
The fit prefers the lowest possible masses for the charged Higgs bosons around 650\,GeV and $\tan\beta$ just above 1.

\subsubsection{Supersymmetric models}
It is not possible to explain the 96\,GeV scalar in the MSSM~\cite{Bechtle:2016kui}. A well-motivated extension of the minimal Supersymmetric Standard Model is given by the Next-to-MSSM (NMSSM), where a mostly singlet-like Higgs that mixes with a doublet component of about 10\% (mixing squared) is a natural candidate to explain the excess~\cite{Heinemeyer:2018wzl,Domingo:2018uim}.
In this kind of models, the decays of such a light Higgs boson depend on the mixing between the different Higgs states.
In particular the couplings to down- or up-type quarks can be suppressed to meet the LEP measurement of $b\bar b$ in Eq.~\eqref{eq:lep}, while the di-photon rate can be moderately enhanced to meet the CMS measurement in Eq.~\eqref{eq:cms}.

The $\mu\nu$SSM is a supersymmetric extension of the Standard Model that incorporates neutrino physics.
In this model the additional light scalar can be interpreted as the CP-even right-handed sneutrino and sizable mixing of the sneutrino and the doublet Higgses are possible.
In general, it is easier to reproduce the LEP observed signal strength in supersymmetric models than the CMS one~\cite{Biekotter:2017xmf}.
This model shares many features with the NMSSM, however, the link to the neutrino sector breaks the $R$-parity and thus avoids Dark Matter bounds, freeing up the parameter space to accommodate the excess.

\subsubsection{Other explanations}
The multiphoton signals have been explained in the context of other models too.
For instance the lightest new particle predicted within the Randall Sundrum (RS) model, the radion, was discussed as possible explanation of the CMS excess in Ref.~\cite{Richard:2017kot}
A model with an extra $U(1)$ and heavy vector-like quarks that also addresses the flavour anomalies was discussed in Ref.~\cite{Liu:2018xsw}. 
The anomaly-free $U(1)_{Y'}$ extension of the SM with two complex scalar singlets were discussed in Ref.~\cite{Aguilar-Saavedra:2020wrj}.

\subsection{Neutrino Anomalies}
\label{sec:neutrino}
In the SM active neutrinos carry isospin charge $\pm {1 \over 2}$ under the weak interaction, where they have left-handed helicity. However, due to the establishment of neutrino oscillations neutrinos are known to have mass, even if small. The observation of solar and very-long-baseline reactor neutrino oscillations and the observation
of atmospheric and long-baseline accelerator neutrino oscillations give  vigorous evidence of three-neutrino mixing~\cite{MINOS:2007ixr,KamLAND:2008dgz}. This poses a problem for the SM. Currently, the origin of mass of neutrinos remains an open question in particle physics. Hence, the neutrino sector is of invaluable significance for particle physics. 

A number of anomalies beyond the three-neutrino mixing framework have emerged in short-baseline neutrino experiments with $L/E_\nu\approx$\,1 m/MeV, where $L$ is the distance that the neutrino traverses and $E_\nu$ is the neutrino energy. These can be classified into two groups: neutrino appearance and disappearance. 
This section covers anomalies in measurements at fixed-target experiments that may point to physics beyond the  standard three-flavour framework.

\subsubsection{Appearance anomalies}
The appearance anomalies pertain to excesses in the production of $\nu_e$ and $\overline{\nu}_e$ via charge-current quasi-elastic production reported by the  LSND~\cite{LSND:2001aii} and MiniBooNE~\cite{MiniBooNE:2018esg} experiments. The MiniBooNE experiment observed a total of 460.5$\pm$99 $\nu_e$ and $\overline{\nu}_e$ excess events in the energy range $200<E_\nu<1250$\,GeV. This excess corresponds to a significance of 4.7$\sigma$ and a recent update confirms previous findings~\cite{MiniBooNE:2020pnu}. 
The two-neutrino oscillation interpretation of the MiniBooNE data is consistent with the corresponding allowed region obtained with data from the LSND experiment. The combined significance of the MiniBooNE and LSND excesses reaches 6$\sigma$. The MiniBooNE excess has been studied in terms of the two-neutrino mixing, where $(\Delta m^2,\sin^2{2\theta})=(0.041$\,eV$^{2},0.96)$~\cite{MiniBooNE:2018esg}. The MiniBooNE excess will be further studied by the Fermilab short-baseline neutrino programme~\cite{MicroBooNE:2015bmn}.

\subsubsection{Disappearance anomalies}
Anomalies pertaining to the disappearance of $\nu_e$ and $\overline{\nu}_e$ are observed in reactor neutrino experiments~\cite{Declais:1994su,CHOOZ:2002qts,Mention:2011rk} 
and Gallium radioactive source experiments anomalies~\cite{Acero:2007su,Giunti:2010zu}. The latter emerge from the GALLEX~\cite{GALLEX:1994rym,GALLEX:1997lja,Kaether:2010ag} and the SAGE~\cite{Abdurashitov:1996dp,SAGE:1998fvr,Abdurashitov:2005tb,SAGE:2009eeu} experiments. Gallium radioactive source experiments are solar neutrino experiments with intense artificial $^{51}$Cr and $^{37}$Ar radioactive sources inside the detectors, where the inverse beta reaction $\nu_e+^{71}$Ga$\rightarrow ^{71}$Ge$+e^-$ is used. These experiments indicate that the ratio of measured to predicted $^{71}$Ge occurrences is less than unity. The significance of the deficit is about 3$\sigma$, and it is referred to as the Gallium anomaly. 

This anomaly has been studied within the framework of two-neutrino mixing, where $\Delta m^2>0.1$\,eV$^2$, also orders of magnitude greater than those of the solar and atmospheric oscillations~\cite{Acero:2007su,Giunti:2010zu}. The compatibility of the Gallium and the reactor anomalies has also been studied within the context the two-neutrino mixing, where reasonable agreement between the two disappearance anomalies is found~\cite{Acero:2007su}.

\subsubsection{Explanation via light neutrinos}
Both types of anomalies indicate that the mass splitting, $\Delta m^2$, has to be much larger than the three-neutrino mixing scheme inferred from the observation of neutrino oscillations in solar, very-long-baseline reactor, atmospheric and long-baseline accelerator experiments. 

Light sterile neutrinos with masses around a few eV have been considered as a possibility to explain the above mentioned anomalies. That said, when interpreting all the anomalies simultaneously in terms of sterile neutrinos through global fits, it is not possible to provide a consistent interpretation~\cite{Collin:2016rao,Gariazzo:2017fdh,Dentler:2017tkw,Dentler:2018sju}. In this light, a comprehensive global programme is under way to study the dependence of the anomalies on distance and energy in order to conclusively ascertain the origin of the anomalies~\cite{DellAcqua:2018lky}.

\subsubsection{Non oscillatory explanations of the appearance anomaly}
Possible non-oscillatory explanations of the MiniBooNE anomaly were discussed in a model-independent way in Ref.~\cite{Brdar:2020tle}.
Most of the existing explanations make use of shortcomings in particle identification of the MiniBooNE detector. Not all explanations address the LSND anomaly, and few produce a good fit to the observed excess spectra. 

The largest class of models explaining the MiniBooNE anomaly introduces heavy neutrinos with masses ${\cal O}(100)$ MeV that are produced in the detector from upscattering of the muon neutrinos, and which subsequently decay into various final states that give rise to electromagnetic radiation~\cite{Gninenko:2009ks,Ballett:2018ynz,Bertuzzo:2018itn,Ballett:2019pyw,Abdullahi:2020nyr,Datta:2020auq,Dutta:2020scq,Abdallah:2020vgg,Abdallah:2020biq}.

Another class of models introduces heavy neutrinos with masses around ${\cal O}(1)$ MeV that are produced at the target and in the decay pipe, which rapidly decay into a final state including an electron neutrino and thus enhance the electron neutrino flux at the detector~\cite{Palomares-Ruiz:2005zbh,deGouvea:2019qre,Dentler:2019dhz,Bai:2015ztj}.

A variation of the above explanations, where heavy neutrinos are produced at the target and travel into the detector to decay into electromagnetic radiation was discussed in Ref.~\cite{Fischer:2019fbw}.
A different explanation was given in Ref.~\cite{Asaadi:2017bhx} where 
short baseline oscillations were explained via a strong medium potential, including a new resonance scattering of neutrinos on the local overly dense relic neutrino background.

\subsection{Astrophysical Anomalies}
\label{sec:astrophysics}
The field of astroparticle physics has been replete with observational anomalies over the past decades.

\subsubsection{Positron excesses}
The first case we will discuss will be the positron excesses observed by multiple experiments and at multiple energies. Observations of consistent excess positrons above the expected backgrounds, at both GeV and TeV energy ranges, by the PAMELA and ATIC instruments respectively~\cite{PAMELA:2008gwm,Adriani:2008zq,PPB-BETS:2008zzu} as well as the HESS telescope~\cite{HESS:2008ibn} (at TeV scales) generated initial speculation that the consequences of BSM physics (dark matter in particular) had been observed~\cite{Hooper:2003ad,Delahaye:2007fr,Nardi:2008ix}. These excesses were strengthened by further data collected by Alpha Magnetic Spectrometer (AMS and AMS-02) collaboration~\cite{AMS:2019rhg} as well as the Fermi and Dark Matter Particle Explorer (DAMPE) satellites~\cite{Fermi-LAT:2017bpc,DAMPE:2017fbg}. This has resulted in strong evidence of unexplained positrons at both the GeV and TeV energy scales. The AMS-02 and DAMPE data have been thoroughly examined for consistency with various dark matter models. In particular the DAMPE excess is considered to favour a leptophilic scenario where dark matter couples to electrons and/or muons via  heavy mediator~\cite{Fan:2017sor,Yuan:2017ysv,Yang:2017cjm}. In the case of PAMELA, Fermi-LAT and AMS-02 the treatment has often been ``model independent" (where annihilation/decay channels are considered independently) such as in Refs.~\cite{Cirelli:2008pk, DiMauro:2015jxa}. In particular, AMS-02 offers one of the most powerful indirect dark matter probes. Authors that study AMS-02 as an observed dark matter signature again find the data favours leptophilic dark matter~\cite{Dev:2013hka, Profumo:2019pob,Li:2017nac}, coupling either directly to electrons and muons or via scalar/gauge mediators. The amplitude of the AMS-02 and DAMPE fluxes typically requires some form of boosting in annihilating cases, such as Sommerfeld enhancement~\cite{Das:2019vxm}, or a local over-density~\cite{Fan:2017sor,Yuan:2017ysv,Yang:2017cjm}. The continued existence of these leptonic excesses in astroparticle physics is clearly fertile hunting ground for new phenomena. For most of the studied models, collider searches remain difficult as the mediator mass is necessarily of the order of few TeV.

\subsubsection{Anti-protons}
Anti-proton spectra also display excesses as observed by PAMELA and then AMS-02~\cite{PAMELA:2011mvy,Aguilar:2016kjl}. The impact of dark matter on the anti-proton content of cosmic-rays had been studied prior to these experimental results in~\cite{Bringmann:2006im}. The significance of the excess and it favouring DM explanations has been extensively argued~\cite{Hooper:2014ysa,Giesen:2015ufa,Cirelli:2014lwa,Ishiwata:2019aet,Cuoco:2016eej,Cuoco:2017rxb,Cui:2016ppb,Cholis:2019ejx}. However, it has recently been shown that the excess significance is almost entirely erased when correlated errors are taken into account~\cite{Heisig:2020nse}. In addition, simple DM models face difficulties maintaining consistency across different targets and frequency bands~\cite{Beck:2015rna,Beck:2019imo}. A common difficulty faced in this regard is that annihilation/decay channels that produce copious anti-protons also result in large gamma-ray fluxes that can be difficult to reconcile with observational data.

\subsubsection{Galactic Centre gamma-ray excess}
An anomalous excess of gamma-rays in the Milky-Way galactic centre, observed with the Fermi-LAT instrument, has long been studied as a potential signature of BSM physics~\cite{Hooper:2010mq,Hooper:2011ti,Abazajian:2012pn,Daylan:2014rsa,Calore:2014xka,Fermi-LAT:2017opo}. This particular excess has been controversial, with competing ``mundane" explanations~\cite{OLeary:2015qpx,Bartels:2015aea} as well as concerns over modelling and the effect of unresolved sources~\cite{Boyarsky:2010dr,Macias:2013vya,Zhou:2014lva,Brandt:2015ula,Lee:2015fea}. Despite these concerns there are still indications that the excess is significant and favours DM explanations~\cite{DiMauro:2021raz}, provided the Milky-Way halo is a contracted Navarro-Frenk-White~\cite{Navarro:1995iw} profile. Similar to the anti-proton case, simple DM models face difficulties in that their indirect emissions in other sources and in the radio frequency range often exceed existing observational data~\cite{Beck:2015rna,Beck:2017wxu,Beck:2019imo}.

\subsubsection{3.5 keV X-ray line}
An unidentified X-ray emission line near 3.5 keV has been detected by a number of groups using XMM-Newton, Chandra and Suzaku datasets~\cite{Bulbul:2014sua, Boyarsky:2014jta, Boyarsky:2014ska, Urban:2014yda, Franse:2016dln, Cappelluti:2017ywp, Hofmann:2019ihc, Sicilian:2020glg}. The most popular explanation so far is in terms of a decaying keV-scale sterile neutrino dark matter~\cite{Abazajian:2017tcc, Boyarsky:2018tvu}. The dark matter interpretation was ruled out with blank-sky observations~\cite{Dessert:2018qih}, but these results have been disputed in Refs.~\cite{Abazajian:2020unr, Boyarsky:2020hqb}. To add to this confusion, there exist various other astrophysical as well as non-astrophysical explanations for the line, and in particular, there are many atomic lines in this energy range, which could mimic the signal~\cite{Jeltema:2014qfa}. Hopefully, the situation will be clarified with more precise data from future X-ray missions, such as Athena~\cite{Nandra:2013jka}. 

\subsubsection{511 keV gamma-ray line}
The INTEGRAL satellite has detected an excess of 511 keV photon emission from electron-positron pair annihilation in the central region of Milky Way relative to astrophysical expectations~\cite{Weidenspointner:2008zz, Siegert:2015knp}. This was recently confirmed by COSI~\cite{Kierans:2019aqz}. The origin of the positrons in the Galactic Centre is still actively debated and it is plausible that it could originate from sources associated with BSM physics, such as pair-annihilation of MeV-scale dark matter particles~\cite{Boehm:2003bt}. Next-generation MeV gamma-ray telescopes such as eASTROGAM~\cite{e-ASTROGAM:2017pxr} with better point source detection sensitivity would provide a clearer picture.

\subsubsection{21-cm absorption line}
The EDGES experiment has reported an
anomalous absorption profile centred at 78 MHz (corresponding to a  redshift $z=17$) in the sky-averaged
21-cm spectrum~\cite{Bowman:2018yin}. This indicates an unexpected cooling of the hydrogen gas during or prior to the so called Cosmic Dawn era, as compared to the $\Lambda$CDM model expectations. The cosmological interpretation of the EDGES anomaly is subject to significant instrumental and systematic uncertainties. Nevertheless,  milli-charged sub-GeV dark matter has emerged as a popular BSM candidate to explain this $3.8\sigma$ anomaly~\cite{Munoz:2018pzp, Barkana:2018lgd}. Some issues regarding this interpretation were raised in Refs.~\cite{Berlin:2018sjs, Barkana:2018qrx, Slatyer:2018aqg, Boddy:2018wzy}, which however can be overcome~\cite{Liu:2019knx}. The new HERA upper limits~\cite{HERA:2021noe} on the power spectrum of 21-cm fluctuations in the early Universe neither support nor disfavour the  EDGES detection, but future measurements from HERA at lower frequencies will be important for further testing this anomaly. 

\subsubsection{XENON1T electron excess}
The XENON1T collaboration has reported an excess of electron recoil events over the known backgrounds in the 2-3 keV recoil energy range~\cite{XENON:2020rca}. There have been several BSM interpretations of this anomaly, such as solar axions~\cite{Takahashi:2020bpq, DiLuzio:2020jjp, Bell:2020bes, Gao:2020wer, Athron:2020maw}, boosted dark matter~\cite{Kannike:2020agf, Fornal:2020npv} and neutrino magnetic moment~\cite{Miranda:2020kwy, Babu:2020ivd}. However, the possibility of a tritium beta-decay background in the detector can neither be confirmed nor excluded so far as the cause of this anomaly~\cite{XENON:2020rca}. 

\subsubsection{DAMA/LIBRA annual modulation}
The DAMA/LIBRA collaboration has reported an annual modulation signal in their NaI detector with a high statistical significance of  $12.9\sigma$~\cite{Bernabei:2018jrt}. Although the signal is compatible with general expectations from halo dark matter in the Milky Way~\cite{Drukier:1986tm}, the DAMA/LIBRA-preferred DM parameter space is now strongly disfavoured by many other direct-detection experiments~\cite{Billard:2021uyg}. There exist a handful of ideas~\cite{Tucker-Smith:2001myb, Feng:2011vu, Foot:2020ehn} to reconcile the DAMA result with other null results; however, it is still a contentious issue whether the DAMA/LIBRA signal is indeed due to dark matter in the first place; see, e.g., Refs.~\cite{Buttazzo:2020bto, Messina:2020pnt}. This needs to be independently verified using the same target material, which is the main goal of two currently running experiments, COSINE~\cite{Adhikari:2018ljm} and ANAIS~\cite{Amare:2021yyu}.

\subsection{Cosmological Anomalies}
\label{sec:cosmology}
There are several cosmological anomalies that have slowly become more robust as available data has increased.

\subsubsection{Dark energy}
The need for dark energy~\cite{SupernovaSearchTeam:1998fmf,SupernovaCosmologyProject:1998vns} may be considered as one of the major anomalies in modern cosmology, as there is no clear candidate to explain this phenomenon. There are many suggested solutions to this problem, from scalar fields, chaplygin gas, holographic dark energy, and modified gravity, cf.~\cite{Li:2012dt}.  

\begin{figure}[htbp]
	\centering
	\resizebox{0.9\hsize}{!}{\includegraphics{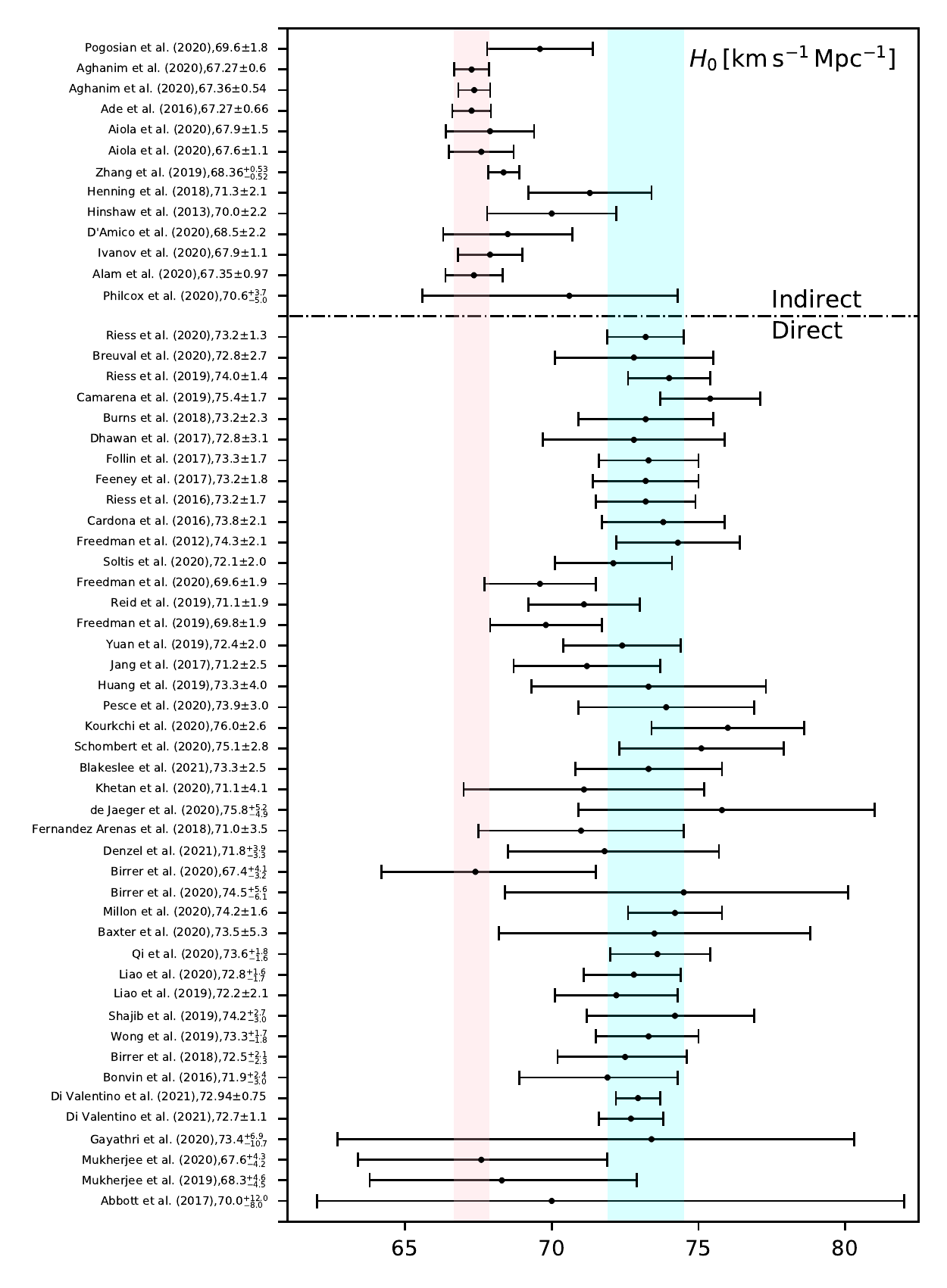}}
	\caption{Hubble tension whisker plot with references from Ref.~\cite{DiValentino:2021izs}. The pink band represents Planck results~\cite{Planck:2018vyg} and the cyan band is for cepheids from Ref.~\cite{Riess:2020fzl}.}
	\label{fig:hubble}
\end{figure}

\subsubsection{Hubble trouble}
A recently emerging anomaly has been the Hubble tension, where late-Universe and early-Universe measurements of the Hubble constant disagree by as much as $3.46\sigma$~\cite{DiValentino:2021izs}. This tension emerges in the comparison of late-Universe measurements such as cosmic shear~\cite{Hildebrandt:2018yau,KiDS:2020suj}, weak lensing~\cite{Joudaki:2016kym,Hildebrandt:2016iqg},  galaxy clustering~\cite{DES:2017myr}, and cepheid distance calibrations~\cite{Riess:2020fzl}. A whisker plot displaying a large selection of both early and late-Universe data from Ref.~\cite{DiValentino:2021izs} is displayed in Fig.~\ref{fig:hubble}. There are many conjectured solutions to this problem, including models with an early dark energy component~\cite{Caldwell:2003vp,Doran:2005sn,Bartelmann:2005fc,Karwal:2016vyq} with alternate models argued to be more natural, achieved via coupling to neutrinos~\cite{Sakstein:2019fmf,Gogoi:2020qif,Fardon:2003eh,Kaplan:2004dq,Peccei:2004sz} or via a $k$-essence scenario~\cite{Tian:2021omz}. Additionally, a time-varying dark energy equation of state in the late-Universe has been argued for in Refs.~\cite{Zhao:2017cud,DiValentino:2017gzb}. A slew of additional models have been suggested in the literature that function by increasing the number of relativistic degrees of freedom at recombination, from sterile neutrinos ~\cite{Carneiro:2018xwq,Gelmini:2019deq,Gelmini:2020ekg}, to thermal axions~\cite{DEramo:2018vss}, to decaying dark matter~\cite{Xiao:2019ccl,Nygaard:2020sow,Blinov:2020uvz} as well as several others~\cite{DiValentino:2021izs}. Additionally, some authors have considered models where dark matter interacts with photons~\cite{Kumar:2018yhh,Yadav:2019jio} or baryons~\cite{Becker:2020hzj}. Otherwise the effect of decaying ultra-light scalars has been considered in Ref.~\cite{Gonzalez:2020fdy} and primordial black holes in Ref.~\cite{Nesseris:2019fwr}. A caveat on this topic is the sensitivity to systematics, as illustrated in Ref.~\cite{Freedman:2020dne}, where the authors use a more robust distance calibration for nearby galaxies to demonstrate that late-Universe Hubble measurements, in this case via extinction in red giant branch stars within the Large Magellanic Cloud (LMC), can be largely reconciled with early-Universe measurements.

\subsubsection{Clustering anomalies}
A second emerging issue is the disagreement in the clustering properties of dark matter between early-Universe measurements from Planck~\cite{Planck:2018vyg} and late-Universe measurements via cosmic shear~\cite{DES:2021bvc} as well as direct clustering measurements~\cite{DES:2021zxv}. These newer measurements confirm earlier results on the same topic~\cite{Lange:2020mnl,Leauthaud:2016jdb}. It is as yet unclear whether this anomaly can be explained via systematics or requires additional physics. Some suggestions are that such data can probe dark energy models~\cite{Huterer:2017buf}, or that the discrepancy can be ameliorated by dark matter decaying into dark radiation~\cite{DES:2020mpv}, or dark matter self interactions~\cite{Buckley:2017ijx}. Finally, this anomaly has been used as a test of modified gravity scenarios~\cite{Chen:2019ftv}, which indicated that general relativity was preferred over the studied forms of modified gravity.

\subsection{Anomalies in ultra-high energy cosmic rays and neutrinos}
\label{sec:cosmics}
The flux of cosmic ray particles impinging on the terrestrial atmosphere provides hadron-nucleus collisions at energies covering many orders of magnitude, including a regime well beyond the LHC's energy.
The interpretation of these observations relies on accurate models of air shower
physics, which is a challenge and an opportunity to test QCD under extreme conditions.

\subsubsection{The muon puzzle}
High-energy cosmic rays can be observed via the extensive air showers in Earth’s atmosphere, which are hadronic cascades that eventually decay into muons.
The muons reaching the detectors at ground-level are the key observable to infer the
mass composition of cosmic rays.
The muon puzzle constitutes an excess of muons at the ground level, measured in extensive air showers stemming from primaries with energies above 10 PeV.
This excess is seen compared to state-of-the-art simulations, and has a significance of $8\sigma$~\cite{Albrecht:2021yla}.
This muon deficit becomes apparent in cosmic ray interactions with centre-of-mass energies around the TeV scale, suggesting that the origin of this excess could be observable at the LHC.

\subsubsection{Earth-emergent EeV events}
At laboratory energies beyond PeV the mean free path of neutrinos becomes smaller than the Earth radius, which implies that beyond this energy, no particle should be able to penetrate the Earth. Upgoing showers are expected as a result of astrophysical tau neutrinos that convert to tau leptons while passing through the Earth, but the observed exit angles are restricted to be small for ultra-high energy neutrinos within the SM. 
Therefore the observations of a couple of Earth-emergent EeV-scale upgoing shower events by the ANITA experiment~\cite{ANITA:2016vrp, ANITA:2020sds} and the Extremely High Energy Northern Track neutrinos by IceCube~\cite{Aartsen:2018vtx} cannot be explained with astrophysical sources~\cite{IceCube:2020gbx} nor within the SM~\cite{Fox:2018syq}.
These anomalies were addressed qualitatively in models involving long-lived particles, leptoquarks and heavy dark matter~\cite{Chauhan:2018lnq,BhupalDev:2020zcy, Collins:2018jpg, Heurtier:2019git}.

%\newpage
\section{Hidden physics at the LHC}
{\it\small Section editors: Kingman Cheung, Rohini Godbole, Zhen Liu and Tao Han\\
Contributions: Shankha Banerjee, Kingman Cheung, Oliver Fischer and Zhen Liu}

\medskip\noindent
After the Run 2 at the LHC, ATLAS, CMS, and LHCb experiments have placed
constraints on many models Beyond the Standard Model, either by direct searches for new particles or interactions, or through precision measurements.
For example, searches for supersymmetric partners have pushed weak-scale SUSY to some uncomfortable corners in the parameter space 
\cite{Sirunyan:2019ctn,Sirunyan:2019mbp,Aad:2020nyj}; 
searches for $W',Z'$ bosons, and leptoquarks have placed limits of
order $3-4$ TeV on their masses; searches for extra scalar bosons have restricted their
masses to be heavier than $600-700$ GeV.

No convincing sign of NP has been detected in form of a resonance, however. This could mean that hypothetical new particles have masses above the LHC energy threshold or tiny production rates, and are therefore inaccessible by the experimental analyses. 
Here we consider the exciting possibility that NP is accessible at the LHC {\it in principle}, but that the corresponding signatures are hidden from experimental detection.

In this section we discuss the properties that the NP has to have in order to be hidden from current analyses.
Last, we provide specific scenarios and models that are considered hidden in the data due to some of these properties.

\subsection{Properties of hidden signatures}
New particles with masses below the LHC energy threshold and non-negligible production rates may not have been covered by experimental searches, despite being testable in principle.
Such new particles may not have been detected or studied for any combination of the following possible reasons:
\begin{itemize}
	\item The triggers employed in Run 1 and 2 don't respond to the final state;
	\item The new particle only decays hadronically and is buried in
	QCD backgrounds;
	\item The new particle is only produced in association with other particles;
	\item The mass of the new particle is very close to an SM particle.
\end{itemize}
Many models exist that are hidden from current searches for any of the above reasons, and it is not a problem to identify signatures and tailor dedicated searches for each of them individually.
In the following we detail and discuss certain properties that render hypothetical NP signatures `hidden' at the LHC.

\subsubsection{Soft Particles in the Final State}
Particles with transverse momentum below the trigger thresholds are called soft particles.
An event that includes only soft particles will typically not be recorded at all.
Therefore, NP with signatures that include only soft particles in the final state constitutes an important class of hidden NP.
Studying this kind of event is extremely challenging as they are drowned in QCD backgrounds, which are the very reason for the triggers and their thresholds.

A generic scenario that can give rise to such signatures includes two or more particles that have almost degenerate masses, i.e.\ the mass spectrum is compressed.
The production of the heavier new particle states, followed by decays to the lightest new particle state, leaves little phase space for the SM particles that are radiated off in this process, such that their transverse momenta are suppressed.
If the lightest new particle state is neutral, it will escape the detector, leaving a signature with missing energy and soft SM particles.
Typical examples are SUSY scenarios with wino- or Higgsino-LSP, where the next-to-lightest SUSY particle is a chargino, which are discussed below.

A practical approach to studying such signatures is to include initial state radiation, e.g.\ a photon, jet, or other SM particles, see, e.g.\ Refs.~\cite{Chen:1996ap,Ibe:2006de}. 
In the case of the lightest new particle being neutral, this leaves a mono-$X$ signature, where $X$ is the radiated-off SM particle. 
The disadvantages of this approach are the reduced signal cross-section, and the limited amount of information that can be used to identify the underlying model.

In the case where the next-to-lightest new particle is charged and has a macroscopic lifetime, its track could be detected.
In this case, even if the decay leaves a soft pion or lepton plus a neutral particle that escapes detection, this decay could be detected via the disappearing track signature, cf.\ Ref.~\cite{ATLAS:2021ttq}, as described in more detail below.

Neutral particles with macroscopic lifetimes may lead to
signatures with soft displaced leptons or jets, which can be detected
with dedicated triggers \cite{Blekman:2020hwr}.

In some scenarios a spectrum may exist that allows for cascade decay of new particles into one another.
An example is given by SUSY, where such cascade decays of gluinos or squarks can give rise to multi-jet and/or multi-lepton events. For instance, in the well-motivated scenarios of stop-bino coannhiliation~\cite{Boehm:1999bj,Ellis:2001nx,Diaz-Cruz:2007ewo,Han:2013gba,Ellis:2014ipa,Keung:2017kot}, the stop decays into soft jets plus LSP bino, one can rely on a hard ISR jet to trigger and select signal events~\cite{An:2021yqd}.

\subsubsection{SUSY without missing transverse energy}
Because of $R$-parity conservation, SUSY particles are usually considered to be produced in pairs, and to promptly decay via chains of varying lengths into the LSP, radiating off SM particles.
Therefore the traditional SUSY analyses search for the missing transverse energy in an event containing a number of SM particles.

Various SUSY scenarios exist, however, where this signature does not apply.
One possible reason is a compressed mass spectrum, which may preclude the event to be registered at all, as described in the previous subsection.
Another plausible reason exists for scenarios where the LSP is not stable, e.g.\ in SUSY scenarios where $R$-parity conservation is violated.
In this case, provided the LSP decays inside the detector, no missing transverse energy is generated from the escaping LSP, and the relevant signature is instead given by the LSP decay products.

\subsubsection{Associated production processes}
New particles that are singly or doubly produced usually give rise to a well understood signature in the detector and can be discovered via a resonance in a given final state.
On the other hand, processes including new particles that are produced in association with additional SM particles could skip the selection filters in experimental analyses.
In such a scenario, a statistically significant resonance may be hidden in the data but invisible to the analyses, e.g.\ because its associated production leads to particles in the final state against which analyses veto.

This situation is theoretically well motivated, for instance in models where the new particles couple to the SM only after mixing of gauge bosons.
This case, where BSM particles are produced in association with $\gamma, W, Z$ gauge bosons has been studied for quite a number of exotic particles beyond the SM, e.g.\ for heavy scalar bosons or axions. 

Associated production with a Higgs boson as a unique window to BSM physics is well motivated, for instance via the Higgs-portal models. Especially the Higgs boson coupling to a dark sector, including DM candidates, can give rise to the mono-Higgs signature.
Furthermore, the heavy Higgs boson(s) in many BSM scenarios could provide new production avenues to discover new particles. For instance, the charged Higgs boson produced in association with top quark and bottom quark can decay into tau slepton and sneutrino, making a discovery possible~\cite{Gori:2018pmk}. 

Associated production with a single top quark, a top quark pair, or the Higgs boson pose further challenges to experimental searches, as the decay products from the heavy SM particles significantly modify the final state, thus obviating present searches for this type of NP.
The heavy SM particles may therefore be a relevant avenue to BSM searches, despite the associated NP production rates often being suppressed. 

One interesting example is given by models where the dark matter couples only to the top quark~\cite{Cheung:2010zf,Haisch:2015ioa} such that it can escape all the existing direct and indirect searches of dark matter. The signature at the LHC would be a top-quark pair plus large missing energies.

\subsubsection{QCD-like final states}
Scenarios exist where new particles decay exclusively into QCD final states, i.e.\ quarks and/or gluons.
Such scenarios are motivated, for instance, with $R$-parity violating (RPV) SUSY models~\cite{Dreiner:2020qbi} or baryonic $Z'$~\cite{Cheung:2011zt}.
Hadronisation of such a final state degrades much of its kinematic structure. Even resonances in invariant mass spectra can be entirely washed out. Combined with the huge QCD backgrounds and pile up this type of signal is notoriously difficult to study at the LHC. 

Searching for large missing energy in this kind of signatures is not very useful, and the possibility to reconstruct a resonance depends crucially on the jet resolution and reconstruction, and is sensitive only to large signal rates.
As traditional analyses are not very effective, other means of detection are necessary.
For instance recent developments in jet substructure can improve identification of fast moving heavy resonances~\cite{Larkoski:2017jix}.
In general, machine learning is a very promising avenue to studying this kind of NP signature.

\subsubsection{Close-by SM resonances}
To avoid contamination from SM resonances, experimental analyses often exclude signal regions that are close to the masses of these known particles.
Therefore, new particles with masses that are very close to known SM particles could have escaped detection up to now, provided that their decay channels are overlapping, and their signal rates and widths would not leave a visible imprint.

So far, we have not encountered such scenarios in collider searches. 
Nevertheless, we cannot easily exclude the possibility that new particles are hidden around the $Z$ boson peak~\cite{Hook:2010tw} or Higgs boson peak~\cite{Gunion:2012he}. 
While all decay modes of the $Z$ boson are tested thoroughly with high precision, those of the Higgs boson still leave plenty of room for the existence of a new particle with a very similar mass and with one or more decay channels the same as the Higgs boson. One then has to carefully scrutinise the shape and the height of the Higgs boson resonance and separate it from the interference effects~\cite{Campbell:2017rke}. 

High-dimensional models can give rise to a non-trivial spectral density of a new particle, which in turn affects its signature in the detector.
The so-called HEIDI models~\cite{vanderBij:2011wy} allow single or double peaks in the invariant mass spectrum, or a continuum of states, which require very high precision measurements in order to be studied.

\subsection{Hidden SUSY scenarios}
The general minimal supersymmetric standard model (MSSM) has more than 100 soft SUSY-breaking parameters. Some GUT-motivated scenarios have greatly reduced the number of parameters, e.g., mSUGRA, and thus a lot of experimental searches were based on such simplified scenarios.  Conventional searches for SUSY rely on the final states with multi-leptons and/or multi-jets plus large missing energies. 
The current searches have pushed these scenarios restrictively to a small part of the parameter space. Nevertheless, there are some scenarios or SUSY breaking models that are still hidden from detection. We are going to describe a few of these scenarios.

\subsubsection{Wino and Higgsino LSP}  
In wino LSP and Higgsino LSP scenarios~\cite{Blekman:2020hwr,Feng:1999fu,Ibe:2006de}, the lightest chargino and neutralino are degenerate or very close in mass, such that the decay of the chargino into the lightest neutralino plus the charged lepton or pion would be very soft.
Nevertheless, some higher order corrections may be able to lift the degeneracy to some extent~\cite{Dine:2007xi,Cheung:2009qk}.
Also in the Compressed SUSY scenario~\cite{Bhattacherjee:2013wna}, all SUSY partners are very close in mass.
In these examples, the decay products are usually too soft to pass detection thresholds, which renders this subclass of SUSY models hidden from detection.

\subsubsection{Stealth SUSY}
In this type of scenarios, the SUSY particles have weak-scale masses that feel SUSY breaking only through couplings to the MSSM, which theoretically motivates the small mass splitting between fermion/boson pairs.

The resulting mass spectrum is compressed, which leaves very little phase space for the missing transverse energy after the decay of the SUSY partner into its SM counter part~\cite{Fan:2011yu}, and the conventional strategy of searching for large missing transverse energies is not effective.
Experimental collaborations instead search for multi-jets, multi-leptons,  and/or multi-photons in the final state, which are expected to be soft~\cite{CMS:2014exa}.

\subsubsection[$R$-parity violation]{\boldmath $R$-parity violation}
In so-called $R$-parity violating (RPV) SUSY~\cite{Dreiner:2020qbi} the $R$-parity is not conserved, which leads to the LSP not being stable.
Depending on the magnitude of the RPV couplings, the LSP decays can be prompt or long-lived. 
In most cases, there is no missing energy in the final state and thus the conventional searches for SUSY fail.  
In addition, if only $U^c D^c D^c$ RPV couplings exist, the decays are totally hadronic and in addition the signature is buried in QCD background. 

\subsubsection{Long-lived NLSP}  
In some SUSY scenarios the next-to-lightest SUSY particle (the NLSP) can have a suppressed decay rate in to the LSP plus some SM particles, which renders its lifetime macroscopic. 
Examples are gauge-mediated SUSY breaking (GMSB) models~\cite{Dimopoulos:1996vz,Giudice:1998bp,Feng:1997zr}, where the LSP is given by the gravitino. The decay of the NLSP, either a neutralino or a stau, can take a long time. The current triggers for SUSY may entirely miss such a scenario. In GMSB models, the gravitino mass is given by~\cite{Dimopoulos:1996yq}
\begin{equation}
m_G = \frac{F}{\sqrt{3} M_p} \simeq  2.5 \left( \frac{F}{ (100 {\rm TeV})^2 } \right) \;\; {\rm eV},
\end{equation}
where $F$ is the SUSY breaking scale and $M_p$ is the reduced Planck mass. If $F$ is of order $10-1000$ TeV, the gravitino is the LSP. Also, the interaction of the gravitino with the NLSP is suppressed by the scale $F$ such that the NLSP has a long decay into the LSP. 

In the case that the lightest neutralino is the NLSP, its decay width into $\gamma G$ is~\cite{Dimopoulos:1996yq}
\begin{equation}
\Gamma (\tilde{\chi}^0_1 \to \gamma G) = \left| 
\cos\theta_W N_{11} + \sin\theta_W N_{12} \right |^2 
\frac{m^5_{\tilde{\chi}^0_1 } }{16 \pi F^2}.
\end{equation}
On the other hand, if the stau is the NLSP, its decay width is given by
\begin{equation}
\Gamma (\tilde{\tau_1} \to \tau G) = 
\frac{m^5_{\tilde{\tau_1} }}{16 \pi F^2}.
\end{equation}
In the former case, if the decay length of the lightest neutralino is too long, the signature would be the same as the conventional missing energy search.
Otherwise, one can search for non-pointing photons in the final state.
In the slepton-NLSP case, the stau carries charge and leaves visible tracks in the inner tracker, which is much easier for detection.

In the scenario of gluino LSP~\cite{Baer:1998pg,Raby:1998xr}, the gluinos will be copiously produced by QCD interactions. Subsequently, they will hadronise into $R$-hadrons, either electrically neutral or charged, or changing charges through nuclear interaction with detector material. When charged, the $R$-hadrons could be detected as stable charged particles, but when neutral, it would be difficult to detect it, because the energy loss in collisions with detector would be small.

\subsubsection[$\nu$CMSSM with a long-lived stau]{\boldmath $\nu$CMSSM with a long-lived stau}
In Ref.~\cite{Banerjee:2016uyt} the constrained minimal SUSY (CMSSM) was extended with right-handed superpartners, a scenario that we call the $\nu$CMSSM. As we are well aware, the CMSSM is strongly affected by the LHC direct searches, the Higgs boson mass constraint as well as from other dark matter experiments. The evidence of neutrino masses  ensuing from neutrino oscillations requires the extension of the SM with at least right-handed neutrions with a Dirac mass term. Thus, in case of the CMSSM, we extend the theory with right-handed sneutrinos. Here, we consider that our next-to-lightest SUSY particle (NLSP) is the $\tilde{\tau}$. Even with this minimal extension of the CMSSM, we get striking signatures of heavy charged metastable particles. We consider several bounds. The most important bounds come from the neutrino mass and from the big bang nucleosynthesis (BBN). 

To be more specific, we extend the MSSM potential by just a single term, for each family
\begin{equation}
W_{\nu}^R = y_{\nu} \hat{H}_u \hat{L} \hat{\nu}^c_R,
\end{equation}
where $y_{\nu}$ is the neutrino Yukawa coupling, the left-handed lepton superfield, $\hat{L}=(\hat{\nu}_L,\hat{\ell}_{\bar{L}})$, and $\hat{H}_u = (\hat{H}^+_u, \hat{H}^0_u)$ is the Higgs superfield. This gives rise to the masses of the $T_3 = +1/2$ fermions. Finally, the superfield for the right handed neutrinos is $\hat{\nu}_R$. From the global fits of the neutrino oscillation parameters from solar, atmospheric, reactor and accelerator neutrino data and from the combination of the Planck temperature and polarisation data, we obtain the following bound
\begin{equation}
y_\nu^H \sin{\beta} \subset [2.8, 4.4] \times 10^{-13},
\end{equation}
where $\tan{\beta} = \big< H_u^0 \big>/ \big< H_d^0 \big>$. Furthermore, if we neglect any inter-family mixing, then the additional mass term for the sneutrinos can be written as
\begin{equation}
-\mathcal{L}_{soft} \supset M_{\tilde{\nu}_R}^2 |\tilde{\nu}_R|^2 + (y_{\nu} A_{\nu} H_u \, \tilde{L}\, \tilde{\nu}_R^c + \, \text{h.c.}),
\end{equation}
where $A_\nu$ is responsible for the left-right mixing in the scalar mass matrix. The left-right mixing in the sneutrino sector can be written as
\begin{equation}
\tan{2\tilde\Theta} = \frac{2 y_{\nu} v \sin{\beta} |\cot{\beta} \mu - A_{\nu}|}{m_{\tilde{\nu}_L}^2 - m_{\tilde{\nu}_R}^2}\,.
\label{eq:mixing}
\end{equation}
The mass eigenstates are
\begin{equation}
m_{\tilde{\nu}_L}^2 = M_{\tilde{L}}^2 + \frac{1}{2} m_Z^2 \cos{2\beta} \;\;\; \textrm{and} \;\;\; m_{\tilde{\nu}_R}^2 = M_{\tilde{\nu}_R}^2,
\end{equation}
where $M_{\tilde{L}}$ ($M_{\tilde{\nu}_R}$) is the soft scalar mass for the left-handed (right-handed) sleptons (neutrinos).

The $\tilde{\tau}_1$ finally decays into the right-handed sneutrinos via $\stau \to W^{(*)} \tilde{\nu}_R$ and the two body decay width (assuming $m_{\stau} > m_{\tilde{\nu}_R} + m_W$) can be written as
\begin{equation}
\Gamma_{\stau} \simeq \Gamma_{\stau \to \tilde{\nu}_R W} = \frac{g^2 \tilde\Theta^2}{32 \pi}|U_{L1}^{(\stau)}|^2 \frac{m_{\stau}^3}{m_W^2}\left[1 - \frac{2(m_{\tilde{\nu}_R}^2 + m_W^2)}{m_{\stau}^2} +  \frac{(m_{\tilde{\nu}_R}^2 - m_W^2)^2}{m_{\stau}^4}\right]^{3/2},
\end{equation}
where $g$ is the $SU(2)_L$ coupling, $m_W$ is $W$-boson mass and $U^{(\stau)}$ is the mixing matrix of the staus ($m_{\stau} \leq m_{\tilde{\tau}_2}$), which relate the mass and the gauge eigenstates as
\begin{equation}
\begin{pmatrix}
\tilde{\tau}_L \\
\tilde{\tau}_R
\end{pmatrix} = U^{(\tilde{\tau})}
\begin{pmatrix}
\tilde{\tau}_1 \\
\tilde{\tau}_2
\end{pmatrix}\,.
\end{equation}
$L1$ indicates the (1,1)$^{th}$ element of this matrix. When the two body decays are forbidden, the dominant three body decays are $\stau \to \tilde{\nu}_R \ell \bar{\nu}, \tilde{\nu}_R q \bar{q}'$. The stau lifetime depends on the decay modes and on the mixing in the stau and the sneutrino sectors. Typical lifetimes vary between a few second to up to 10$^{11}$ seconds.

The NLSP's lifetime is not long enough to ensure that its decay occurs well after its freezeout. However, $\tilde{\nu}_R$ contains all good properties of cold dark matter. It is stable due to $R$-parity conservation and because it evades direct detection constraints owing to suppressed interactions due to tiny Yukawa coupling. The density parameter of $\tilde{\nu}_R$ can be written as
\begin{equation}
\Omega_{\tilde{\nu}_R} = \frac{m_{\tilde{\nu}_R}}{m_{\stau}} \Omega_{\stau},
\end{equation}
where $\Omega_{\stau}$ is the present density parameter of the $\tilde{\tau}_1$ NLSP, assuming it to be stable.

\begin{figure}[t]
	\begin{center}
		{
			\includegraphics[width=10cm,height=10cm]{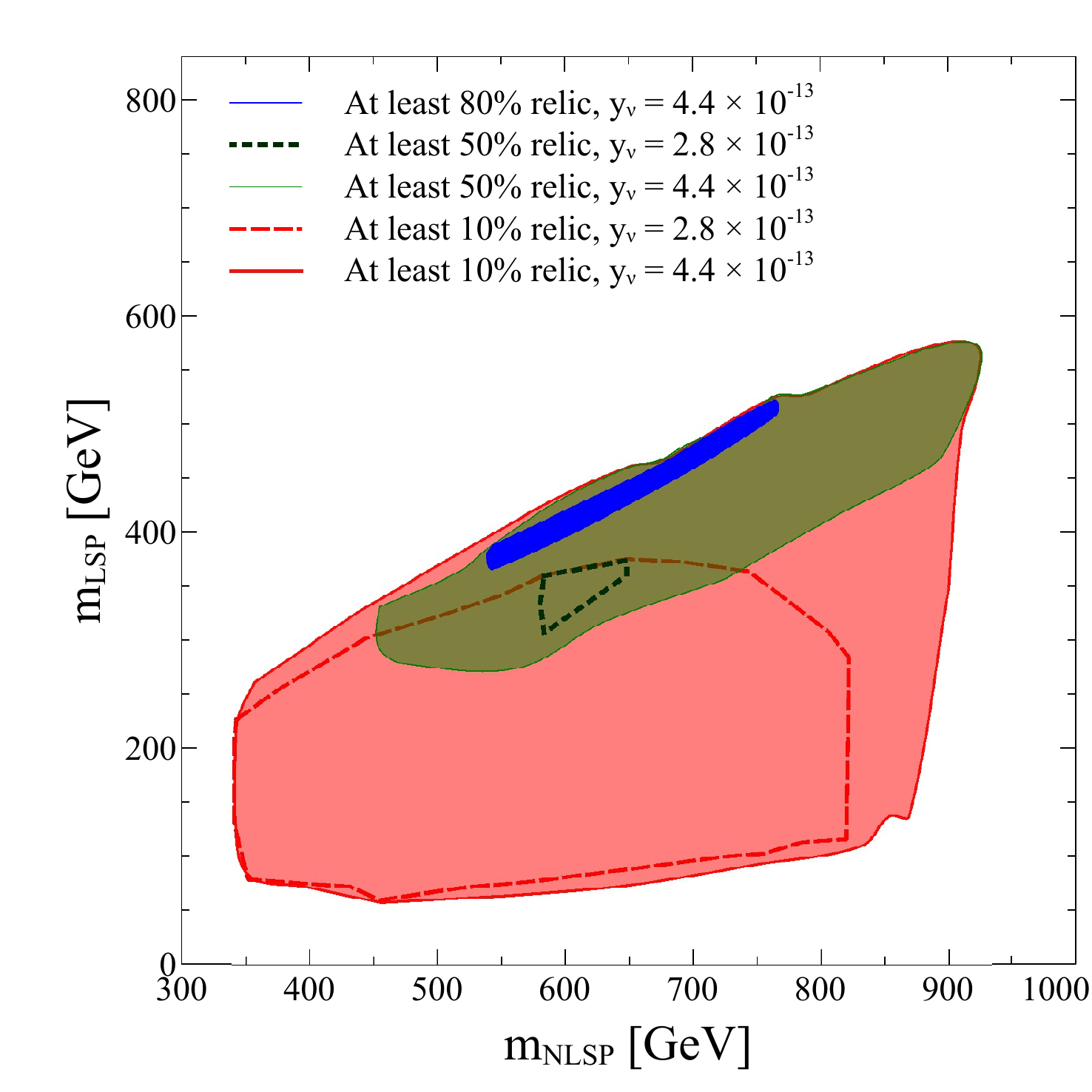}
			\caption{Allowed parameter range with percentage relic abundance in the $m_{\stau}-m_{\tilde{\nu}_R}$ $(m_{\textrm{NLSP}}-m_{\textrm{LSP}})$ space for two different Yukawa couplings corresponding to the degenerate and `hierarchical'  neutrino masses. }
			\label{fig:DM_per}}
	\end{center}
\end{figure}

\begin{figure}[!htb]
	\begin{center}
		{
			\includegraphics[width=10cm,height=10cm]{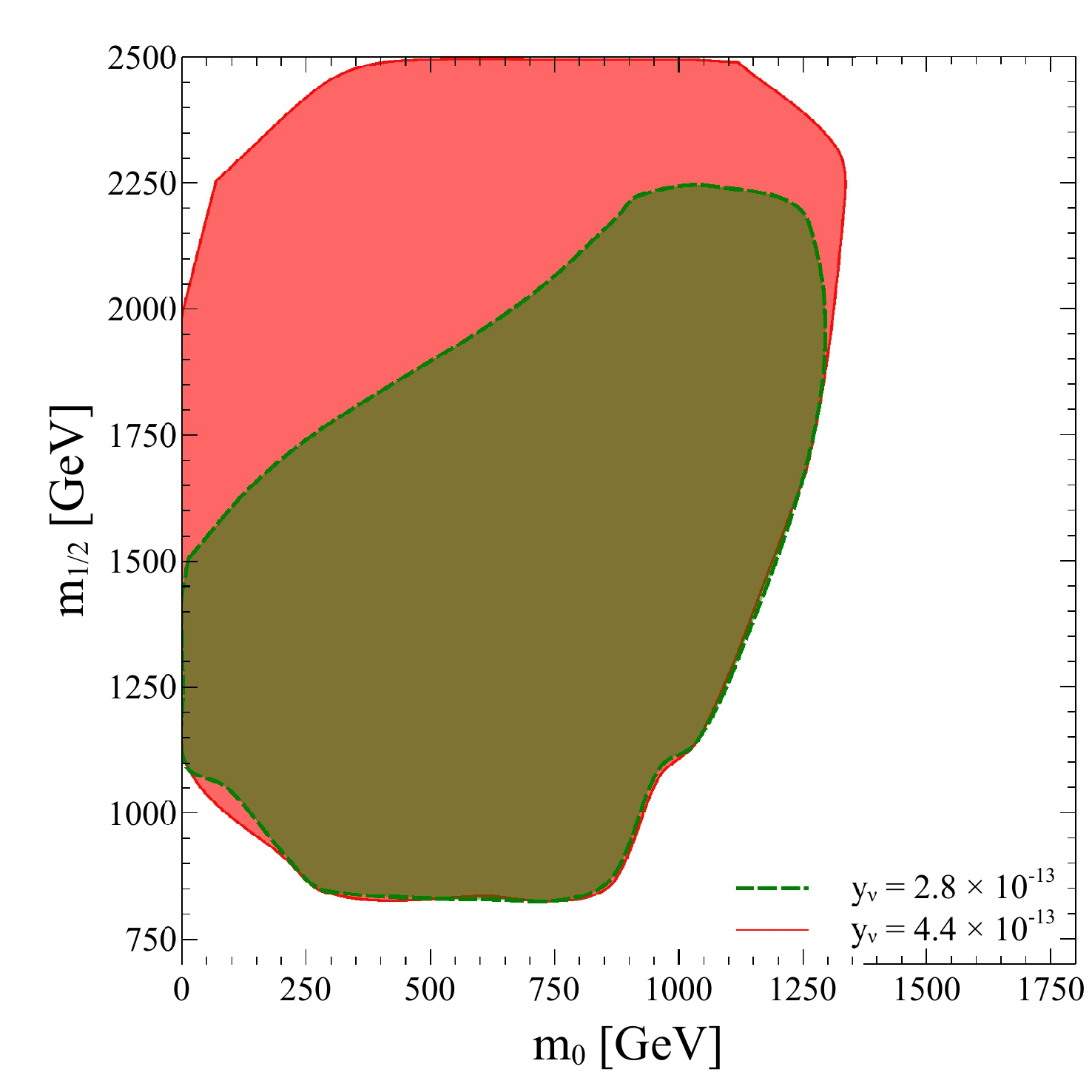}
			\caption{Allowed parameter region in the $m_0-m_{1/2}$ plane satisfying existing collider, low energy, relic and BBN constraints for the `hierarchical' (green) and degenerate (red) neutrino masses. Here, $m_{0,1/2} < 2500$ GeV, $|A_0| < 3000$ GeV, $5 < \tan{\beta} < 40$, $0 < m_{\tilde{\nu}_R} < m_0$ and sign$(\mu) > 0$.}
			\label{fig:m0m12}}
	\end{center}
\end{figure}

\begin{figure}[t]
	\hspace*{-0.3cm}
	\begin{center}
		{
			\includegraphics[width=10cm,height=10cm]{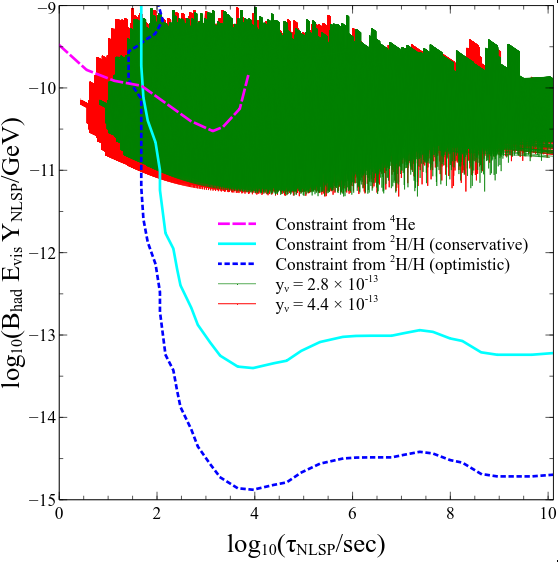}
			\caption{Allowed parameter region in the lifetime-injected hadronic energy plane which satisfies every existing constraint for the `hierarchical' (green) and degenerate (red) neutrino masses. The two curves are for the constraint from $^4$He (magenta dashed) and $^2$H/H (cyan solid) abundance. The dotted (blue) curve denotes the impact of assuming a tightened $^2$H/H determination.}
			\label{fig:BBN}}
	\end{center}
\end{figure}

Below, in Fig.~\ref{fig:DM_per} we show the allowed parameter space for two different Yukawa values assuming at least 10\% relic contribution.
We also show the allowed parameter region abiding all other collider and cosmological constraints in Fig.~\ref{fig:m0m12}.
We explicitly show the constraints ensuing from the BBN in Fig.~\ref{fig:BBN}, where the visible energy is $E_\text{vis}=\frac{m_{\stau}^2+m_W^2-m_{\tilde{\nu}_R}^2}{2 m_{\stau}}$ and $B_\text{had} = 2/3$ which corresponds to the hadronic branching ratio of the $\tilde{\tau}_1$ for two body decays. Lastly, $Y_\text{NLSP}$ is the ratio of the number density to the entropy density at the $\tilde{\tau}_1$ freeze-out.

Before concluding this section, we wish to mention the LHC prospects of this model. We studied the potential of the following channels:

\begin{itemize}
	\item $2 \, \stau + N \,\textrm{hard}\, \textrm{jets}\: (N\ge 2)\,,$
	\item $2 \, \stau$ (two stable charged tracks)\,,
	\item passive detection of highly-ionising (slow) particles.
\end{itemize}

\begin{figure}
	\begin{center}
		{
			\includegraphics[width=12cm,height=8cm]{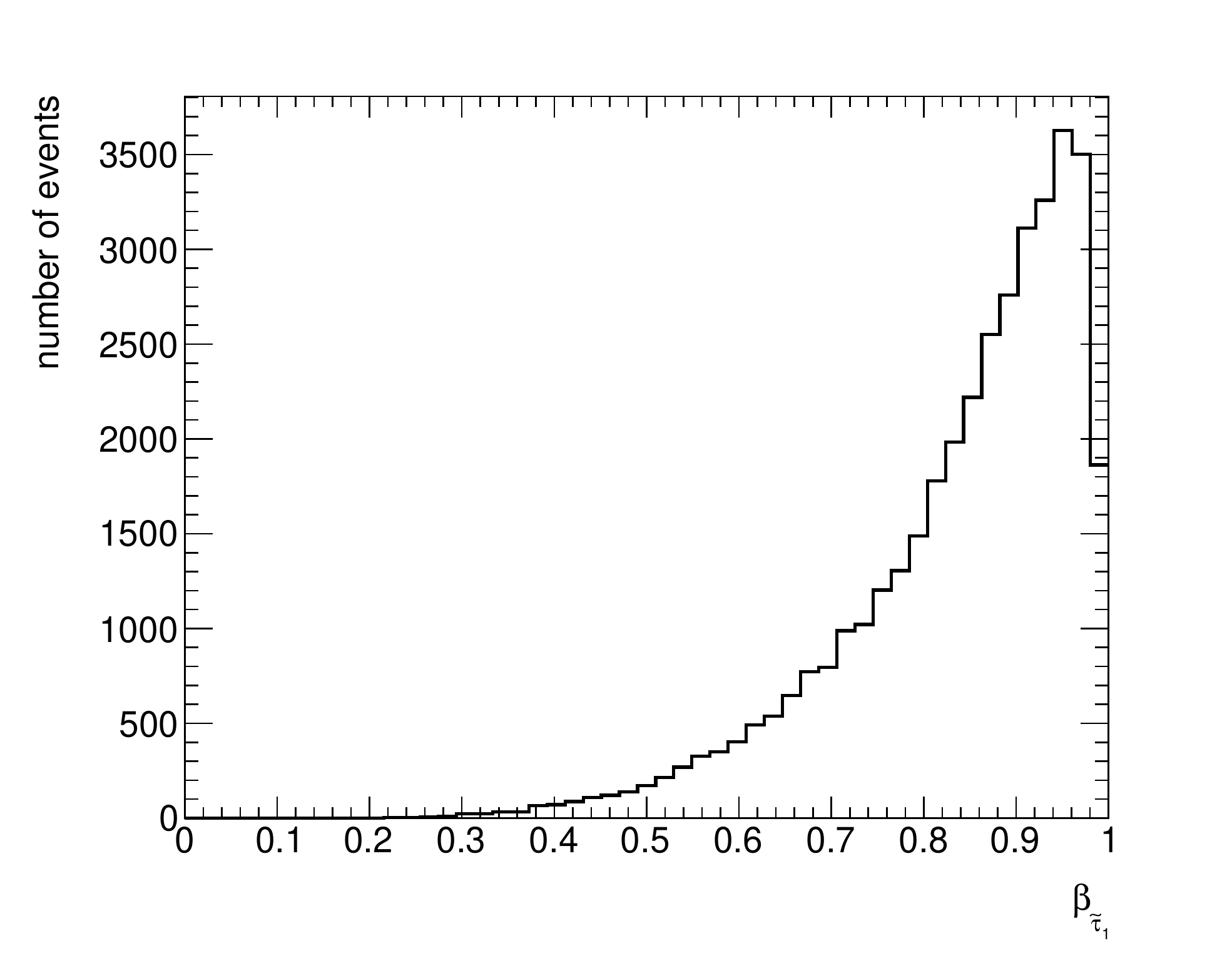}
			\caption{Velocity distribution of the $\stau$-NLSP for BP3.}
			\label{fig:vel-dist-BP1}
		}
	\end{center}
\end{figure}

For this, we consider BPs following the trend

\begin{equation}
m_{\tilde{\nu}_{R}} < m_{\stau} < m_{\chi_1^0} < m_{\tilde{e}_1, \tilde{\mu}_1} < \ldots < m_{\tilde{g}} \nonumber
\end{equation}
The "\textit{stable}" $\tilde{\tau}_1$ will behave like a slow muon, with velocity $\beta = p/E$ much less than 1 (as can be seen in Fig.~\ref{fig:vel-dist-BP1}).

The $2 \, \stau + N \,\textrm{hard}\, \textrm{jets}\: (N\ge 2)\,,$ comes with $\slashed{E}_T$ and is of no concern in this section. The category with 2 stable charged tracks is something that is worth saying a few words in this section. The most dominant final state is with two muons. For the four BPs listed in Ref.~\cite{Banerjee:2016uyt} we list the results in Table~\ref{tab:sig}.

\begin{table}[!ht]
	\centering
	\begin{tabular}{|c|c|c|c|c|c|}
		\hline 
		Cut set   & Benchmark point &  $N_S$ & $N_B$ & $N_S/N_B$ & $\mathcal{S}$ \\
		\hline 
		\multirow{3}{0.5cm}{A}  & BP1 & 526 &      & 0.09 & 6.7 \\ 
		& BP2 & 358 & 5684 & 0.06 & 4.6 \\
		& BP3 & 258 &      & 0.05 & 3.3 \\
		& BP4 & 47  &      & 0.01 & 0.6 \\
		\hline
		\multirow{3}{0.5cm}{B}  & BP1 & 1337 &       & 0.10 & 11.3 \\ 
		& BP2 & 1069 & 12772 & 0.08 & 8.9 \\
		& BP3 & 826  &       & 0.06 & 7.0 \\
		& BP4 & 232  &       & 0.02 & 2.0 \\
		\hline
		\multirow{3}{0.5cm}{C}  & BP1 & 1543 &      & 0.44 & 21.8 \\ 
		& BP2 & 1014 & 3481 & 0.29 & 15.1 \\
		& BP3 & 715  &      & 0.21 & 11.0 \\
		& BP4 & 211  &      & 0.06 & 3.5  \\ 
		\hline
	\end{tabular} 
	\caption{Table shows the number of signal and background events after the selection cuts (Table~\ref{tab:cuts}), the ratio $N_S/N_B$ and the statistical significance. $\mathcal{L} =$ 3000 fb$^{-1}$.}
	\label{tab:sig}
\end{table}

\begin{table}
	\centering
	\begin{tabular}{|c|c|c|c|}
		\hline
		Cut on  & Cut set A  & Cut set B & Cut set C  \\ 
		\hline \hline
		$\beta$ &  $> 0.85$ & $-$ & $ < 0.95$ \\
		$p_T^{\mu_{1,2}}$ & $> 200$\,GeV & $> 200$\,GeV & $> 70$\,GeV \\
		$\sum{|p_T^{vis.}|}$ & $ > 700$\,GeV & $ > 500$\,GeV & $-$ \\
		$|y(\mu_{1,2})|$ & $ < 2.4$ & $ < 2.4$ & $ < 2.5$ \\
		$M_{\mu_1, \mu_2}$ & $> 1200$\,GeV & $ > 1000$\,GeV & $-$ \\ 
		$\Delta R(\mu_1,\mu_2)$ & $> 0.2$ & $> 0.2$ & $-$ \\
		$\Delta R(\mu,j)$ & $> 0.4$ & $> 0.4$ & $-$ \\
		$\Delta R(j,j)$ & $> 0.4$ & $> 0.4$ & $-$ \\
		\hline 
	\end{tabular}
	\caption{Three sets of selection cuts applied in the $\stau$ pair analysis.}
	\label{tab:cuts}
\end{table}

\subsection{Hidden portal models}
Portal models consider NP that interacts with the SM via one specific (class of) SM particles, namely the neutrinos, the Higgs boson, and the vector bosons. Usually, also pseudoscalar particles are considered, and often new fermions are introduced.
This approach allows for an effective categorisation of observed signatures and for some degree of model independence.
For a recent discussion, see for instance Ref.~\cite{Agrawal:2021dbo}.
Below we highlight the portals' properties that are relevant here.

\subsubsection{The Neutrino portal}
The observation of neutrino oscillation has firmly established the non-vanishing masses of the active neutrinos. New physics is required to invoke certain mechanisms for neutrino mass generation. 
One of the most celebrated theoretical means is the so-called type I seesaw mechanism, which generates small neutrino mass by introducing right-handed neutrinos, which couple to the active neutrino and the Higgs fields through a Dirac mass term, plus a Majorana mass term, and can be described in a simple way as
\begin{equation}
{\cal L}_Y = - Y_D \bar{L} H N - M_N \bar N^c N + \text{h.c.},
\end{equation}
where $H$ is the SM Higgs field, $N$ are the right-handed neutrinos with Majorana mass $M_N$ and Yukawa coupling $Y_D$. 
In the type I seesaw approximation the small neutrino mass is given by $ M_D^2 / M_N$, where $M_D= Y_D \langle H \rangle$ is the Dirac mass term.

The right-handed and active neutrinos mix, thus creating a number of light and heavy mass eigenstates, with dominant active and sterile neutrino components, respectively.
The smallness of the light neutrinos' masses can be achieved either by a very large Majorana mass of order $10^{11-13}$\,GeV, by a very small Yukawa coupling $Y<10^{-5}$, or by invoking additional symmetries~\cite{Mohapatra:1986bd}. 

These kinds of models are often called `heavy neutral leptons', `sterile neutrinos', or `neutrino-portal models', which have been discussed often and in many different settings, cf.\ e.g.\ Refs.~\cite{Dasgupta:2021ies,Cottin:2018kmq,Antusch:2018svb,Helo:2018qej,deVries:2020qns,Chiang:2019ajm}.

At the LHC, heavy neutrinos are produced predominantly in Drell Yan processes, and they have been searched for traditionally via lepton number violating signatures, which are striking because of absent SM backgrounds~\cite{Atre:2009rg}.
Symmetry protected models allow for in principle large production cross-sections, but even their most promising signatures are not very effective due to the signal rates being suppressed by other constraints, and the ubiquitous backgrounds~\cite{Antusch:2018bgr}.
Heavy neutrinos that are lighter than the $W$ boson have long lifetimes.
They may give rise to displaced vertex signatures in any of the detector components thus be hidden from the current searches, which will be discussed below.

\subsubsection{The Higgs portal} 
Here we describe a simple Higgs-portal model with an additional real SM-singlet scalar field $X$ that mixes with the SM Higgs doublet field $\Phi$ in the presence of a new $Z_2$ symmetry.
The new scalar field $X$ is odd under the $Z_2$ such that no $X$ or $X^3$ terms appear, while all the SM fields are even.  The Lagrangian is given by
\begin{equation}
{\cal L} = \frac{1}{2}\partial_{\mu}X\partial^{\mu}X
+\frac{1}{2}\mu^{2}_{X}X^{2}-\frac{1}{4}\lambda_{X}X^{4}
-\frac{1}{2}\lambda_{\Phi X}(\Phi^\dagger \Phi)X^{2} +  {\cal L}_\text{SM} \;. \label{L'}
\end{equation}
After electroweak symmetry breaking both the SM Higgs doublet field $\Phi$ and the new scalar singlet field $X$ are expanded around their vacuum-expectation values $\langle\phi\rangle\approx 246$\,GeV and $\langle\chi\rangle$:
\begin{eqnarray}
\Phi (x) &=& \frac{1}{\sqrt{2}} \left( \begin{array}{c}
0 \\
\langle \phi \rangle + \phi(x) \end{array} \right ) \;,  \\
X(x)    &=& \langle\chi\rangle + \chi (x) \;.
\end{eqnarray}
The mass matrix of the two scalar fields is 
\begin{equation}
{\cal L}_m =  - \frac{1}{2} \left( \phi \; \chi \right )\,
\left( \begin{array}{cc} 
2\lambda\langle\phi\rangle^2 & \lambda_{\Phi X}\langle\phi\rangle\langle\chi\rangle  \\
\lambda_{\Phi X}\langle\phi\rangle\langle\chi\rangle & 2\lambda_{X}\langle\chi\rangle^2 \end{array} \right )\,
\left( \begin{array}{c} 
\phi \\
\chi \end{array} \right ) \;.
\end{equation}
It is possible to rotate $(\phi\; \chi)^T$ to $(h \; h_{s})^T$ 
through an angle $\theta$
\begin{equation}
\left( \begin{array}{c}
h \\ 
h_{s}  \end{array} \right ) 
= 
\left( \begin{array}{cc} 
\cos\theta & \sin \theta \\
- \sin \theta & \cos\theta  \end{array} \right )\,
\left( \begin{array}{c}
\phi \\ 
\chi  \end{array} \right ) \;,
\end{equation}
where $h$ is the scalar Higgs boson observed at 125\,GeV while $h_s$ 
is the new scalar boson of the model. The mixing angle $\theta$ is
constrained to be very small due to various experimental datasets
including the Higgs boson strength data~\cite{Chang:2016lfq}. The new
scalar boson $h_s$ originating from a hidden sector can decay 
back into SM particles via the mixing angle, which is suppressed
by $\sin^2\theta$. For example, the partial width into a pair 
of SM fermions is given by 
\begin{eqnarray}
\Gamma ( h_{s} \to f \bar f) &=& N_f \, \sin^2{\theta} \, 
\frac{m_\ell^2 m_{h_{s}}}{ 8 \pi \langle \phi \rangle^2 } \left( 1 - 
\frac{4 m_f^2}{m_{h_{s}}^2 } \right )^{3/2}\;,  \label{eqn:Gammahs2ll}
\end{eqnarray}
where $N_f$ is the colour of the fermion. The total decay width of $h_s$ 
can be obtained by summing over all kinematically allowed fermion pairs. 
For small enough $\theta$ and light $h_s$ the lifetime of $h_s$ can travel a macroscopic distance before decay, which may complicate its detection.

To detect a new physical scalar through mixing with the Higgs boson at the LHC is more complicated than one would think. Depending on the mass of the new particle, most of its decays leave hadronic final states and have to contend with towering backgrounds involving top quarks and multiple vector boson production, cf.\ e.g.\ Ref.~\cite{CMS:2015hra}.

The four-lepton final state is often referred to as the `golden channel' due to small and controllable SM backgrounds, and used by the ATLAS~\cite{ATLAS:2017tlw} and CMS~\cite{CMS:2018mmw} collaborations  to search for heavy scalars. 
However, the tiny total cross-section of this process is suppressed further by the necessarily small scalar mixing and by additional decay channels, such as di-top, or Higgs $Z$, and possible decays involving (invisible) new particles.

\subsubsection{The vector portal} 
Generic BSM theories often contain additional gauge symmetries, especially $U(1)^\prime$ symmetries. 
Additional spin-$1$ or vector particles commonly arise from the breakdown of a larger gauge symmetry factor, or when the vector is a composite state. 
Such vector particles may have interactions with SM particles, and also with possible new particles.
In addition, these vectors may mix kinetically with the $U(1)$ factor in the SM gauge group. 
For example, after the electroweak symmetry breaking and diagonalisation of the gauge kinetic terms, a dark photon ($A^\prime$) theory may have the following Lagrangian:
\begin{equation}
\mathcal{L}_{A^\prime} \supset -\frac 1 4 F^\prime_{\mu\nu} F^{\prime \mu\nu} + \frac 1 2 m_{A^\prime}^2 A^\prime_\mu A^{\prime \mu} + \epsilon e A^\prime_{\mu} J_{\rm EM}^\mu,
\end{equation}
where $m_{A^\prime}$ is the dark photon mass, $\epsilon$ the small kinetic mixing parameter, $J^\mu_{\rm EM}$ the standard model electromagnetic current, and $F^\prime_{\mu\nu}$ the standard field strength operator for $A^\prime$. While $A^\prime$ could have more interactions with the SM particles, the Lagrangian above can be considered the minimal dark photon scenario for phenomenological purposes. 

In this case, the partial width of $A^\prime$ to SM fermions follows:
\begin{equation}
\Gamma_{A^\prime\to f\bar f}=\frac {\epsilon^2 \alpha_{\rm EM} \kappa} {3} m_{A^\prime} \left(1+ \frac {2m_f^2} {m_{A^\prime}^2}\right) \sqrt{1-\frac {4 m_f^2}{m_{A^\prime}^2}}.
\end{equation}
Here $\kappa\equiv 3 Q^2$ for SM quarks with charge Q and $\kappa\equiv 1$ for SM charged leptons. The kinetic mixing parameter $\epsilon$ is expected and constrained to be tiny. 

Dark photons can be produced at the LHC via their mixing with the SM photon. The strong experimental limits on the mixing parameter $\epsilon$ render the production rates tiny, and thus difficult to test.
The limits on the mixing also necessitate a small decay rate, and hence enforce a long lifetime for the dark photon, further complicating the discovery of its signatures.

\subsection{Long-Lived Particles}
The negative results at the LHC have raised the question of whether there is a systemic shortcoming in detecting new physics. 
Indeed, one of the possibilities is that new physics may manifest itself in the form of long-lived particles (LLP), which might have escaped from detection in the current design of the experimental triggers or due to the size of detectors or negligible interactions with the detectors.  

This particular class of NP models raised strong interest in the HEP community. 
Experimentalists and theorists are working hard to overcome the challenge of detecting LLP signatures, which include disappearing tracks, emerging jets and leptons, kinks in tracks and which depend strongly on the explicit considered model and particle content.

\subsubsection{The experimental view}
In recent years, there are rising interests in searches for LLPs in both theoretical and experimental communities to search for LLP signatures at the LHC which is, however, a very challenging task.
Current hardware and software triggers and analyses have been focused on 
promptly decaying new particles, such as squarks or gluinos in supersymmetry, top partners in composite models, and leptoquarks in GUT models.   

Efforts by the collaborations in recent years led to tremendous progress to cover a large variety of signatures from long lived particles, see e.g.\ Ref.~\cite{ATLAS:2019kpx}.
The LHC collaborations have developed a broad programme of LLP searches~\cite{Alimena:2019zri}, which grew from discussions with theorists, cf.\ e.g.\ Refs.~\cite{Liu:2015bma,Lee:2018pag,Alimena:2019zri}.

It is worthwhile to comment on the scope of the programme for LLP searches, and its future perspective.
Improved triggers and new components in the detector have been suggested, discussed, and implemented at the ATLAS and CMS experiments to accommodate the searches for LLPs~\cite{Gershtein:2017tsv,Liu:2018wte,CMS-PAS-FTR-18-018,Liu:2020vur,Gershtein:2020mwi}. 
New search strategies have been suggested and discussed, and will be implemented already in the next run~\cite{Liu:2015bma,Liu:2018wte,Liu:2020vur,Lee:2018pag,Curtin:2018mvb,Alimena:2019zri}. 
Specific experiments are planned, under construction, or being commissioned, to add complementary search capacity to the big LHC collaborations. Examples are MilliQan~\cite{Haas:2014dda}, FASER~\cite{Feng:2017uoz}, MoeDAL~\cite{MoEDAL:2009jwa}, and MATHUSLA~\cite{Curtin:2018mvb}.

\subsubsection{Signatures}
Particles with long lifetimes give rise to distinct signatures in the LHC detectors, which crucially depend on their electric charge, lifetime, velocity, and interactions with the detector. 
The search for the LLPs at the LHC can make use of the tracker detector, calorimeters, muon spectrometer, and/or the new timing detector, depending on the decay length and decay products of the LLP. 

Charged LLPs (cLLP) interact with the detector components, and in particular they leave a track in the tracker detector, possibly with a very characteristic ionisation signature.
A cLLP's signature can be very different, depending on its lifetime:
In case of the decay taking place outside the detector, one has a muon-like ionised track throughout all the detector components;
When a cLLP decays into a number of soft final states it can give rise to a disappearing track, provided the charged daughter particle is missed;
A cLLP that decays into one charged daughter and one (or more) neutral particles can give rise to a charged track with a kink.

Neutral LLP can be observed when they decay, possibly via a chain, into charged particles inside the detector.
If the decay takes place in the tracker, i.e.\ if the tracks of the charged daughter particles can be reconstructed, it is in principle possible to also reconstruct the point of decay, the so-called displaced vertex.

In general, displaced decays make misinterpretation of the daughter particles possible, in particular if the LLP decay takes place outside the tracker, and only partial detector information on the daughter particles (in particular their charges) is available. Then the reconstruction of the displaced vertex is not always possible but, depending on the LLP mass, the decay products could be delayed, which would make them observable via the time delay feature, as proposed in Ref.~\cite{Liu:2018wte}.

\subsubsection{Theoretical motivation}
Lifetimes are inversely proportional to the total decay rate of the decaying particle, and can therefore become larger when the coupling constant(s) are tiny and/or when the phase space is suppressed.
In SUSY models with a very small $R$-parity-violating (RPV) coupling, the lightest supersymmetric particle (LSP) can travel a macroscopic distance 
before decay, thus gives rise to LLP signatures. One can identify displaced vertices~\cite{Liu:2015bma,Csaki:2015uza,Wang:2019orr} in the tracker detector or make use of the timing detectors to detect less relativistic and heavy LLPs~\cite{Liu:2018wte,Cheung:2021utb}.
In gauge-mediated SUSY breaking models, the photino-NLSP case is the more
difficult scenario. It decays into a non-pointing photon and missing energy.
One has to rely on the EM calorimeter to determine the displaced vertex. 
If the decay falls outside the EM calorimeter, the event would be easily lost.

In the Higgs-portal models hadronic LLP signatures can generically be produced~\cite{Craig:2015pha,Csaki:2015fba,Liu:2018wte,Alipour-Fard:2018rbc,Liu:2020vur}. If $m_{h_s} \sim O(1)$\,GeV the $h_s$ can decay into
a pair of muons or pions with a displaced vertex~\cite{Chang:2016lfq}. Such muon pairs appear as energetic collimated muons (so-called ``muon-jets''), and collimated pion pairs appear as so-called ``fat-jets''. One can also make use of muons and jets with a displaced vertex detected in the inner tracker or the muon spectrometer~\cite{ATLAS:2018tup,CMS:2021juv}. 

Other light portals have similar LLP signatures but with different kinematic distributions. For instance, searches for Axion-Like-Particles (ALPs) have posed a great challenge at the LHC due to its low production rate and large hadronic background. It is possible to cover new ground for ALPs through a displaced vertex search~\cite{Hook:2019qoh}, or through new production and decay modes that are beyond the minimal model~\cite{Bauer:2017nlg}. Note that LHCb has certain advantage in searching for low mass prompt and displaced dilepton resonances~\cite{Ilten:2016tkc,Ilten:2018crw}, allowing for complementary coverage in the low mass regime with respect to the ATALS and CMS experiments. 
In models of sterile neutrinos, once they are produced, they can travel a macroscopic distance before they decay into leptons and/or hadrons~\cite{Liu:2019ayx,deVries:2020qns}, thus giving rise to
displaced vertex or emergent leptons or jets. The signals can be detected at the tracker and/or calorimeters.

In addition, the hidden strong dynamics~\cite{Strassler:2006im,Knapen:2016hky,Pierce:2017taw,Chacko:2020zze,Yuan:2020eeu,Knapen:2021eip} have gained increasing attention due to the distinct feature of dark showers. Such dark shower signatures require close examination of the experimental capabilities and they are under active development. 

Complementing the searches with the existing detectors and upgrades, new peripheral experiments and low energy particle physics experiments are being evaluated and under construction to look for these exotic long-lived signatures~\cite{Beacham:2019nyx}.

\subsubsection{Backward moving objects from TeV LLP}
Of particular interest is a less-studied signature of backward moving objects (BMOs)~\cite{Banerjee:2017hmw}. While we are more used to studying fast moving particles whose decay products are collimated along the direction of the parent, for slow moving particles, such decay products have a wider distribution. With a few examples, we show that for heavy LLP (around TeV mass scale) searches at the LHC, the particles ensuing from the secondary vertex can be at large angular separations with respect to the direction of motion of the LLP. A fraction of such particles can even go in the backward direction, giving rise to striking signatures as these particles traverse the various layers of the detector \textit{outside-in}, towards the direction of the beam pipe. This can be translated to the energy deposited in the tracker. The particles can come from as far as the hadron calorimeter (HCAL) or even ones that can come from outside the detector and into the muon chamber. We see that the most prominent effect comes from LLPs which come to rest inside the detector, where the example being studied is the $R$-hadrons. We also see similar results when the LLPs are lighter than the TeV scale or when some of the available energy is carried away by a massive invisible daughter. The four benchmarks studied are the following:
\begin{itemize}
	\item $X \to qq$, where $X$ is the LLP and $q$ is a massless quark. Such decays can be seen in $R$-parity violating (RPV) supersymmetric models. We can have processes like $\tilde{q} \to qq$ or $\tilde{l} \to qq$. We classify this channel as $2BM0$.
	\item $X \to qqq$: This is another example of a RPV process where an example process is $\tilde{\chi}_1^0 \to qqq$. We call this category as $3BM0$.
	\item $X \to q DM$, where $DM$ is a heavy invisible daughter. Such processes are possible through $R$-parity conserving scenarios with channels such as $\tilde{q} \to q \chi_1^0$ or $\tilde{g} \to g \chi_1^0$, where $\chi_1^0$ is the lightest neutralino. We categorise this channel as $2BM$.
	\item $X \to qq DM$: This can also come about in an $R$-parity conserving scenario and the following three body decay encapsulates such a process, $\tilde{g} \to q\bar{q} \chi_1^0$. We call this the $3BM$ category.
\end{itemize}

\begin{figure}[tbhp]
	\includegraphics[height=4cm,width=6.95cm]{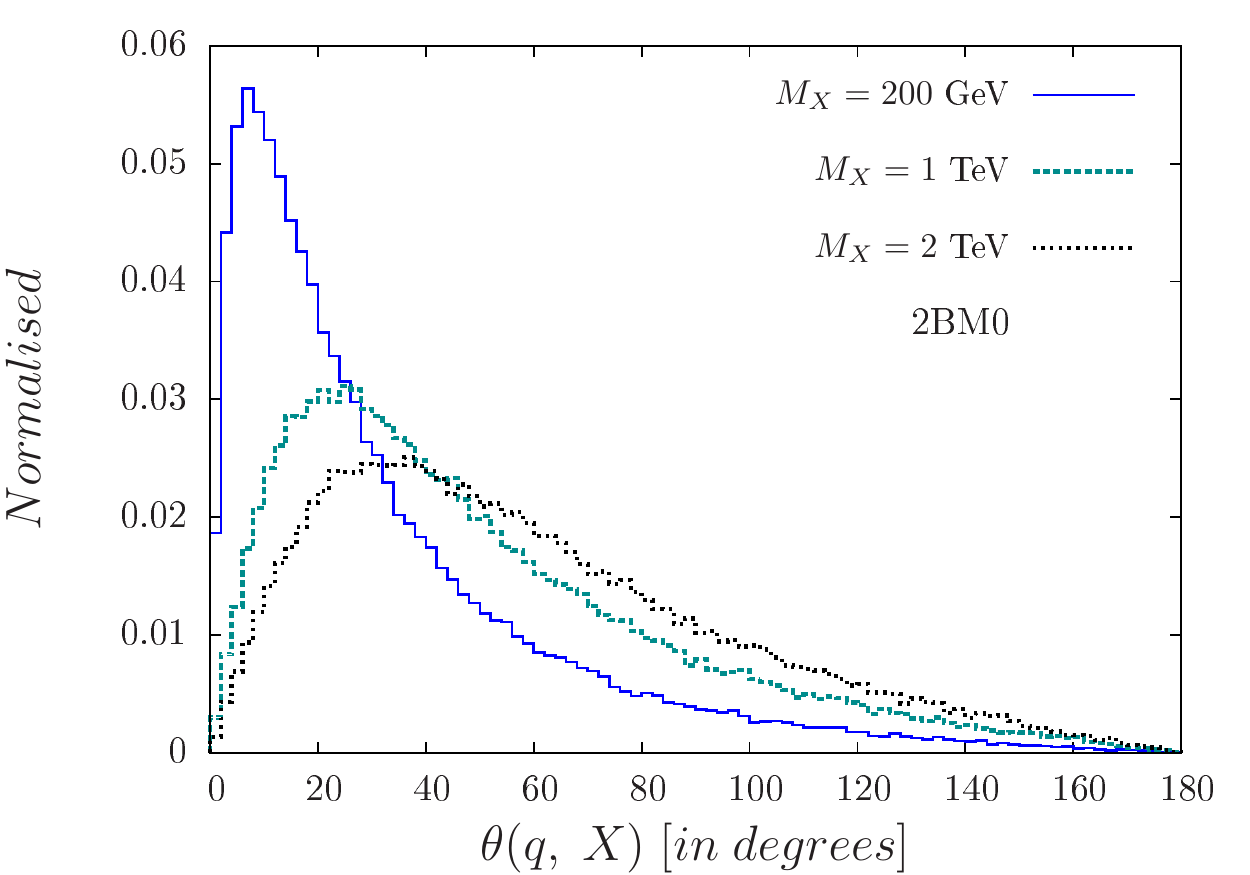}
	\includegraphics[height=4cm,width=6.95cm]{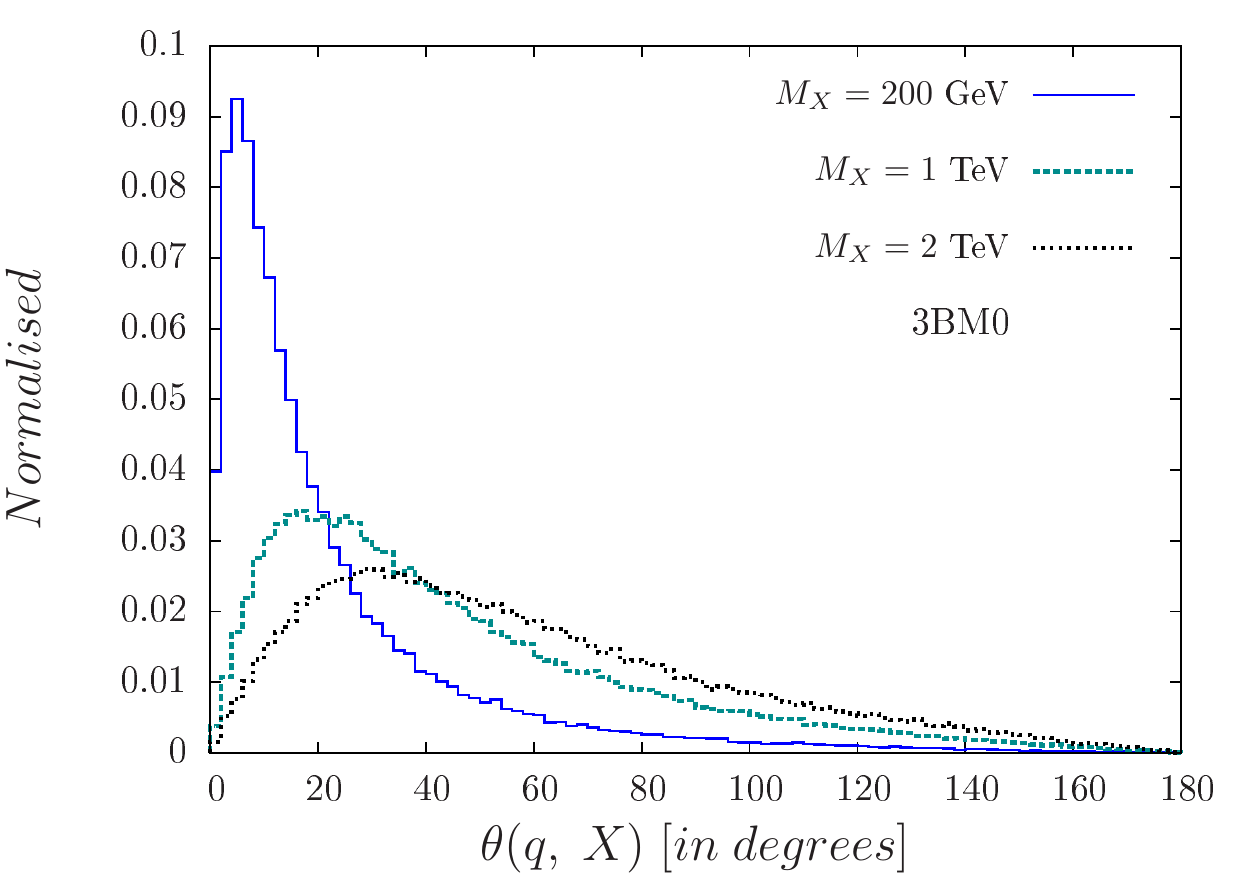}\\
	\includegraphics[height=4cm,width=6.95cm]{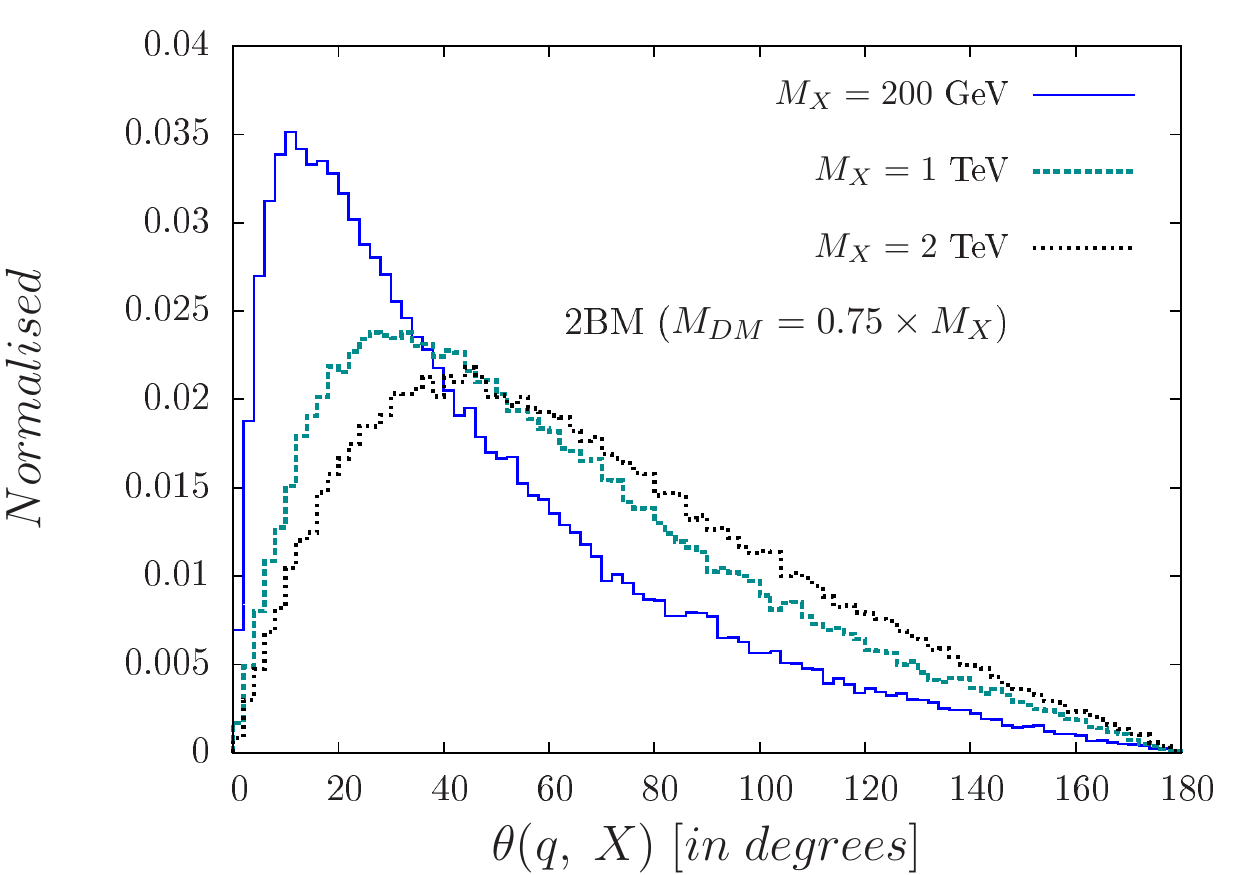}
	\includegraphics[height=4cm,width=6.95cm]{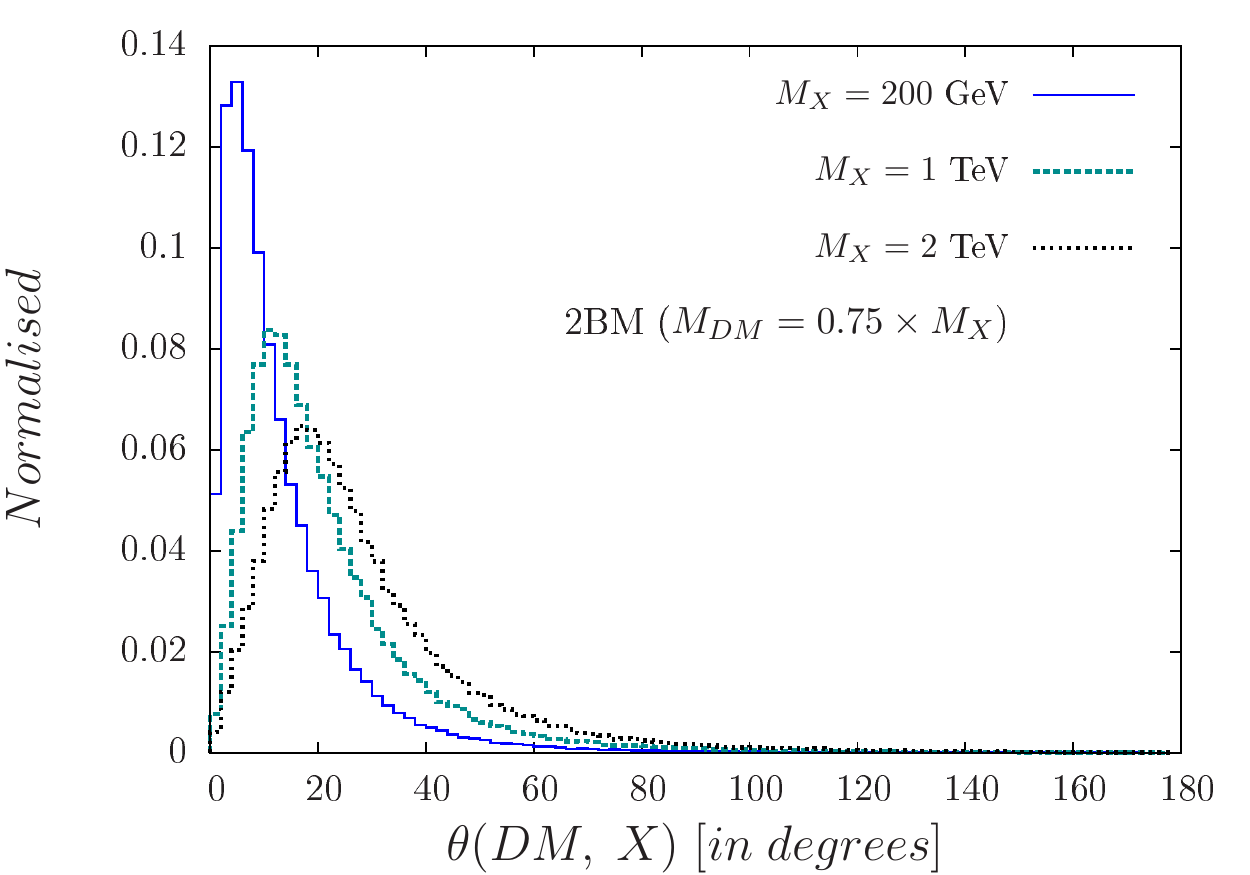}\\
	\includegraphics[height=4cm,width=6.95cm]{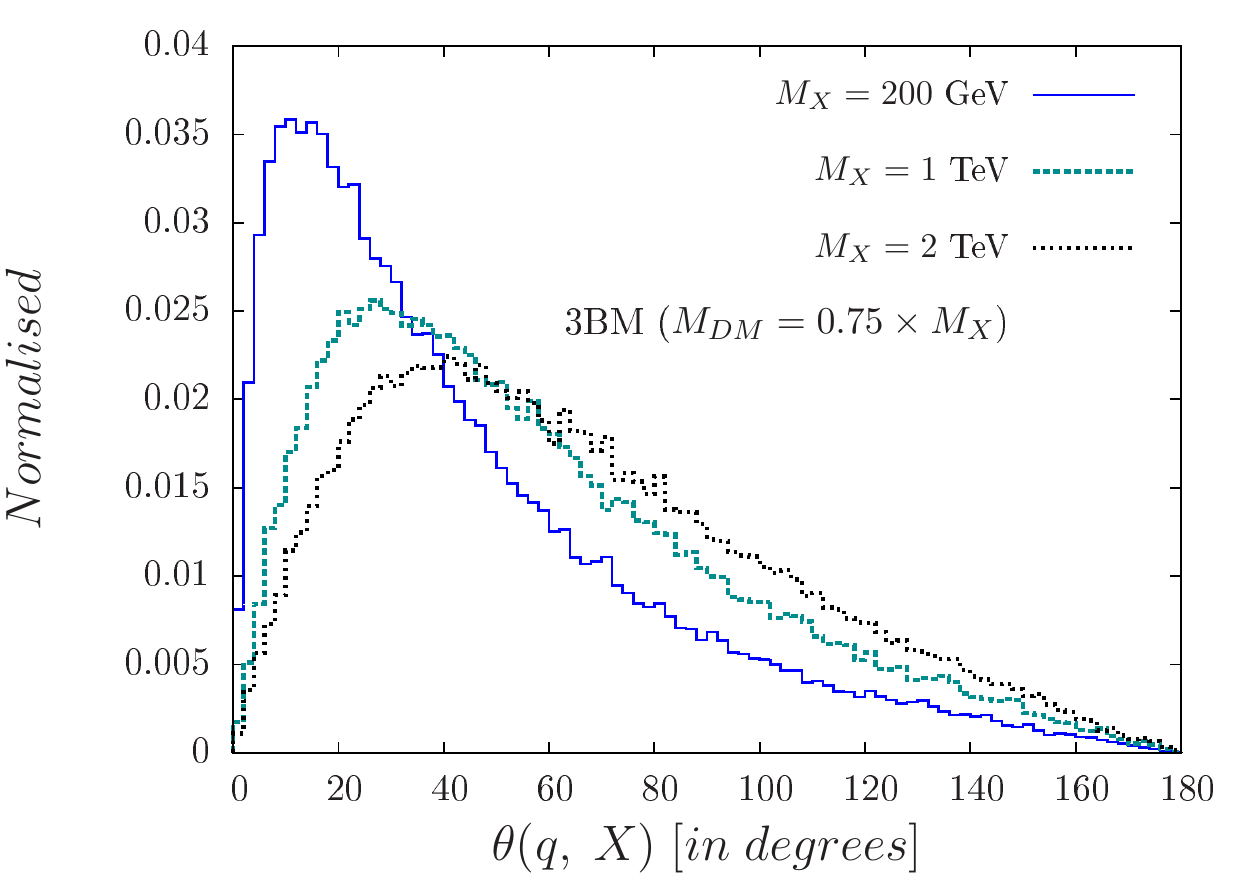}
	\includegraphics[height=4cm,width=6.95cm]{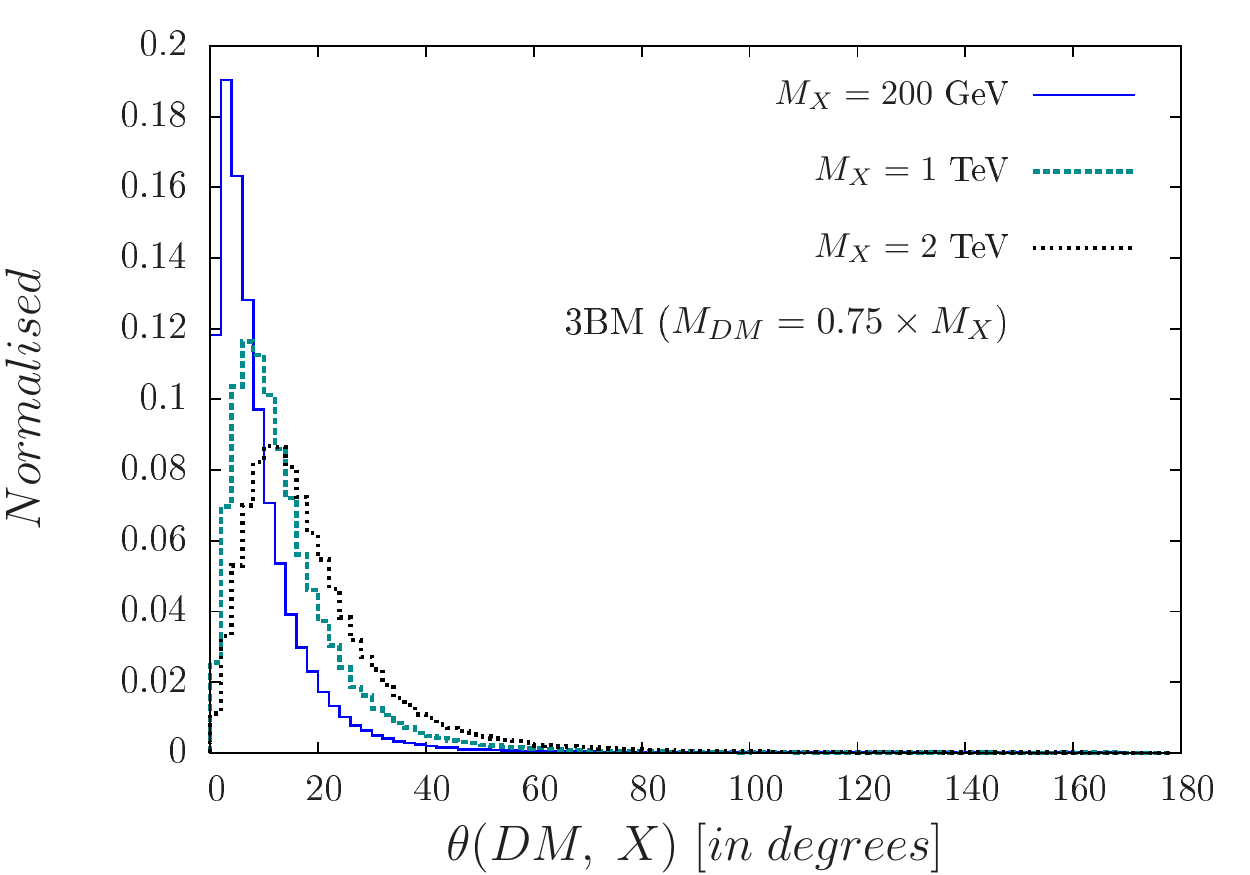}\\
	\includegraphics[height=4cm,width=6.95cm]{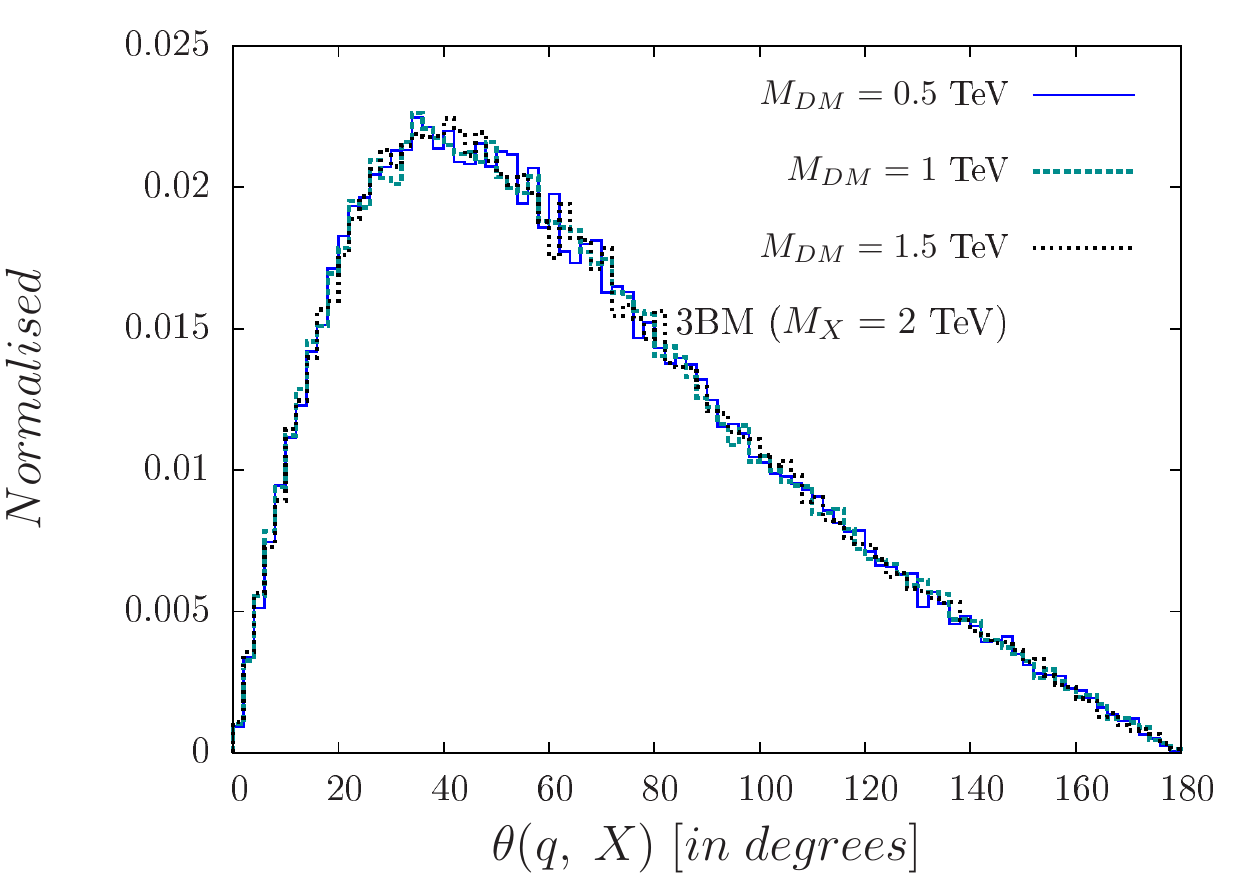}
	\includegraphics[height=4cm,width=6.95cm]{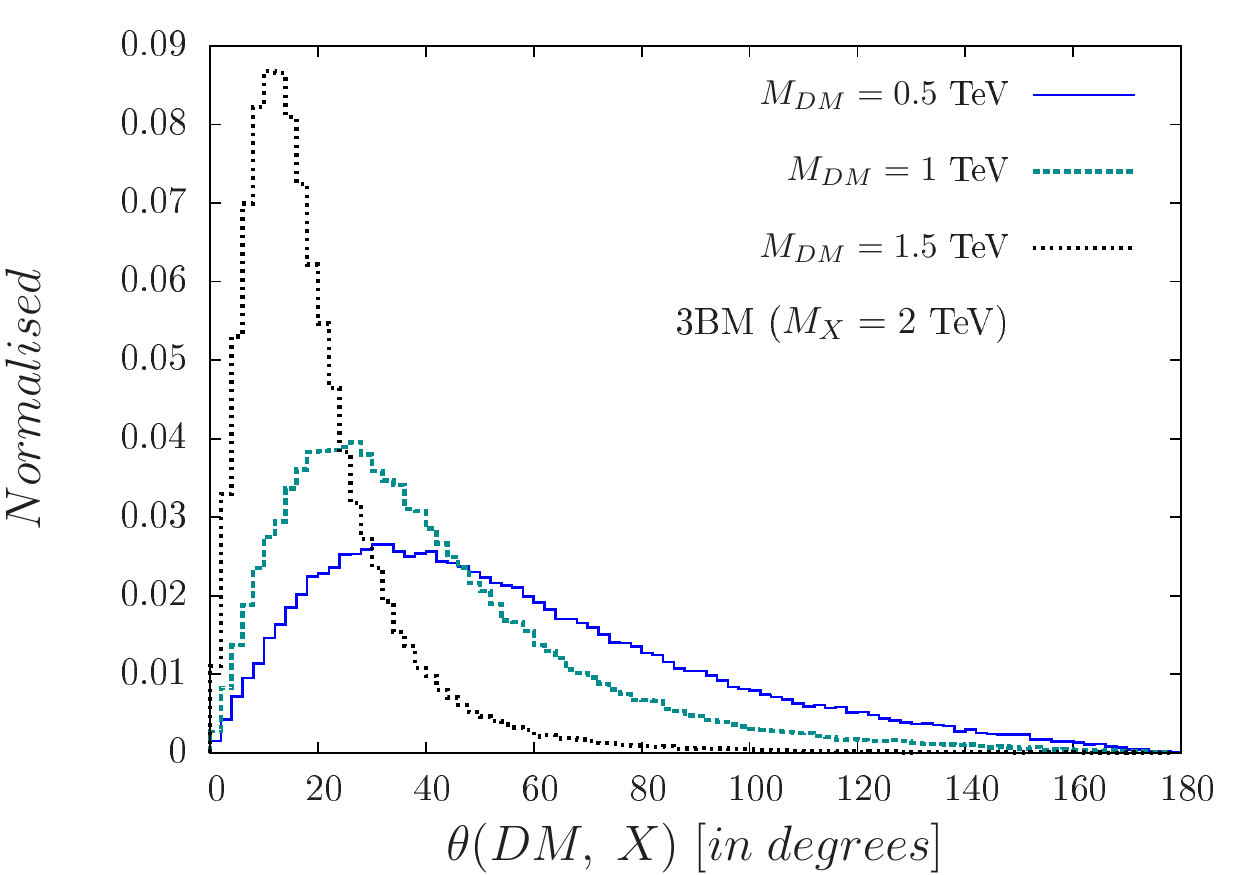}\\
	\caption{Angle $\theta$ between the direction of $X$ and one of the quarks, $q$ or the massive daughter (DM).}
	\label{fig:LLP_BMO_1}
\end{figure}

In Fig.~\ref{fig:LLP_BMO_1}, we show the angle $\theta$ that $1$ or $DM$ makes with the direction of the mother LLP, for several benchmark points (BPs). We choose various values for the LLP, $M_X$ and the mass of the invisible particle, $M_\text{DM}$. We can clearly see that for many of these BPs, the fraction of (light) particles going in the backward direction, is significant.

\begin{figure}[!h]
	\centering
	\includegraphics[height=4cm,width=6.95cm]{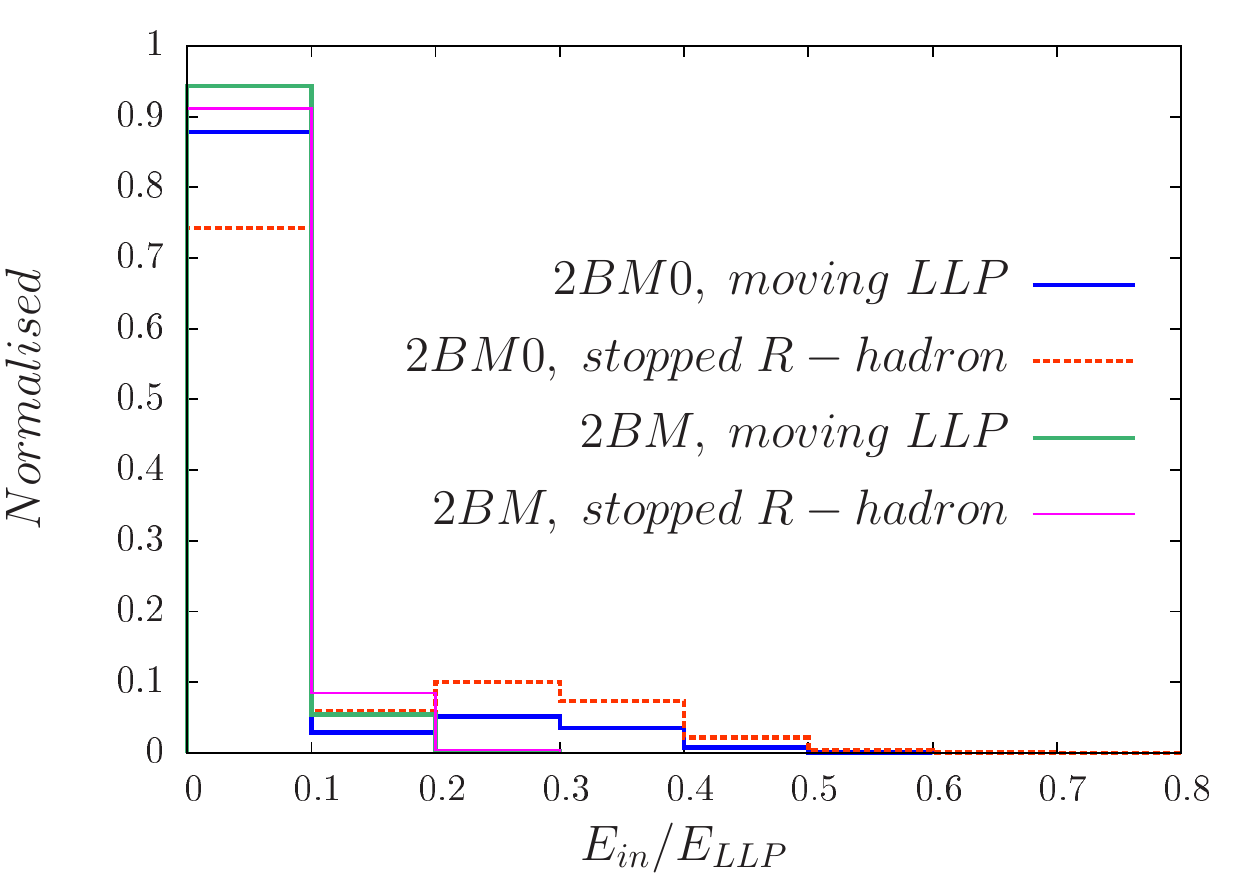}~\includegraphics[height=4cm,width=6.95cm]{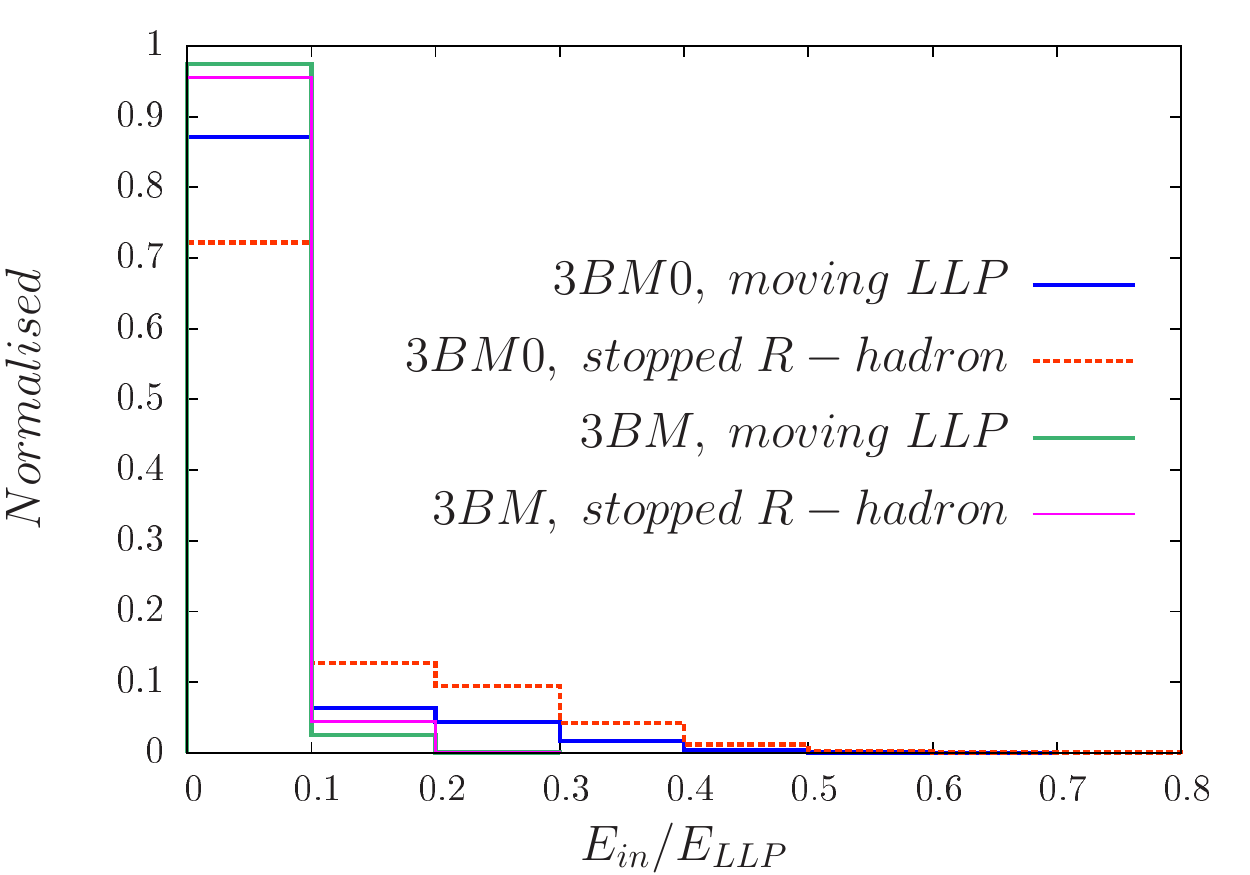}
	\caption{Normalised distribution of $E_{\textrm{in}}/E_{\textrm{LLP}}$; $M_{X}=2$ TeV, $M_{\textrm{DM}}=1.5$ TeV. For the definition of the 2BM/3BM decays, see the text. For the first bin ($E_{\textrm{in}}/E_{\textrm{LLP}}< 0.1$) $E_{\textrm{in}}=0$.}
	\label{fig:LLP_BMO_2}
\end{figure}

We now consider two scenarios to show the energy fraction that traverses back. In the first scenario we consider particles that can decay within the HCAL and traverse towards the tracker. We perform our analysis by assuming a very simple geometry with $L_\text{tracker} = 600$ cm along the $z$-axis, $R_\text{tracker} = 100$ cm and the transverse distance of the last layer of the HCAL to be 300 cm from the $z$-axis. We show the energy fraction $E_\text{in}/E_\text{LLP}$ provided the LLP decays somewhere between 100 cm and 300 cm in the transverse direction, somewhere between the tracker and the HCAL. Figure~\ref{fig:LLP_BMO_2} shows this energy fraction for the 4 categories. For the $2BM0$ and $3BM0$ scenarios, the observations are striking. For the $2BM0$ case, the fraction of energy coming back inside the tracker for the massless two-body decay scenario are 25.9\% for the stopped $R$-hadrons and 12.2\% for the moving LLP. For the $3BM0$ scenario, these numbers respectively become 34.2\% and 14.2\%.

In a similar vein we consider the scenario when a particle decays just outside the muon chamber and comes back inside. For this, we again follow a simplified geometry with $R_\text{muon-chamber} = 750$ cm and $L_\text{muon-chamber} = 1300$ cm along the $z$-direction. The energy fractions, at least for the massless cases, are similar to the one that we discussed for the tracker scenario.

\begin{figure}[!h]
	\centering
	\includegraphics[height=4cm,width=6.95cm]{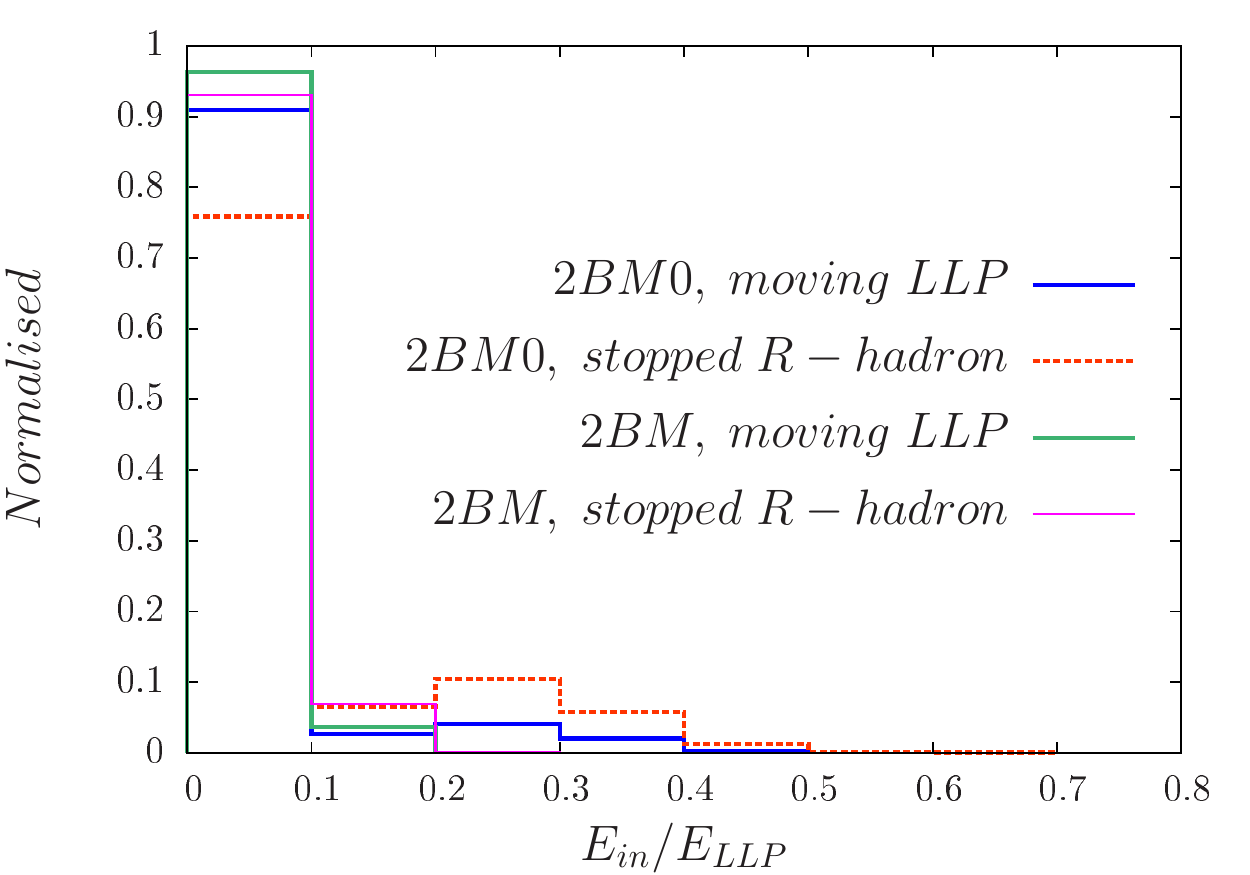}~\includegraphics[height=4cm,width=6.95cm]{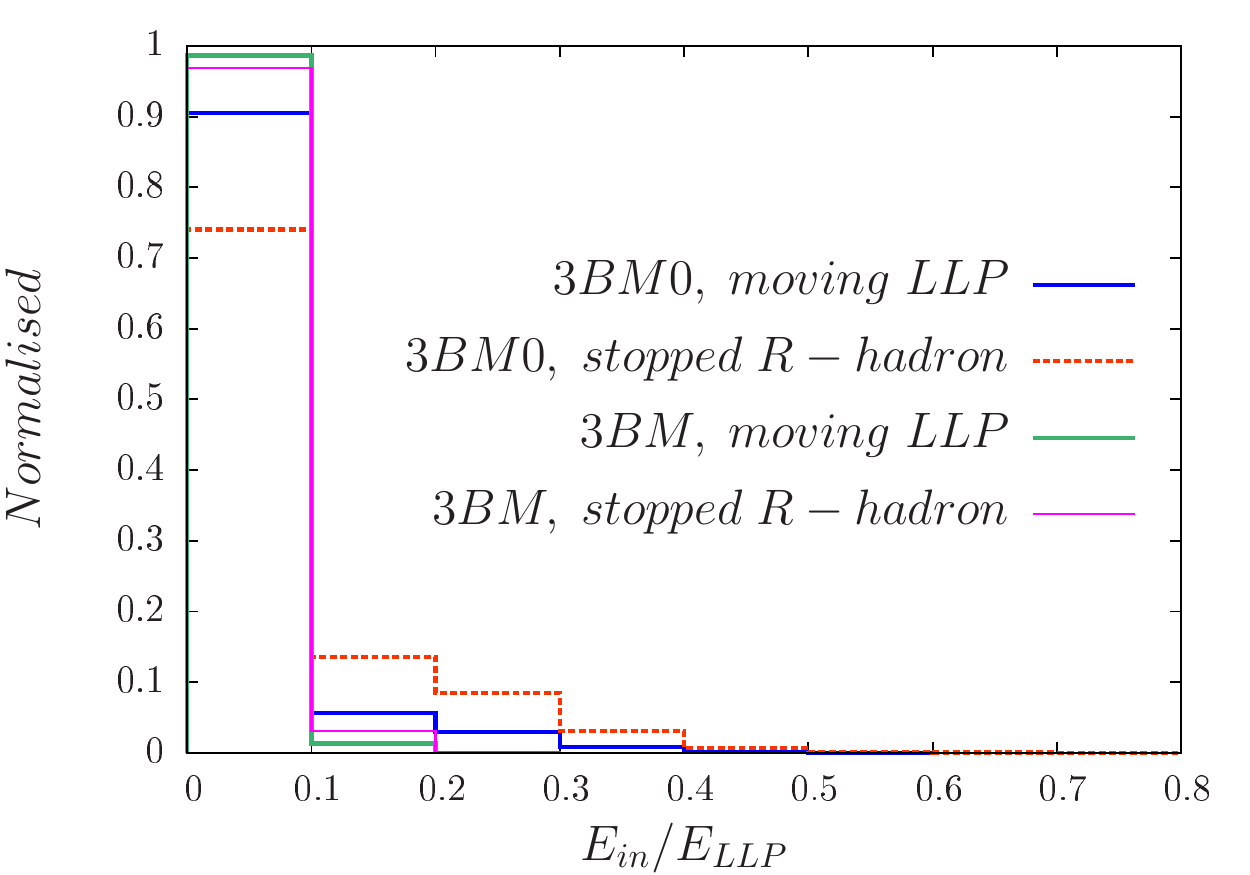}
	\caption{As in Fig.~\ref{fig:LLP_BMO_2} but for the case of the muon chamber.}
	\label{fig:Efac-mu}
\end{figure}

Before concluding this part, we will briefly touch upon the experimental considerations as what we have discussed until now is mostly at the parton-level analysis. We have not discussed the backgrounds but the dominant backgrounds would ensue from cosmic ray events. One way of suppressing such backgrounds is to tag the backward moving objects only in the lower half of the detector. Moreover, there can be backgrounds ensuing from beam-induced noise, overlapping events as well as instrumental noise. Furthermore, shower shapes for the ECAL would be a good identifier as it is expected that the shapes would be different for the inside-out conventional jets versus the outside-in jets. Using ECAL timing information can help in drastically reducing such backgrounds. When the LLP decays in one of the outer layers of the HCAL, we will be more interested in the HCAL shower shape variables. Provided the upgrade of the HCAL has depth segmentation, the energy, $E(D_i)$ deposited in the $i^{th}$ layer of the HCAL, can be used as inputs in a BDT to discriminate between backward moving signal jets and forward moving background jets. The proposed high-granularity calorimeter within CMS might help us address several of these issues. Furthermore, gaining timing information from the muon chamber as well as the upgrades in the tracker (with the inclusion of additional timing layers) will be important for understanding such signatures, better. $BMOs$ which are heavily displaced with respect to the primary vertex can have large impact parameters and will mostly be not recognised by the present jet algorithms. The jet algorithms can be tuned to catch such displaced jets with large impact parameters but this can be extremely resource intensive. Ideas like data scouting and parking can be used to improve this situation. Moreover, reconstructing the $BMOs$ in the tracker can be very challenging and one has to make modifications in the track reconstruction algorithms by relaxing the requirements on the impact parameter.

\begin{figure}
	\centering
	\includegraphics[width=0.49\textwidth]{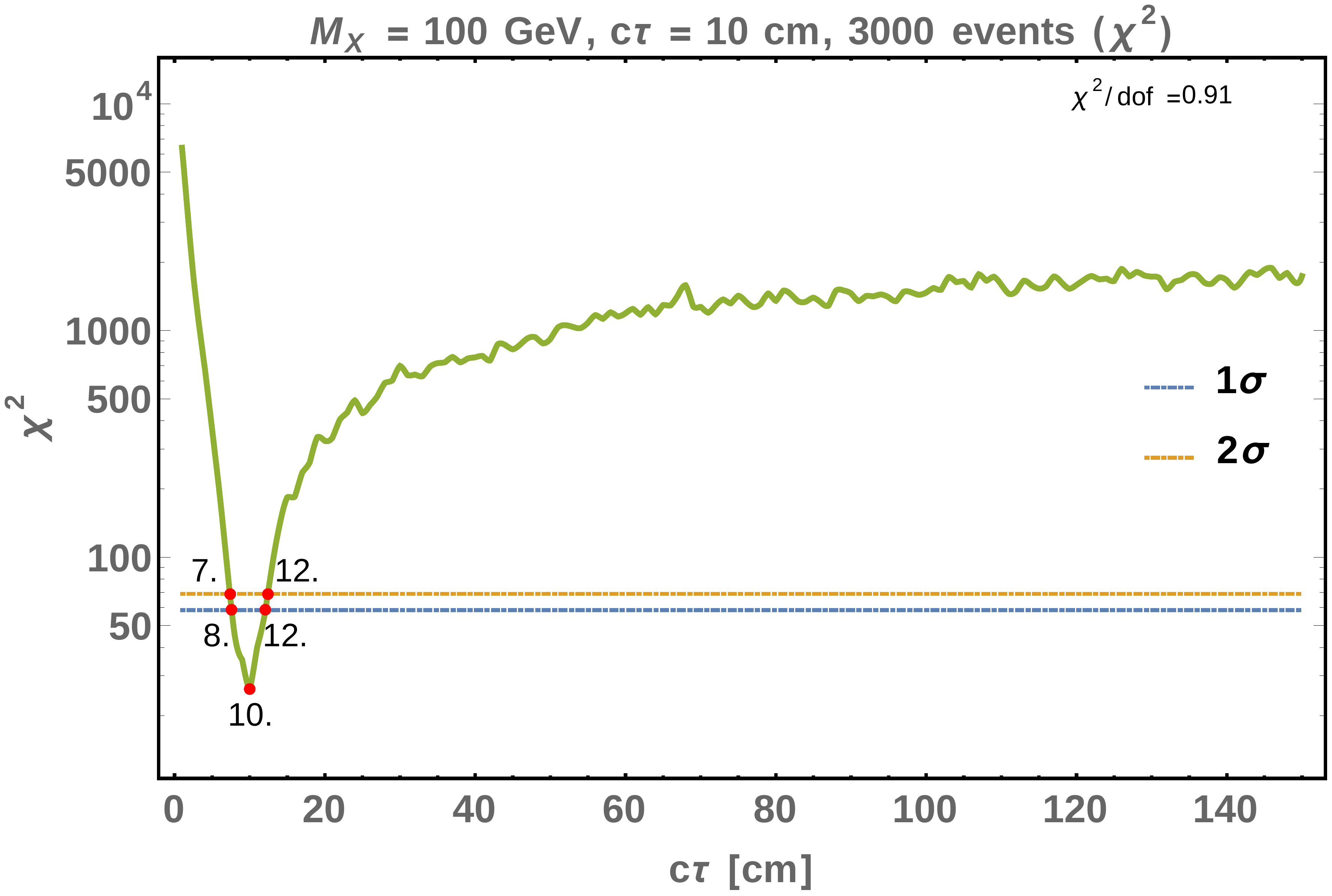}~
	%\label{fig:chi2_M100_DL10_d}
	\includegraphics[width=0.49\textwidth]{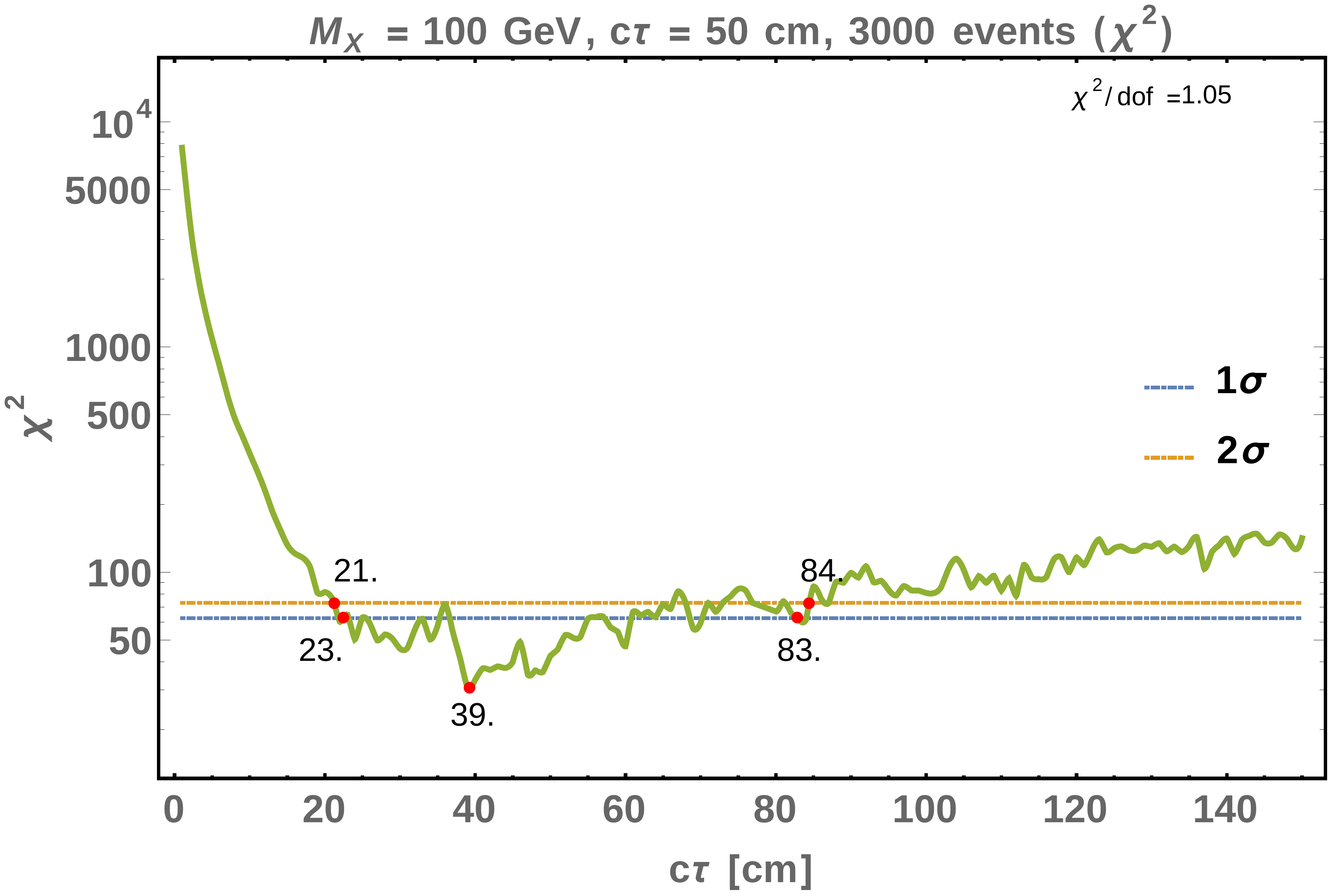} \\[2mm]
	% \label{fig:chi2_M100_DL50_d} 
	\includegraphics[width=0.49\textwidth]{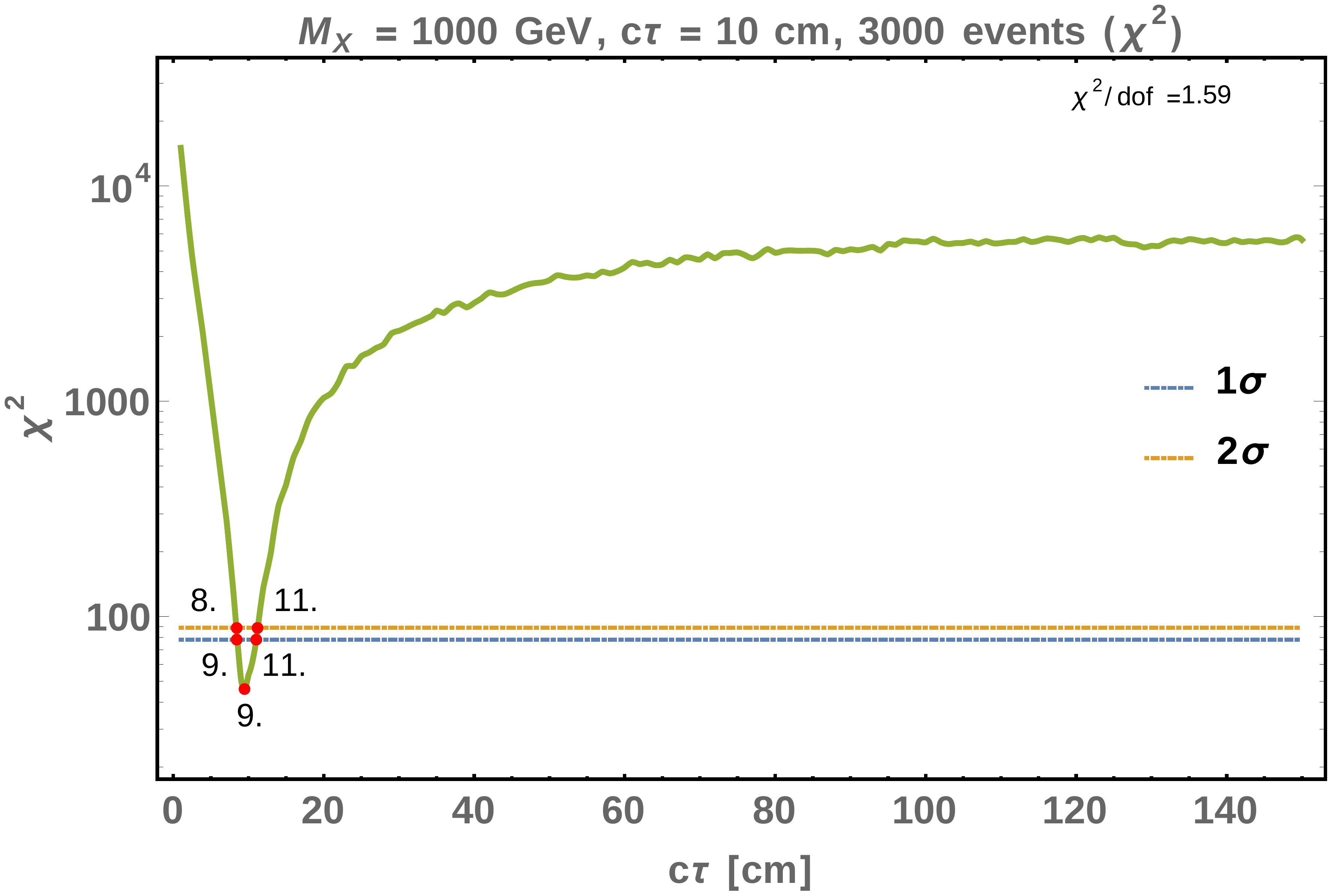}~
	% \label{fig:chi2_M1000_DL10_d} 
	\includegraphics[width=0.49\textwidth]{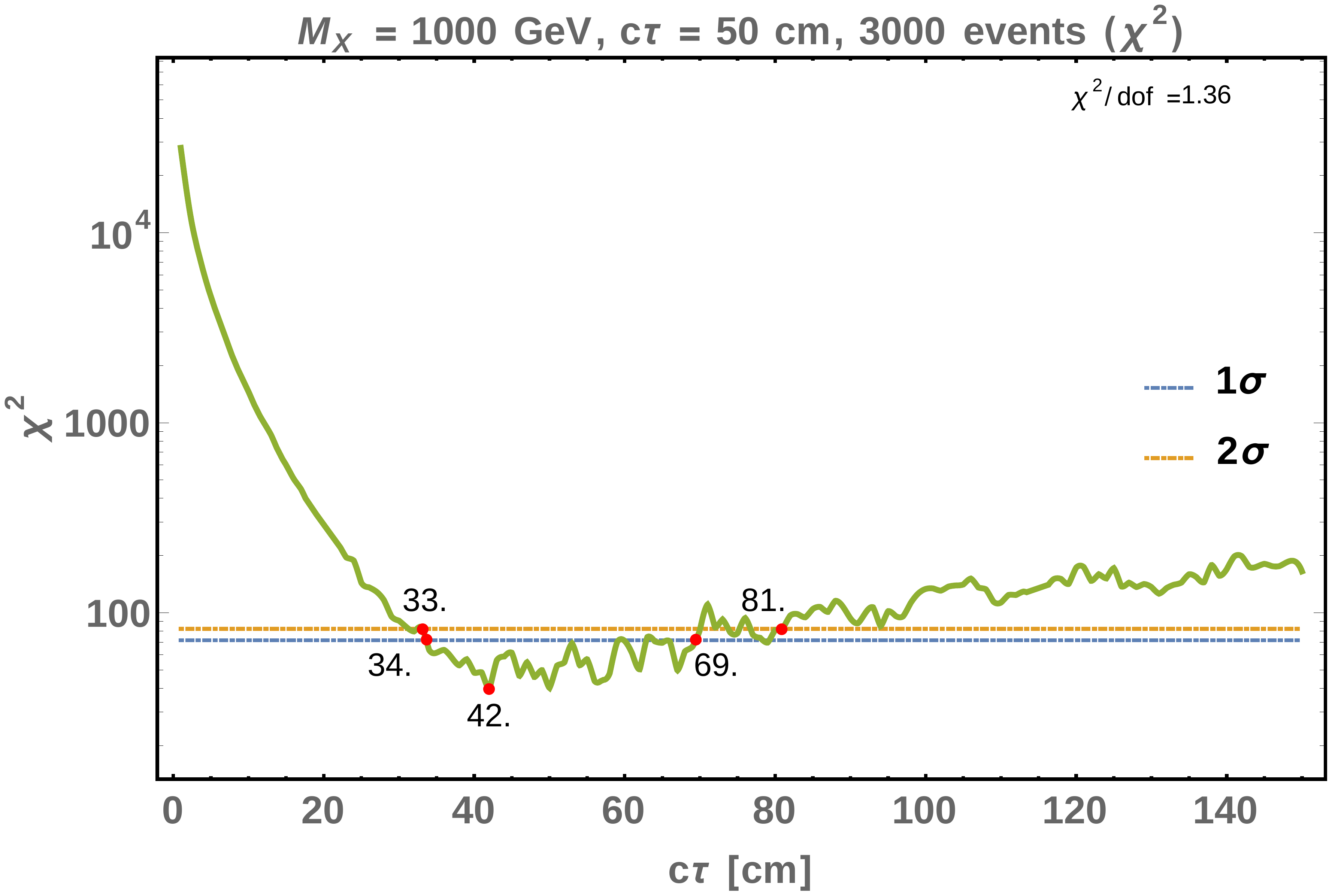}
	%\label{fig:chi2_M1000_DL50_d}
	\caption{Model-dependent $\chi^2$ as a function of the reconstructed decay length $c\tau$.}
	\label{fig:chi2_dep}
\end{figure}

Finally, before concluding this section, we want to refer to a study~\cite{Banerjee:2019ktv} that shows the potential of constraining the proper lifetimes of LLPs provided they are discovered. This study considers the prospects of the high-luminosity runs of the LHC. High pile-up is considered with the various upgrades that are proposed. Model-dependent and model-independent methods are utilised and machine learning algorithms employed to reconstruct the proper lifetimes of neutral LLPs decaying into leptons (may be also accompanied with missing energy). The proper lifetimes of charged LLPs decaying into leptons and missing energy is also considered. Neutral LLPs decaying into displaced jets is discussed along with the challenges faced in high PU environments. As an example, we show the lifetime estimates for the model-dependent displaced lepton category in Table~\ref{tab:chi2_dep} and Fig.~\ref{fig:chi2_dep}.

\begin{table}
	\centering
	\begin{tabular}{|c|c|c||c||c|c||c|c|}
		\hline
		$M_X$ & DL & cross-section & Rec.\ DL & $1\sigma$ LL & $1\sigma$ UL & $2\sigma$ LL & $2\sigma$ UL \\
		(GeV) & (cm) & (fb) & (cm) & (cm) & (cm) & (cm) & (cm) \\
		\hline\hline
		\multirow{3}{*}{100} & \multirow{3}{*}{10} & 1 & 10 & 8 & 12 & 7 & 12\\
		& & 0.1 & 11 & 4 & 23 & 4 & 27\\
		& & 0.05 & 7 & 3 & 30 & 2 & 35\\    
		\hline
		\multirow{3}{*}{100} & \multirow{3}{*}{50} & 1 & 39 & 23 & 83 & 21 & 84\\
		& & 0.1 & - & 9 & $>$150 & 8 & $>$150\\
		& & 0.05 & - & 5 & $>$150 & 4 & $>$150\\
		%100 & 100 cm & & & & & \\
		\hline\hline
		\multirow{3}{*}{1000} & \multirow{3}{*}{10} & 1 & 9 & 9 & 11 & 8 & 11\\
		& & 0.1 & 10 & 7 & 16 & 6 & 17\\
		& & 0.05 & 11 & 6 & 22 & 5 & 25\\    
		\hline
		\multirow{3}{*}{1000} & \multirow{3}{*}{50} & 1 & 42 & 34 & 69 & 33 & 81\\
		& & 0.1 & - & 19 & $>$150 & 17 & $>$150\\
		& & 0.05 & - & 10 & $>$150 & 9 & $>$150\\    
		%100 & 100 cm & & & & & \\
		\hline
	\end{tabular}
	\caption{Lifetime estimates by model-dependent $\chi^2$ fitting of the $d_T$ distribution for the displaced leptons signature.}
	\label{tab:chi2_dep}
\end{table}

\subsection{Summary}
For many known and unknown reasons, BSM physics may be hidden under the current searches at the LHC. There are many notoriously difficult scenarios in some well-known frameworks, such as supersymmetry and hidden-sector models. Known reasons include non-prompt decays of the new particles, very soft particles in the final states, overlapping with existing particles, and/or very small production rates. 

For these difficult scenarios with known reasons, theorists and experimenters have been working together to formulate useful strategies to cover them. New triggers have been put in the trigger system to identify non-prompt or long-lived particles, covering various signatures for the LLPs in various parts of the detector. New techniques including 
machine-learning or initial-state radiation are employed to improve the possibility of detecting such soft particles in the final state and for 
those signals with small production rates.
For those scenarios that are hidden due to unknown reasons or unexplored theories, unconventional ideas such as anomaly detection with machine learning would help.

%\newpage
\section{Open data in particle physics}
{\it\small Section editors: Nishita Desai and Suchita Kulkarni\\
Contributions: Matthew Bellis, Philip Harris, Clemens Lange, Kati Lassila-Perini and Jesse Thaler}

\medskip\noindent
The goal of discovering new physics at the LHC rests on our ability to collect and analyse all interesting collision data.  The detectors at the LHC are intricate machines that require extraordinary levels of technical sophistication.  Although the abstract concept of a ``collision event'' is clear enough to theorists and experimentalists alike, the actual data readout and its translation into a ``reconstructed'' event from which physics may be extracted is extremely complicated and the algorithms employed to do this are perpetually under improvement.  The question of data preservation therefore is also extremely complicated and fraught.  Indeed for nearly all particle physics experiments before now, no one outside of the respective experimental collaborations has had either access or ability to make any use of data collected at the said experiment.  The reliance on computer software to store and analyse the data further means that it is usable only as long as the software is runnable (which unsuprisingly is no longer viable after a couple decades in most cases).  

The High Energy Physics community is historically strongly in support of initiatives to keep all scientific research available to the public at no cost (i.e.\,the cost of publication and the infrastructure to maintain access is borne not by the individuals who perform or access research but by participating educational institutes, governmental or non-governmental organisations).  The SCOAP3 initiative, for example, is a partnership of over three thousand libraries, major funding agencies and research centres in 44 countries and 3 intergovernmental organisations to ensure open access to research published as journal articles.  Nearly all articles published in research journals are also voluntarily and independently uploaded by authors on the arXiv preprint server.  The LHC experimental collaborations have also increasingly made  the data from published papers ---  histograms, figures and tables --- available in human readable and machine readable formats via the HepData portal. In this context the LHC reinterpretation forum and the Les Houches recommendations have been important~\cite{LHCReinterpretationForum:2020xtr,Kraml:2012sg}.  All major software used in scientific computation is made available under a public licence (most commonly the GNU Public licence, but occasionally Creative Commons or similar licences may be used.)  Zenodo~\cite{zenodo} is another sister repository, also maintained by CERN with the aim to provide citable DOIs to preserve research software and data products and in this sense is complementary to HEPData.  The idea that data from publicly funded experiments might still remain behind a curtain is therefore highly unpalatable to most high-energy physicists.  However, the question of what to release as data and how to ensure integrity in its future use remain questions that are under active and vigorous discussion.

As a first step towards open data, CERN launched the Open Data portal in November 2014~\cite{CODP}. At the moment, this portal contains rich content involving collision and simulated datasets for research, derived datasets for education, configuration files and documentation, virtual machines and container images and software tools and analysis examples. As of March 2021, this data contains over 7600 bibliographic records and over 900k files or with 2.4 petabytes of data.

The availability of Open Data has also enabled novel theoretical research.  There are currently over thirty papers citing the open data framework and multiple new studies in the realm of new searches, QCD jet studies, and Machine Learning~\cite{Cesarotti:2019nax, Andrews:2018nwy, Larkoski:2017bvj} have been performed.  At least one such work has then prompted further work by the publishing experiment~\cite{CMS:2017qlm}, completing the experiment to theory and back to experiment cycle and providing evidence to the claim that Open Data would result in genuine scientific advance.

\subsection{Challenges}
Open data efforts at the LHC need to overcome several technical and philosophical challenges. Some of these are related to the fact that only a subset data is stored and hence the lost data can never be recovered. Data collected at the LHC is also released with some delay allowing collaborations to exploit first. Challenges relating to preserving technology used to process this data by collaborations, data documentation and necessity for data validation mechanisms also need to be considered.

\subsubsection{The problem with triggers}
What shall be deemed an ``interesting'' collision event at the LHC is determined based on our current understanding of the Standard Model.   The unprecedented number of proton collisions per second  --- or luminosity --- means that not every ``event'' can be stored to the disk.  The experiments therefore choose what to store using certain criteria called {\it triggers}.  These triggers --- combining hardware level and software level decision making --- aim to select events that either have large enough energy signatures (e.g. by requiring high-momentum particles) or unusual combination of detectors firing at the same time, with the assumption that anything that ``happens'' would be caught by one of these.  The first of these strategies has been honed over time.  Over the last few years, considerable work has also been done to improve triggers to include unusual events with long-lived particles or radiation from hidden sectors.  This second kind of triggering, however, necessarily needs physicists to know beforehand what kind of non-standard signature is expected.  Therefore, it is obvious that  only theories that the field already knows are visible in standard triggers, or are popular enough to have a trigger designed for that specific signature.

\subsubsection{The role of the collaborations}
Experimental physics collaborations currently have not only the privilege of data access but also the responsibility of ensuring the accuracy of their interpretation.  Every result announced by an experimental collaboration is painstakingly cross checked by several independent internal groups who each use their own algorithms and strategies to ensure that the final calibration and statistical analysis is accurate.  

Once data is publicly released, this quality control is out of the hands of the experimental collaborations and therefore it would be impossible for the experiments to ensure the quality of results claimed based on their data. This has for a long time been the primary reason cited to avoid releasing data publicly.   
Furthermore, the resulting publicity from a spectacular claim often proceeds under its own steam and fraudulent or over-enthusiastic claims that are then proven to be false would then result in erosion of public trust in the scientific process and further endanger funding of future fundamental physics experiments which clearly cannot be afforded by any single university, or even country. It is important to point out that this concern is largely acknowledged by both experimental and theory community. The theory community is hence especially careful in using open data.

\subsubsection{Technical challenges}
First, the experiments need to decide whether they intend to release raw or processed data, both of these choices would come with their own issues. In general, if we wish to have a reliable and usable open data framework, much thought and work is needed in at least three directions:
\begin{itemize}
	\item Reliable storage and access technology.  Preservation of software used for data analysis and hardware capable of running said software.
	\item Detailed meta-data and documentation: \begin{itemize}
		\item In case of raw data, all relevant calibration information, preservation of associated algorithms etc.\ to obtain the reconstructed events
		\item In case of releasing reconstructed events only, extensive documentation and internal information characterising the reconstruction and explaining limitations in its usability
	\end{itemize} 
	\item Mechanisms for validation of results. How to ensure sanity of results derived using this data?  Educating the public on what is ``real'' and what fake.
\end{itemize}

\subsubsection{The ultimate aim}
When the signatures of new physics have been captured and saved, long-term access to data by the scientific community could use it to test all possible theories.  The case for an Open Data platform is therefore motivated not just by arguments of democratic access to data obtained by a publicly funded experiment, but makes solid scientific sense. 

Such open data policy has long existed in the field of astrophysics, where e.g.\,data from publicly funded telescopes is released regularly.  The main challenge in doing this for particle physics experiments comes from the complexity in triggering and calibrating the data.

\subsection{Existing framework}
Experimental data being openly available is a well established philosophy in HEP. However, making large amounts of data available is not the same as it being useful. This scientific data management and stewardship is taken care of by the FAIR principles, which are also embraced by CERN open data. Moreover, keeping in mind the needs of different users, data are made available in different levels via different platforms as well. We will briefly review implementation of FAIR principles in CERN open data portal and associated efforts to make data more user-friendly.

\subsubsection{The FAIR principles}
This independent effort is the articulation of principles needed to publish and maintain the quality of data released.  This is summarised by the FAIR~\cite{FAIR} guiding principles for scientific data management, which emphasises that all data published should be \ul{F}indable, \ul{A}ccessible, \ul{I}nteroperable, and \ul{R}eusable.  The CERN Open Data effort aligns with the FAIR principle by ensuring: \begin{itemize}
	\item Findability: Assigning a “record ID” and optionally a DOI to every data product.  Rich context description and associated documentation is provided.  All histograms and numerical data are required to be machine readable.  A search interface is provided for searching and identifying correct datasets.
	
	\item Accessibility: A graphical user interface for manually downloading the data and an automated cern opendata client is available.  This command line client supports downloading the dataset and metadata via HTTP and XRootD access protocols.
	
	\item Interoperability: CERN open data portal offers several data formats and vocabularies as a community standard. In addition, data is also offered under common classification rather than formal vocabularies. The latter assists in physics interpretation, while the former is applicable in designing appropriate data processing chains. In addition, the data variables are equipped with detailed semantics description. This helps identifying the variables which are of utmost importance for analysis design. For some examples of analysis development based on the principle of interoperability see Refs.~\cite{Wunsch:dimuon, Wunsch:higgs}.
	
	\item Reusability: A detailed record of data provenance i.e.\ how the data was generated is available in JSON format.  Instructions for processing of both RAW and reconstructed datasets (e.g. AOD format used by CMS) are provided and lastly, computing environment is preserved in the form of docker and singularity images. 
\end{itemize}

\subsubsection{Available data from the LHC collaborations}
The CERN Open data portal currently catalogues data from six experiments ---  ATLAS, ALICE, CMS, LHCb,  OPERA and PHENIX.  Focusing on the LHC experiments, of the four, CMS has the largest repository of research-level data available (currently at over 1.1k datasets).  ATLAS currently only has outreach-level data.  ALICE has released 15 datasets and documentation from 2.76\,TeV and 7\,TeV runs.  Whereas LHCb has 4 datasets of very limited events to be used with some published analysis software.  Several data releases for research-level data are available. The latest release in this series was in December 2020 and contains data from the CMS experiment including first 2010–11 heavy-ion data samples and reference proton-proton datasets (214 TB).  We will discuss the CMS Open data format in detail in the coming sections, however, we outline general principles here.

\subsubsection{CERN open data policy}
The CERN open data policy recommends data releases at four different levels:
\begin{itemize}
	\item {\bf Level 1} consists of results which are released in the form of publications. Access to these publications is enabled under the SCOAP3 agreement for high energy physics.
	\item {\bf Level 2} is aimed at outreach and education and access is enabled via CERN open data portal.
	\item {\bf Level 3} consists of reconstructed data which will be useful for reproduction of physics analysis as well as for new physics analyses to be designed later if needed.  
	\item {\bf Level 4} contains raw collision data which may not be suitable for external consumption as  experiment-specific knowledge of calibration or resolution may be needed to correctly interpret this data.
\end{itemize}
In addition, all data published by CERN experiments will try to abide by FAIR requirements.  However, aside from publicly available data, some amount of internal or ``restricted" analysis knowledge needs to be preserved. This is also done in accordance of the FAIR principles, on the CERN Analysis Preservation portal~\cite{restricted}. It should be remembered that FAIR principles do not necessarily require open access.  

\subsubsection{CERN Open Data Portal}
\label{sssec:opendataportal}
The CERN Open Data Portal\footnote{\url{https://opendata.cern.ch/}} is the gateway to accessing the CMS open data. For CMS open data users, it provides access to the data themselves, the software needed access these data, and a number of examples and mini-tutorials on how to use these tools both for analysis and education. It also has an incredibly detailed amount of information about the provenance of these datasets, both collision data and Monte Carlo. In a sense, the ``documentation" provided is complete in the same way a dictionary is complete. But while a dictionary will help you define any word you run across, it is perhaps not the best resource if you want to compose a sonnet in the style of Shakespeare. For a more concrete example, the CMS Open Data Guide will show a user how to access the data and explore the different types of ``physics objects", but it does not give much explanation on these objects or why there might be multiple definitions of a "Muon". 

As described in Sec.~\ref{sec:cms:external}, in a 2017 paper from Tripathee et al.~\cite{Tripathee:2017ybi}, the non-CMS-member authors used CMS open data to explore jet substructure. That paper contains an appendix, ``Advice to the community", in which the authors detailed their experience with the open data, both good and bad, and provided advice for improving access for other users. While the authors were able to produce new scientific measurements with these data, it was not without challenges. The success of their group motivated the CMS open data team to provide a better experience for open data users by significantly improving the documentation. 

It should be pointed out that this did {\it not} mean getting rid of the information on the Portal. That information is complete and necessary for users who want the details of what triggers were used for collision datasets or the details of generator parameters for the Monte Carlo. Instead, the decision was made to develop a {\it guide} for new users and to organise a {\it workshop} to provide a hands-on experience for these users.

\subsubsection{REANA and HEPdata}
In the context of open access to analysis information, an excellent resource has also been developed in the form of the REANA framework~\cite{reana}. This is a platform for reproducible analysis and aims at providing an integrated access to the data, computing environment, and recipes. The framework can be deployed by using containerised workflows on Kubernetes, HTCondor, Slurm back-ends.  It is possible to process both CMS and ATLAS data via the REANA framework.

Another very notable resource is the HEPData repository. HEPData~\cite{hepdata} hosts publication level data. It also provides an interactive interface to explore and download publication-level data behind plots and tables. So far HEPData hosted only results in the form of plots, however now it has started to also host likelihood information, which is a very welcome step towards creating a central place for repository of a publication level data. Along with HEPData,  it also offers a bridge to GitHub and hence preserves older software releases. Zenodo is widely used by the machine learning community among other users at the LHC. Finally, some amount of internal or “restricted” analysis knowledge still needs to be preserved. This is also done in accordance of the FAIR principles, on this portal~\cite{restricted}. It should be remembered that FAIR principles do not incorporate open access

\subsection{CMS open data}\label{sec:cms}
The CMS experiment established a data preservation, re-use and open access policy in 2012~\cite{cmspolicy} and started regular releases of research-quality data in 2014. All proton-proton data collected during 2010 -- 2011 and a half of those from 2012 are now in the public domain, and the latest release in December 2020 includes heavy-ion data from 2010--2011. This is a substantial amount of data, resulting, together with associated artifacts, in a total volume of 2.3 PB.

\subsubsection{CMS data releases}\label{sec:cms:releases}
The CMS data releases take place regularly, after an embargo period of six years following data taking. The releases include 50\% of the collision data and the corresponding simulated datasets. The collision data release is completed within 10 years. However, the amount of open  data will be limited to 20\% of data with the similar centre-of-mass energy and collision type while such data are still planned to be taken.

CMS releases a full reprocessing of data at Level 3 (reconstructed data good for physics analysis) from each data-taking period and the data are released in the same format and with the same data quality requirements from which analyses of the CMS collaboration start. For the Run 1 data (data taking 2010 --2012) the format is the Analysis Object Data (AOD) format, based on the ROOT framework~\cite{root} and processed through the CMS software CMSSW~\cite{cmssw}. This format contains reconstructed “physics objects” such as electrons, muons, jets, and their properties, and keeps the most relevant lower-level information such as hits in the tracking system and calorimeter clusters corresponding to the physics objects.

The CMS open data follow FAIR principles and are provided with rich associated metadata.  Due to the complexity of experimental particle physics data, the FAIR principles alone do not guarantee the re-usability of these data, and additional effort is needed to pass the knowledge needed to interpret them. The interplay between the CMS experiment open data team and the CMS open data users is of utmost importance in this context. 

\begin{figure}
	\centering
	\includegraphics[width=\textwidth]{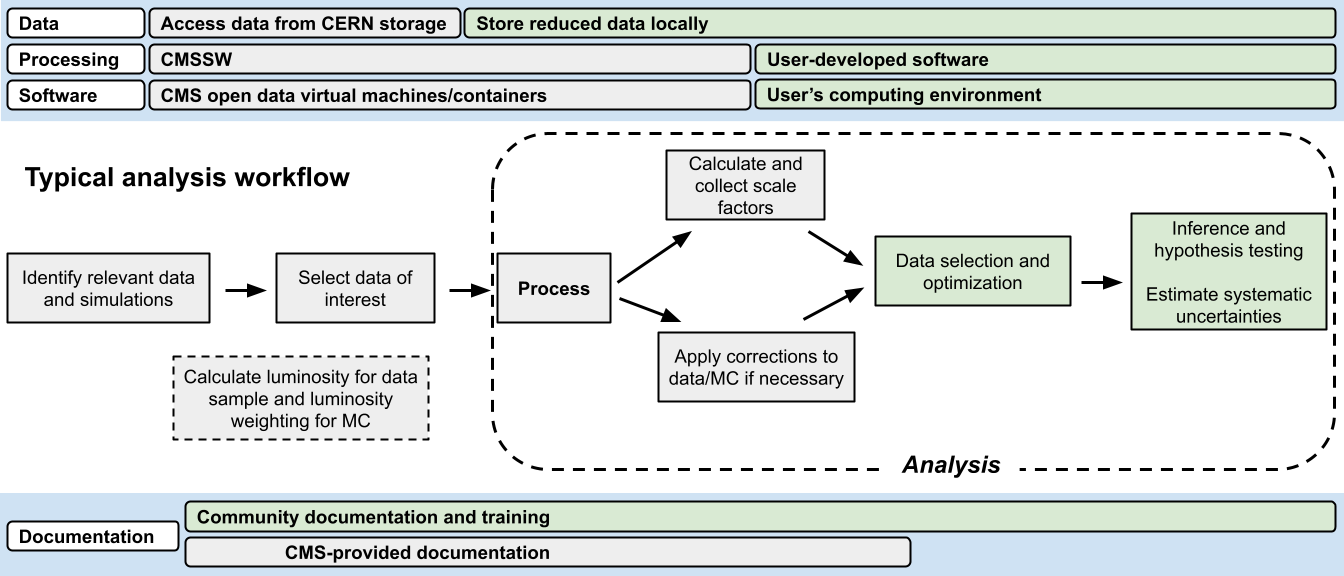}
	\caption{A coarse flowchart for a typical analysis taken from Ref.~\cite{LassilaPerini2021}. The upper part of the figure shows the hardware and software that an analyst might use for different stages of the analysis and the lower part of the figure shows where CMS’s documentation efforts span and end. Grey boxes indicate procedures, software, or hardware and storage are provided by the CERN or the CMS open data group. Light green boxes indicate steps where the procedures, software, or hardware and storage is left to the individual analyst.}
	\label{fig:flowchart}
\end{figure}

Figure~\ref{fig:flowchart} provides a simplified flowchart of an analysis, as well as what
hardware, software, and documentation resources are needed/provided for the CMS open data. Grey boxes indicate the resources that are provided by CERN open data portal and CMS open data team. Light green boxes indicate steps where the procedures, software, or hardware and storage is left to the CMS open data user.

The  recommendations from users of CMS open data who are
external to the CMS collaboration (see Section \ref{sec:cms:external}) are valuable
%The following section gives recommendations from users of CMS open data who are external to the CMS collaboration. Their experience and feedback is valuable 
to the CMS open data team, and acts a guideline for the future directions, in the limits of possible. The challenges and measures taken to address them by the CMS open data team are discussed in Secs.~\ref{sec:cms:complexity}--\ref{sec:cms:usability}.

\subsubsection{Data complexity}
\label{sec:cms:complexity}
The CMS data are released in the format that allows their widest possible use, without special preparation for their use in the public domain. This is also the format from which the collaboration members start the analysis. However, as underlined in the preceding section, the complex data format is a challenge for open data users.

The complexity originates from several factors. First, there is no single definition of a physics object, there can be multiple instances of an object, such as jets defined with different algorithms, and the object type is chosen to match the requirements of the analysis. Furthermore, a signal can be interpreted as several different object candidates, for example, a signal interpreted as a muon can also appear in the list of electron candidates. Therefore, physics object identification criteria are always applied in the data analysis, and the criteria are adapted balancing between the efficiency of selecting an object and the purity of the selected object, and depend on the analysis being carried out.

Final corrections and fine-tuning to the objects are often applied in the analysis phase. This happens because corrections and algorithms are developed at the same time as the first analysis of the data, and do not necessarily make their way to the final reprocessed data. Some of these corrections are available from the condition database, and some others as separate ``recipes'' to be applied to the objects.

Another challenge are the triggers and procedures related to data taking. A single dataset consists of events passing one or more of hundreds of different trigger paths. In some cases triggers are prescaled so that only a predefined fraction of events passing the trigger selection is recorded, to enable collecting data for processes that occur often and otherwise would fill the entire data-taking bandwidth. One event may also end up in two different datasets, and this eventual overlap needs to be accounted for in the analysis. All this needs to be handled in the data analysis to properly scale the number of selected events with the cross-sections being measured.

The use of CMS data also requires some knowledge of the CMS experiment software CMSSW, built on top of the HEP-specific ROOT data structure. The software is openly available, and provided as software containers to open data users. The efforts for facilitating its use are discussed in Sec.~\ref{sec:cms:examples}.

Part of these difficulties will be overcome when the slimmer miniAOD and nanoAOD formats will be made available. These formats are both based on the ROOT data structure, but the nanoAOD format does not require using CMSSW software, and will therefore be of particular interest to CMS open data users. These formats are routinely produced for Run~2 data, the nanoAOD format starting from the data taking of 2016. However, this slimming has a price and it is estimated that roughly half of the analyses done in CMS can be done using nanoAOD. Therefore, miniAOD format will also be made available.

\subsubsection{Examples and documentation}
\label{sec:cms:examples}
At the time of publication of this document, users of the CMS Open Data have had three options for learning about how to access and analyse these data: the information provided on the CERN Open Data Portal, CMS Open Data Guide, and the (to-date) two workshops run by the CMS open data team. The Open Data Portal is the original source of documentation, but suffers from a lack of any sort of ``roadmap" for new users, cf.\ Sec.~\ref{sssec:opendataportal}. The CMS Open Data Guide is {\it very} much a work-in-progress at the time of this document (Summer 2021). The workshops have been very successful and have provided person-power and feedback to improve the documentation, in addition to being a pinned source of information.

\paragraph{CMS Open Data Guide:}
Within CMS, one of the most widely used onboarding tools is the {\bf CMS Offline Workbook},\footnote{\url{https://twiki.cern.ch/twiki/bin/view/CMSPublic/WorkBook}} a series of wiki pages that walk users through all the software requirements of becoming a CMS member. From getting your computing accounts to how to start data analysis to generating your own Monte Carlo data. Much of it is public, but it also links to internal pages that require CMS membership to access. It also tends to be focused on the latest data releases, and not older datasets like are released on the open data portal, though the documentation is still there. 

The Workbook is laid out like a roadmap, walking the user from the first steps of finding and accessing data to more complicated procedures like jet energy corrections. It is a great resource for CMS users and the CMS open data team is currently working on an analogous site for the open data, to be referred to as the {\it CMS Open Data Guide}. It is very much a work-in-progress but the overall structure is there, and it mirrors the Workbook in that it tries to provide a starting point for a brand new user, eager to work with the open data. 

It would be a waste of person-power to try and completely rewrite the documentation that exists on the Open Data Portal and it would be near impossible to keep them in sync, should things change. Instead, the goal is to have the Open Data Guide {\it link to} documentation on the Portal that already exists, while providing contextualizing information and guidance on how that documentation can be used. What additional information is necessary is informed by the paper from Thaler~et~al.~\cite{Tripathee:2017ybi} and feedback from participants in the 2020 and 2021 workshops.

\paragraph{CMS Open Data Workshops:}
The CMS open data group has conducted two workshops in 2020 and 2021: ``{\it CMS Open Data Workshop for Theorists at the LPC}"\footnote{\url{https://indico.cern.ch/event/882586/}} and ``{\it CMS Open Data Workshop}",\footnote{\url{https://indico.cern.ch/event/1031398/}} which ran for 3 and 4 days, respectively. There were about 25 engaged participants in the first and about 50 in the second. Because of the pandemic, both were held virtually and attracted scientists from all over the world. These workshops built on the model of a hands-on workshop where participants were required to actually {\it do} the exercises and run the code themselves. The group opted to build the workshop lessons on a framework developed by the Software Carpentry\footnote{\url{https://software-carpentry.org/}} organisation, a framework also in use by other CERN workshops. 

The workshops followed a similar format on a coarse level. Participants were shown how to find the data of interest on the Portal. Lectures and examples were given about to locate and apply different triggers and how to access different physics objects, as well as what those objects meant. Examples were also given of how to apply different energy corrections (e.g. jets) and how to use those in systematic uncertainty calculations. Both workshops ended with a hands-on example of how users could leverage Google Cloud computing to process the open data to scale and store the skimmed data on Google's Cloud infrastructure. 

Time was also allotted to gather feedback from the participants on their experience with the workshop and their interest in using the open data, and a survey was sent out after the workshop to supplement this real-time feedback. This feedback has been and will be used to improve the next versions of these workshops as well as to inform the development of the Open Data Guide. For example, some pages of the Guide might point to the lessons from the Workshops, all of which are accessible on the web still along with recordings of the lectures. 

\subsubsection{Usability}
\label{sec:cms:usability}
There are several challenges when it comes to the usability of open data.
The complexity of the data on its own as discussed in Sec.~\ref{sec:cms:complexity} often makes it difficult to get started in the first place.
In addition, the original data sets are huge and require substantial computing power for processing.
Derived and therefore simplified data sets make it easier to obtain a first meaningful result.
The examples, documentation, and workshops discussed in Secs.~\ref{sssec:opendataportal} and~\ref{sec:cms:examples} lower the barrier of entry.
However, the examples provided are far from a realistic and complete physics analysis.

In order to e.g.\ generate new Monte Carlo simulation samples to test a new physics model, change the underlying reconstruction to evaluate new reconstruction methods, or assess systematic uncertainties one needs to use the experimental software, i.e.\ CMSSW in the case of CMS, and for the Run 1 data the AOD format.
For a few events, this can be done on a local computer, but the data set required for an analysis including simulation samples typically consist of millions to billions of events.
In order to be able to process these, one needs to have access to high-throughput computing resources.
In addition, these need to provide the possibility to execute CMSSW.
The use of software containers make this significantly easier as long as the underlying processor architecture is the same.
Thus, CMS open data releases are accompanied by software container images that contain the respective CMSSW release needed for processing the data.

Besides using these software images for Monte Carlo event generation, simulation and reconstruction, they can be used to extract desired event information and convert it into CMSSW-independent data formats as discussed in Sec.~\ref{sec:cms:external}.
Furthermore, by extending the existing software, new algorithms can be developed and integrated.
The updated software can be added to the container image and shared via container registries such as Docker Hub~\cite{dockerhub}.

There are currently no computing resources provided by the CMS Collaboration for the use of CMS open data by non-CMS members.
These therefore need to be provided by the open data users themselves, e.g.\ at their institution or by using commercial public cloud computing services.
To help open data users make use of public cloud compute resources, tutorials for ``CMS open data in the cloud'' have been conducted at the CMS open data workshops, cf.\ Sec.~\ref{sec:cms:examples}.

Since individual data and simulation events are independent of each other, processing of the data sets can be parallelised.
A typical analysis workflow therefore splits or scatters the files belonging to a data set into smaller chunks, which are then merged or gathered after the processing step.
The input data can hereby be streamed directly from the open data servers or first copied to local disks.
Since public cloud computing services often provide the possibility to quickly scale up computing resources and then scale them down immediately after processing, this scatter-gather step can be sped up significantly.
These steps can be implemented as a simple workflow, for example using Argo Workflows~\cite{argo}, for which examples are provided at the CMS open data workshops.
After this first processing step, the results can usually be analysed further on smaller computing clusters or even local computers.

The use of software containers has also been instrumental for the development and validation of CMS open data workflows in general since they provide a possibility to use CMSSW independently of the standard high-energy physics computing environments as well as in continuous integration systems.
The latter are particularly important to ensure that the open data remain usable.
However, while in particular Docker has made it significantly easier to use software containers, there are still some technical hurdles that need to be overcome, which often depend on the operating system used.
Feedback from the CMS open data workshops was very useful to further improve documentation and usability.

\subsection{Recommendations from External Users}
\label{sec:cms:external}
In the first application of CMS Open Data, based on the release of 2010 data, the authors of Ref.~\cite{Tripathee:2017ybi} highlighted a number of challenges and recommendations for the community.
Some of these issues have been resolved by subsequent CMS Open Data releases, such as the inclusion of simulated Monte Carlo data sets.
Other issues are a challenge not only for external users, but also for internal CMS application, such as the lack of centralised documentation.

\subsubsection{Missing information}
Because the CMS Open Data is stored in the AOD format, it is in principle possible to reproduce any CMS analysis that does not rely on lower-level information.
In practice, though, there are numerous technical challenges to outsiders using the AOD format as well as important knowledge about the CMS experiment that is not fully archived.
Despite these challenges, there have been a number of successful uses of the CMS Open Data in the literature, cf.\ Refs.~\cite{Larkoski:2017bvj,Tripathee:2017ybi, Madrazo:2017qgh, Andrews:2018nwy, Bhaduri:2019dfa, Cesarotti:2019nax, Andrews:2019faz, Lester:2019bso, Facini:2019rgg, Apyan:2019ybx, Komiske:2019jim, Moreno:2019neq, Bhaduri:2019zkd, Felser:2020mka, Bhaduri:2020yfn, Pata:2021sud, Andrews:2021ejw}.

\subsubsection{Validation samples}
The first recommendation is to provide reference validation examples that includes all steps of published CMS analyses.
For example, the authors of Ref.~\cite{Cesarotti:2019nax} attempted to repeat the measurement of the $Z$ boson cross-section to validate their treatment of the di-muon final state; a reference $Z$ boson measurement from CMS would have been helpful in this context.
Progress towards establishing some benchmarks were made in Ref.~\cite{Apyan:2019ybx}, though the code for that study is not yet public.

\subsubsection{Industry standard file format}
The second recommendation is to release data in an industry standard file format.
The current pipeline involves running the CMS software framework on a virtual machine, which can become unwieldy for analyses that require a large number of events or need to run in a cluster/cloud environment.
The authors of Ref.~\cite{Komiske:2019jim} translated a subset of the information from select AOD files into the standard HDF5 format~\cite{hdf5} and posted them on Zenodo~\cite{MOD:ZenodoCMS, MOD:ZenodoMC170, MOD:ZenodoMC300, MOD:ZenodoMC470, MOD:ZenodoMC600, MOD:ZenodoMC800, MOD:ZenodoMC1000, MOD:ZenodoMC1400, MOD:ZenodoMC1800} along with example code~\cite{EnergyFlow,MODDemo,MODRepo}.
This format allows the use of external analysis software, and it provides a benchmark data sample for the jet physics community.
This issue may eventually be resolved when CMS releases data in the nanoAOD format.

\subsection{Summary}
\subsubsection{Current status}
The particle physics community has spent much time and effort to articulate and design principles and practices to enable publication of data at various levels of openness.  The SCOAP3 agreement provides platforms in terms of journals, books and a document repository that ensures all material published under this agreement will be available as open access, without cost to individuals who would wish to access it.  Open access to publication-level material however does not automatically mean open access to data and this has to be negotiated separately.  Although releasing data publicly has long been the norm in fields like astrophysics, it is not common in particle physics due to the extremely complex nature of data and the extraordinary amount of processing needed to bring it to a level from which physics analyses can be done.  Two main efforts in this regard are the publication of FAIR principles for scientific data and the CERN Open Data project.   All CERN experiments also encourage that published analyses are accompanied by digitised tables and plots uploaded to the HepData repository.   The implementation of this last requirement is still patchy, however compliance is improving.

The premise and usage of open data is multi-fold, therefore data releases correspond to four different levels ranging from publication-level results up to raw collision data.  The two intermediate levels, one that allows for education and outreach, and one that enables real data preservation and re-analysis, are the levels that require the most thought and sophistication in implementation.  

It is not enough just to have open data available, it is important that it is findable, accessible, interoperable and reusable. This is ensured by means of the FAIR principles on basis of which open data portals are designed.  Currently, we find only the CMS experiment has research-level data available on the CERN OpenData portal.

\subsubsection{Future improvements}
Despite advances of open data in particle physics, its usage remains challenging for any person who does not have training in the vocabulary used by the experimental collaboration.  The data format, the methods for selecting and downloading datasets and the preprocessing needed are quite opaque and require extensive documentation to be made usable.  Publication of step-by-step validation of analyses and using industry-standard file formats (as opposed to home-grown ones) are two suggestions we find from current non-CMS users of CMS Open Data. The availability of simulated datasets (not just collision data) would also help in disambiguation of complicated processes.  
It is also worth noting that apart from CMS, data from no other experimental collaboration is currently at a level that can be used for physics studies.  It will be much easier to have a fruitful discussion about improvements once a few more implementations of the open data principles become available that can provide illustrations of best practices (or lack thereof).

Another important factor that hampers usability, at least for theorists, is the non-availability of computing resources.  Given the size of the data-files and associated frameworks, it is nearly impossible to run it on a laptop or a personal computing machine. Access to high-performance computing clusters necessary to process data is expensive and currently limits accessibility to individuals who belong to universities or institutes that already have some provision for computing.  Simplifying access to powerful computing machinery will improve open data usage beyond experiments and a handful of theory groups. 
Therefore, while open data signals the beginning of a very important journey towards open science, there is still some progress to be made before it is widely used and exploited.

\section{Summary}
The field of particle physics is at the crossroads. The discovery of a Higgs-like boson, a major accomplishment for the field almost fifty years in the making, completes the predictions for fundamental particles and interactions in the SM.
At present no clear guidance is available as to how it will break down.
On the other hand, the motivation for the existence of New Physics, given by, for example, the established existence of Dark Matter and Neutrino Oscillations, has not diminished since the discovery of the Higgs-like boson. 

The immense and far-reaching scope of the LHC physics programme promises to ultimately unveil the New Physics.
The exploration of the phase-space by the LHC experiments is far from exhausted even with the available data sets, where Run 3 and the HL-LHC promise to deliver up to 4\,ab$^{-1}$ of integrated luminosity. 
Unfortunately, the exploration of the LHC data by way of inclusive or model dependent searches performed to date indicates that no striking resonances have been observed in the accessible dynamic range. 
In this paper, we posit that New Physics is accessible at the LHC in principle, but that it is inaccessible by current analysis strategies for a variety of reasons, and that its signatures are hidden in the data. 

We summarise the status of the most significant anomalous experimental results in particle physics, including the most recent results for the flavour anomalies, the multi-lepton anomalies at the LHC, the Higgs-like excess at around 96\,GeV, and anomalies in neutrino physics, astrophysics, cosmology, and ultra-high energy cosmic rays.
The anomalies corroborate the need for extensions of the SM, and we provide overviews over possible BSM models.
The fact that many anomalies can be explained within the same theoretical framework is pivotal, as it stimulates model building, which already gave rise to a plethora of BSM models and classes of models. 

The known systemic shortcomings of the LHC and its search strategies allowed us to identify some of these reasons, including: final states consisting in soft particles only, associated production processes, QCD-like final states, close-by SM resonances, and SUSY scenarios where no missing energy is produced.
We find that new strategies are necessary to unveil the hidden NP signatures, which have to strike a careful balance between the model-centric and less model-dependent approaches.
It is generally understood that Machine Learning can play a significant role here, with unsupervised and semi-supervised learning and a wide range of algorithms becoming invaluable assets.
Another very promising avenue is presented by CERN's open data policy, which provides a testing ground for new search strategies with a quick turnaround. 

We discussed the challenges for open data in particle physics, including preservation and access of data and software, documentation, and validation mechanisms. 
The CERN open data policy and the independent FAIR principles are meant to ascertain that open data will be useful to the community.
A specific example of open data is given by the CMS collaboration, which releases reconstructed data after six years to be used widely for analyses and is so far the only collaboration sharing its data in this way.
CMS open data can be accessed via the CERN Open Data Portal, the CMS Open Data Guide, and two workshops run by the CMS open data team.
We summarised CMS users' and external users' recommendations on data complexity and gave an overview over the challenges and measures taken by the CMS open data team to address them.

We conclude that wide access to open data by individuals is necessary to fully exploit the potential of the LHC and that, despite the advances in CERN open data, its public usage remains challenging for individuals.
Improvements of data formats, the documentation, and availability of computing resources, are required to enable the community.
We find that individuals using public data for their own research does not imply competition with experimental efforts, but rather provides unique opportunities to give guidance for further NP searches by the collaborations. 
The communication between theorists and experimentalists is paramount, possibly now more than ever.

\subsection*{Acknowledgements}
We thank S. Kraml for useful comments. SK is supported by the Austrian Science Fund Elise-Richter grant project number V592-N27. 
ND acknowledges the support of Department of Science and Technology of the Government of India via the Ramanujan Fellowship SB/S2/RJN-070/2018.
BB is supported by the ERC research grant NEO-NAT no. 669668.
ZB is supported in part by the MIUR grant PRIN 2017X7X85K 
and in part by the SRNSF grant DI-18-335. 
TH is supported in part by the U.S.~Department of Energy under grant No.~DE-FG02-95ER40896. 
%...
KC is supported in part by Taiwan Ministry of Sciences and Technology with grant number MoST-110-2112-M-007-017-MY3.
JT is supported by the National Science Foundation under Cooperative Agreement PHY-2019786 (The NSF AI Institute for Artificial Intelligence and Fundamental Interactions, http://iaifi.org/), and by the U.S. DOE Office of High Energy Physics under grant number DE-SC0012567.
A.C. and C.A.M. acknowledge financial support by the Swiss National Science Foundation, Project No.\ PP00P2\_176884.
M.H.\ is supported by the Swiss National Science Foundation, Project No.\ PCEFP2\_181117.
MB is supported by the Deutsche Forschungsgemeinschaft (DFG, German Research Foundation) under grant  396021762 -- TRR 257.
B.C. is supported by the Italian Ministry of Research (MIUR) under the Grant No. PRIN 20172LNEEZ.
A.P. is supported by the Spanish Government and ERDF funds from the EU Commission [grant FPA2017-84445-P] and by the Generalitat Valenciana [grant Prometeo/2017/053].
BM and XR are grateful for support from the South African Department of Science and Innovation through the SA-CERN programme and the National Research Foundation for various forms of support.
MK was supported by MIUR (Italy) under a contract PRIN 2015P5SBHT and by INFN Sezione di Roma La Sapienza and partially supported by the ERC-2010 DaMESyFla Grant Agreement Number: 267985.
Contribution by MB is based upon work supported by the National Science Foundation under Grant No. PHY-1913923.
DM acknowledges support by MIUR grant PRIN 2017L5W2PT and the INFN grant SESAMO.
The work of BD is supported in part by the U.S. Department of Energy under Grant No. DE-SC0017987.

\bibliographystyle{utphysmod}
\bibliography{main.bib}

\end{document}